\documentclass[a4paper,11pt]{article}
\usepackage[latin1]{inputenc}
\usepackage{graphicx}
\usepackage{indentfirst}
\usepackage{amsmath}
\usepackage{amssymb}
\usepackage{amsfonts}
\usepackage{latexsym}
\usepackage{epsfig}
\begin{document}
\title{\bf Seismic motion in urban sites consisting  of blocks in welded contact with a soft
layer overlying a hard half space: II. large and infinite number
of identical equispaced blocks\rm\\}
\author{Jean-Philippe Groby\thanks{CMAP, UMR 7641 CNRS/\'Ecole Polytechnique,
91128 Palaiseau cedex,
 France ({\tt groby@cmapx.polytechnique.fr})} and Armand Wirgin\thanks{LMA, UPR 7051
 CNRS, 31 chemin Joseph Aiguier, 13402
Marseille cedex 20, France, ({\tt wirgin@lma.cnrs-mrs.fr})} }
\date{\today}
\maketitle

\begin{abstract}
We address the problem of the response to a seismic wave of an
urban site consisting of  a large and infinite number ($N_{b}$) of
identical, equispaced blocks overlying a soft layer underlain by a
hard substratum. The results of the theoretical analysis,
appealing to a space-frequency mode-matching (MM) technique, are
compared to those
 obtained by a space-time  finite element
(FE) method. The two methods are shown to give rise to much the
same prediction of  seismic response for $N_{b}=10$. The MM
technique is also applied to the case $N_{b}=\infty$, notably to
reveal the structure and natural frequencies of the vibration
modes of the urban site.  The mechanism of the interaction between
blocks and the ground, as well as that of the collective effects
of the blocks, are studied. It is shown that the presence of a
large number of blocks modifies the seismic disturbance in a
manner which evokes, and may partially account for,  what was
observed during many
 earthquakes in Mexico City. Disturbances at a
much smaller level, induced by a small number of blocks  are
studied in the companion paper.
\end{abstract}
\newline
\newline
Abbreviated title: Seismic response in  quasi-periodic and
periodic urban sites
\newline
\newline
Corresponding author: Armand Wirgin, tel.: 33 4 91 16 40 50, fax:
33 4 91 16 42 70, e-mail: wirgin@lma.cnrs-mrs.fr
\newpage
\tableofcontents
\newpage
\section{Introduction}\label{intro}
The Michoacan earthquake that struck Mexico City in 1985 presented
some particular characteristics which have since been encountered
at the same site and at various other urban sites
\cite{segu03,iwata,maeda,haghshenas}, but usually at a smaller
scale. Other than the fact that the response in downtown Mexico
varied considerably in a spatial sense \cite{flno87,sior93}, was
quite intense and of very long duration at certain locations (as
much as $\approx$3min \cite{Perezrocha,sior93,fuke98}), and often
took the form of a quasi-monochromatic signal with beatings
\cite{mateos,sior93}, a remarkable feature of this earthquake
(studied in
\cite{fuke98,Cardenassato,grobyetwirgin2005,grobyetwirgin2005II})
was that such strong motion could be caused by a seismic source so
far from the city (the epicenter was located in the subduction
zone off the Pacific coast, approximately 350km from Mexico City).
It is now recognized \cite{Cardenassato,cardenas} that the
characteristics of the abnormal response recorded in downtown
Mexico were partially present in the waves entering into the
 city (notably $60$km from the city as recorded by the authors of
\cite{fuke98}) after having accomplished their voyage
 from the source, this being thought to be due to the excitation
 of Love and generalized-Rayleigh modes by the irregularities of the
 crust
\cite{Cardenassato,chavezgarciasalazar,fuke98}).

In the present investigation (as well as in the companion paper),
we focus on the  the presence of the built features of the urban
site as a complementary explanation of the abnormal response. In
the companion paper, we treat the case of one or two (different or
identical) built features in the form of cylindrical blocks.
Herein, we study the case of many (10, 20, 40,...., $\infty$)
identical blocks. Such a configuration is a more realistic
representation of a real city due to the large number of blocks it
incorporates, but the assumptions that the blocks are: i)
cylindrical, ii) identical and iii)  periodically arranged, are
rather far removed from reality, except in restricted portions of
modern cities and megacities. Nevertheless, these assumptions are
not more unrealistic than random dispositions
 of blocks with random  sizes \cite{clau01,tswi03,grts05}
and compositions, and have the advantage of enabling a theoretical
analysis which can shed some light on the physical origins of the
above-mentioned exotic phenomena.

The periodic model of  cities built on sites with a soft layer
overlying a hard substratum solicited by seismic waves originated
in the work of  Wirgin and Bard \cite{wiba96}. Unfortunately, the
theoretical apparatus underlying the numerical results was not
given in this paper (only references were made to studies that
treat this issue, but as most of these references were relative to
problems of electromagnetic waves, they have escaped the attention
of the civil engineering and geophysical communities) and the
distance between blocks in the computational results was taken to
be 2000m, which, in our present opinion, is unrealistically large
(unless the blocks represent skyscrapers, of which there are
usually few, and largely-distant one from the other).

The object of the present investigation is thus twofold: i) give
the missing theoretical foundations of the results of
\cite{wiba96}, and ii) treat cities that are more  realistic than
those in \cite{wiba96} and in the companion paper. More
specifically, we shall address the following questions:

(i) how should one account for the principal features of the
seismic response in the cases of a relatively-large number of
blocks (the case of a relatively small number of blocks being
treated in the companion paper)?

(ii) what are the vibrational modes of the global structures (i.e.
the superstructure plus the geophysical structure) and what are
the mechanisms of their excitation and interaction?

(iii) what are the repercussions of resonant phenomena on the
seismic response?

(iv) what are the differences in seismic response between
configurations with a small and a large number of blocks?
\clearpage
\section{Candidate sites}
\begin{figure}[ptb]\begin{center}
\includegraphics[scale=0.35] {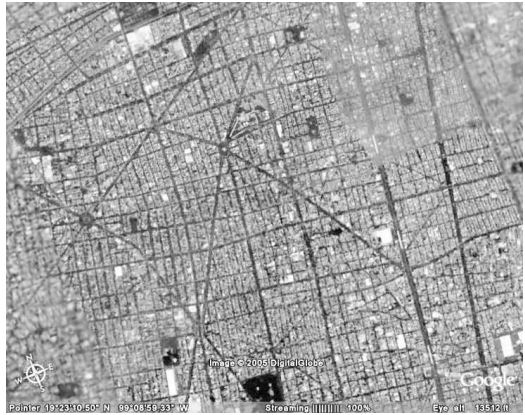}
 \caption{Satellite picture of a portion of Mexico City,
   Mexico.}
   \label{figsite3}
  \end{center}
  \end{figure}

Many  earthquake-prone cities and megacities (New Delhi, Tokyo,
Mexico City, Istanbul, San Francisco, Basel, etc.) are built on
soft soil underlain by a hard substratum and contain districts
with periodic, or nearly-periodic arrangements of blocks or
buildings. An attempt will be made in the present investigation to
analyze the seismic response of districts of this type in the two
cities, Mexico City and Nice (satellite pictures of which are
depicted in figs. \ref{figsite3} and \ref{figsite11}
respectively).

\begin{figure}[ptb]\begin{center}
\includegraphics[scale=0.35] {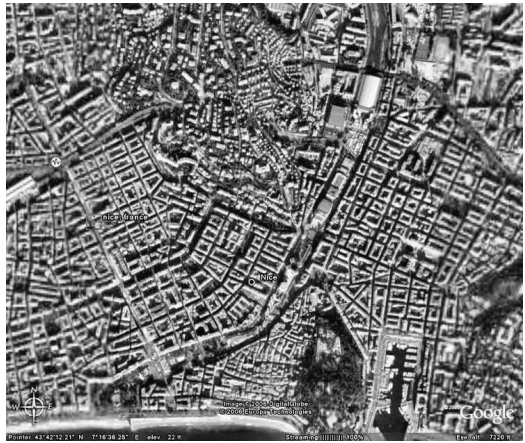}
\caption{Satellite picture of  a portion of Nice, France.}
\end{center}\end{figure}
  \label{figsite11}

\section{Description of the configurations}\label{config}
\begin{figure}[ptb]
\begin{center}
\includegraphics[scale=0.4] {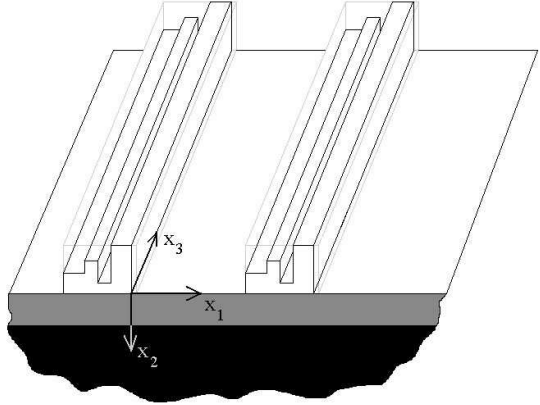}
\end{center}
\caption{View of the 2D city (only two of the blocks are represented).}
\label{fig1}
\end{figure}

\begin{figure}[ptb]
\begin{center}
\includegraphics[scale=0.4] {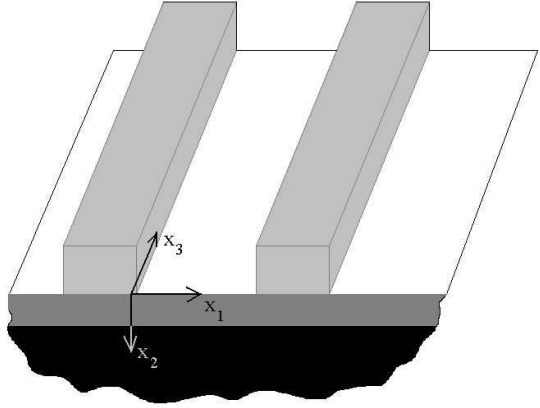}
\caption{View of the 2D city with homogenized blocks (only two of
the blocks are represented).} \label{fig2}
\end{center}
\end{figure}

\begin{figure}[ptb]
\begin{center}
\includegraphics[scale=0.4] {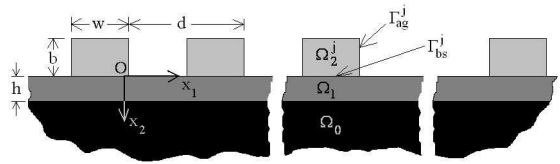}
\caption{Sagittal plane view of the 2D city with a set of identically,
equally-spaced, homogenized blocks.}
\label{fig3}
\end{center}
\end{figure}

We focus on a portion of a city characterized by a periodic or
quasi-periodic set of identical blocks, assumed to have 2D
geometry, with $x_{3}$ the ignorable coordinate of a
$Ox_{1}x_{2}x_{3}$ cartesian coordinate system (see fig.
\ref{fig1}).
 The blocks are in welded contact with the ground underneath
which is located a horizontal soft layer underlain by a hard half
space (see fig. \ref{fig1}).  Each block is composed of one or
more buildings, all the blocks have the same shape (rectangular in
the cross-section plane), are of the same size (characterized by
two constants,  their height $b$ and width $w$) and composition,
and their separation (center-to-center spacing) is a constant $d$.
For the purpose of the analysis, each block is homogenized, so
that the final geometry of the city is as represented in  fig.
\ref{fig2}.
Fig. \ref{fig3} represents a cross-section (sagittal plane) view
of the city.
$\Gamma_{f}$ is the stress-free surface composed of a
ground portion $\Gamma_{g}$, assumed to be flat and horizontal,
and a portion $\Gamma_{ag}$, constituting the reunion of the
above-ground-level boundaries $\Gamma_{ag}^{j}~;~j\in \mathbb{Z}$
of the blocks. The ground $\Gamma_{G}$ is flat and horizontal, and
is the reunion of $\Gamma_{g}$ and the base segments
$\Gamma_{bs}^{j}~;~j\in \mathbb{Z}$ joining the blocks to the
underground.

The medium in contact with, and above, $\Gamma_{f}$ is air,
assumed to be the vacumn (i.e., $\Gamma_{f}$ is stress-free). The
medium in contact with, and below $\Gamma_{G}$ is the
mechanically-soft layer occupying the domain $\Omega_{1}$, which
is laterally-infinite and of thickness $h$, and whose lower
boundary is $\Gamma_{h}$, assumed to be flat and horizontal. The
soft material in the layer is in welded contact across
$\Gamma_{h}$ with the mechanically-hard material in the
semi-infinite domain (substratum) $\Omega_{0}$.

The domain of the $j$-th block is denoted by $\Omega_{2}^{j}$ and
the reunion of all the $\Omega_{2}^{j}~;~j\in \mathbb{B}$ is
denoted by $\Omega_{2}$. The material in each block is in welded
contact with the material in the soft layer across the base
segments $\Gamma_{bs}^{j}~;~j\in \mathbb{Z}$.

The origin $O$ of the cartesian coordinate system is on the
ground, $x_{2}$ increases with depth and $x_{3}$ is perpendicular
to the (sagittal) plane in figs. \ref{fig1}-\ref{fig3}. With
$\mathbf{i}_{j}$ the unit vector along the positive $x_{j}$ axis,
we note that the unit vectors normal to $\Gamma_{G}$ and
$\Gamma_{h}$ are $-\mathbf{i}_{2}$.

The media filling $\Omega_{0}$, $\Omega_{1}$ and $\Omega_{2}^{j}$
are $M^{0}$, $M^{1}$ and $M^{2j}=M^{2}~;~\forall j\in\mathbb{B}$
respectively and the latter are assumed to be initially
stress-free, linear, isotropic and homogeneous (thus, each block,
which is generally inhomogeneous, is assumed to be homogenized in
our analysis). We assume that $M^{0}$ is non-dissipative whereas
$M^{1}$ and $M^{2}$ are dissipative.

The seismic disturbance is delivered to the site in the form of a
shear-horizontal (SH)  plane pulse wave, initially
propagating in $\Omega_{0}$.
The SH nature of the {\it incident wave} (which fact is indicated
by the superscript $i$ in the following) means that the motion
associated with it is strictly transverse (i.e., in the $x_{3}$
direction and independent of the $x_{3}$ coordinate). Both the SH
polarization and the invariance of the incident wave with respect
to $x_{3}$ are communicated to the fields that are generated at
the site in response to the incident wave. Thus, our analysis will
deal only with the propagation of 2D SH waves (i.e., waves that
depend exclusively on the two cartesian coordinates $x_{1},~x_{2}$
and that are associated with motion in the $x_{3}$ direction
only).

We shall be concerned with a description of the elastodynamic
wavefield on the free surface (i.e., on $\Gamma_{f}$) resulting
from plane seismic wave solicitation of the site.
\section{Governing equations}\label{goveqs}
\subsection{Space-time framework wave equations}\label{xtframe}
In a generally-inhomogeneous, isotropic elastic or viscoelastic
medium $M$ occupying $\mathbb{R}^{3}$, the space-time framework
wave equation for SH waves is:
\begin{equation}\label{w32.7}
 \nabla\cdot (\mu(\mathbf{x},\omega)\nabla u(\mathbf{x},t))-
 \rho(\mathbf{x})\partial_{t}^{2}u(\mathbf{x},t)=-\rho(\mathbf{x})
 f(\mathbf{x},t) ~,
\end{equation}
wherein $u$ is the displacement component in the $\mathbf{i}_{3}$
direction,  $f$ the component of applied force density in the
$\mathbf{i}_{3}$ direction, $\mu$ the Lam\'e descriptor of
rigidity, $\rho$ the mass density, $t$ the time variable, $\omega$
the angular frequency, $\partial^{n}_{t}$ the $n-$th partial
derivative with respect to $t$, and $\mathbf{x}=(x_{1},x_{2})$.
Since our configuration involves three homogeneous media and the
solicitation takes the form of a plane  wave initially propagating
in $\Omega_0$ (thus, the source density is nil for this type of
wave), we have
\begin{equation}\label{w33.1}
 \left ( c^{m}(\omega)\right ) ^{2}\nabla\cdot\nabla u^{m}(\mathbf{x},t)-
 \partial_{t}^{2}u^{m}(\mathbf{x},t)=0
 ~~;~~\mathbf{x}\in\Omega_{m}~;~m=0,1,2j~~,~j\in\mathbb{B}~,
\end{equation}
wherein $m$ superscripts designate the medium (0 for $M^{0}$,
etc.), and
$c^{m}$ is the generally-complex velocity of shear body waves in
$M^{m}$, related to the density and rigidity by
\begin{equation}\label{w33.2}
  \left ( c^{m}(\omega)\right ) ^{2}=\frac{\mu^{m}(\omega)}{\rho^{m}}~,
\end{equation}
it being understood that
$\rho^{m},\mu^{m}(\omega)~;~m=0,1,2j,~j\in\mathbb{B}$ are
constants with respect to $\mathbf{x}$. In addition, the densities
are positive real, and we assume that the substratum is a
dissipation-free solid so that the rigidity therein is a positive
real constant with respect to $\omega$, i.e.,
$\mu^{0}(\omega)=\mu^{0}>0$.
\subsection{Space-frequency framework wave
equations}\label{xomegaframe}
The space-frequency framework versions of the wave equations
(\ref{w33.1}) are obtained by expanding the displacements in
Fourier integrals:
\begin{equation}\label{w33.3}
 u^{m}(\mathbf{x},t)=\int_{-\infty}^{\infty}
  u^{m}(\mathbf{x},\omega)e^{-\mbox{i}\omega t}d\omega~,\forall t\in
  \mathbb{R}~,
\end{equation}
(wherein $\mbox{i}=\sqrt{-1}$, $\omega$ the angular frequency and
$t$ the time) so as to give rise to the Helmholtz equations
\begin{multline}\label{w34.1}
 \nabla\cdot\nabla
  u^{m}(\mathbf{x},\omega)+\left ( k^{m}(\omega)\right ) ^{2}
  u^{m}(\mathbf{x},\omega)=0~~;~~\forall\mathbf{x}\in
  \Omega_{m}~~;~~m=0,1~,2j,~j\in\mathbb{B}~,
\end{multline}
wherein
\begin{equation}\label{w34.2}
  k^{m}(\omega):=\frac{\omega}{c^{m}(\omega)}=
  \omega\sqrt{\frac{\rho^{m}}{\mu^{m}(\omega)}}~.
\end{equation}
is the generally-complex wavenumber in $M^{m}$.

We shall deal with constant quality factors $Q^{m}~;~m=1,2$ over
the frequency range of solicitation, so that the
frequency-dependent rigidities take the form given in \cite{kj79},
\begin{equation}\label{dissmat.16}
 \mu^{m}(\omega)=\mu^{m}_{ref} \left( \frac{-\mbox{i} \omega}
 {\omega_{ref}} \right) ^{\frac{2}{\pi}
 \arctan\left ( \frac{1}{Q^{m}}\right ) }~;~m=1,2~,
\end{equation}
wherein: $\omega_{ref}$ is a reference angular frequency, chosen
herein to be equal to $9\times 10^{-2}$Hz. Hence
\begin{equation}\label{dissmat.17}
 c^{m}(\omega)=c^{m}_{ref} \left( \frac{-\mbox{i} \omega}
 {\omega_{ref}} \right) ^{\frac{1}{\pi}
 \arctan\left ( \frac{1}{Q^{m}}\right ) }~;~m=1,2~,
\end{equation}
with $c^{m}_{ref}:=\sqrt\frac{\mu^{m}_{ref}}{\rho^{m}}$. Even
though $Q^{m}~,~m=1,2$ are non-dispersive (i.e., do not depend on
$\omega$) under the present assumption, the phase velocities
$c^{m}~;~m=1,2$ {\it are} dispersive.

Due to the assumptions made in sect. \ref{config}:
\begin{equation}\label{w34.3}
  k^{0}(\omega):=\frac{\omega}{c^{0}}=
  \omega\sqrt{\frac{\rho^{0}}{\mu^{0}}}~,
\end{equation}
(i.e., $Q^{0}=\infty$ so that $k^{0}$ is real).
\subsection{Space-frequency and space-time framework expressions of
the incident plane wave}\label{incwave}
As mentioned above, we shall be concerned with plane wave excitation of the city.

Actually, a plane wave satisfies the homogeneous wave equation in
the space-time framework and a homogeneous Helmholtz equation in
the space-frequency framework.

The field is chosen to take the form of a pseudo Ricker-type pulse
in the space-time framework, whose shape is  directly connected
with the site we will consider in the numerical application
(either a Nice-like site, or a Mexico city -like site) .
\subsubsection{Space-frequency-framework representation of the plane,
impulsive, incident wave}\label{xomegaimpuls}
The plane wave nature of the incident wave is embodied in the
choice
\begin{equation}\label{1.6}
   u^{i}(\mathbf{x},\omega)=S(\omega)\exp\left [\mbox{i}k^{0}
   \left ( x_{1}s^{i}-x_{2}c^{i}\right)
   \right]~,
\end{equation}
wherein  $s^{i}=\sin \theta^{i}$, $c^{i}=\cos \theta^{i}$ and
$\theta^{i}$ is the angle of incidence in the $x_{1}-x_{2}$ plane
with respect to the $x_{2}$ axis.

The fact that the incident wave is a pseudo Ricker pulse means
that the amplitude spectrum $S(\omega)$ is given by
\begin{equation}\label{1.7}
   S(\omega)=\frac{1}{\sqrt{\pi}}\frac{\omega^{2}}{4\alpha^{3}}
   \exp\left (\mbox{i}t_{s}\omega-\frac{\omega^{2}}{4\alpha^{2}}\right) ~,
\end{equation}
to which corresponds the temporal variation (Fourier inverse of
$S(\omega)$):
\begin{equation}\label{1.8}
   S(t)=-\left[ 2\alpha^{2}(t_{s}-t)^{2}-1\right]
   \exp\left [ -\alpha^{2}(t_{s}-t)^{2}\right] ~,
\end{equation}
for Nice-like site solicitation and by
\begin{equation}\label{1.7}
   S(\omega)=-\frac{2 \alpha^2}{\sqrt{\pi}}\frac{\omega^{2}}{4\alpha^{3}}
   \exp\left (\mbox{i}t_{s}\omega-\frac{\omega^{2}}{4\alpha^{2}}\right) ~,
\end{equation}
to which corresponds the temporal variation (Fourier inverse of
$S(\omega)$):
\begin{equation}\label{1.8}
   S(t)=2\alpha^{2}\left[ 2\alpha^{2}(t_{s}-t)^{2}-1\right]
   \exp\left [ -\alpha^{2}(t_{s}-t)^{2}\right] ~,
\end{equation}
for the Mexico-like site solicitation.

In both cases, $\alpha=\pi/t_{p}$, $t_{p}$ is the characteristic
period of the pulse, and $t_{s}$ the time at which the pulse
attains its maximal value. In particular we will chose
$t_p=t_s=\frac{1}{\nu_0}$, where $\nu_0$ is
 the central frequency (in Hz) of the spectrum of the pulse.
\subsection{Boundary and radiation conditions in the
space-time framework}\label{xtboundrad}
Since our finite element  method \cite{grobyettsogka,grts05} for
solving the wave equation in a heterogeneous medium $M$ (in our
case, involving three homogeneous components, $M^{0}$, $M^{1}$ and
$M^{2}$) relies on the assumption that $M$ is a continuum, it does
not appeal to any boundary conditions except on $\Gamma_{f}$ where
the vanishing traction condition is invoked.  The latter is
modeled with the help of the fictitious domain method
\cite{becachejolyettsogka}, which allows us to model diffraction
of waves by a boundary of complicated geometry, not necessarily
matching the volumic mesh. Furthermore, since the essentially
unbounded nature of the geometry of the city cannot be implemented
numerically, we take the geometry to be finite and surround it
(except on the $\Gamma_{f}$ portion) by a PML perfectly-matched
layer \cite{collinoettsogka} which enables closure of the
computational domain without generating unphysical reflected waves
(from the PML layer). In a sense, this replaces the radiation
condition of the unbounded domain.
\subsection{Boundary and radiation conditions in the
space-frequency domain}\label{boundrad}
The translation of the stress-free (i.e., vanishing traction)
nature of $\Gamma_{f}=\Gamma_{g}\bigcup\Gamma_{ag}$, with
$\Gamma_{ag}:=\bigcup_{j\in\mathbb{B}}\Gamma_{ag}^{j}$, is:
\begin{equation}\label{boundrad.1}
\mu^{1}(\omega)\partial_{n}
u^{1}(\mathbf{x},\omega)=0~;~\mathbf{x} \in \Gamma_{g},
\end{equation}
\begin{equation}\label{boundrad.2}
\mu^{2}(\omega)\partial_{n}
u^{2j}(\mathbf{x},\omega)=0~;~\mathbf{x} \in
\Gamma_{ag}^{j}~,~j\in\mathbb{B}~,
\end{equation}
wherein $\mathbf{n}$ denotes the generic unit vector normal to a
boundary and $\partial_{n}$ designates the operator
$\partial_{n}=\mathbf{n}\cdot\nabla$.

Since $M^{1}$ and $M^{2}$ are in welded contact across
$\Gamma_{bs}:=\bigcup_{j\in\mathbb{B}}\Gamma_{bs}^{j}$, the
displacement and traction are continuous across $\Gamma_{bs}$:
\begin{equation}\label{boundrad.3}
u^{1}(\mathbf{x},\omega)-u^{2j}(\mathbf{x},\omega)=0~;~\mathbf{x}
\in \Gamma_{bs}^{j}~,~j\in\mathbb{B}~,
\end{equation}
\begin{equation}\label{boundrad.4}
\mu^{1}(\omega)\partial_{n}
u^{1}(\mathbf{x},\omega)-\mu^{2}(\omega)\partial_{n}
u^{2j}(\mathbf{x},\omega)=0~;~\mathbf{x} \in
\Gamma_{bs}^{j}~,~j\in\mathbb{B}~.
\end{equation}
Since $M^{1}$ and $M^{0}$ are in welded contact across
$\Gamma_{h}$, the displacement and traction are continuous across
this interface:
\begin{equation}\label{boundrad.5}
u^{1}(\mathbf{x},\omega)-u^{0}(\mathbf{x},\omega)~;~\mathbf{x} \in
\Gamma_{h}~,
\end{equation}
\begin{equation}\label{boundrad.6}
\mu^{1}(\omega)\partial_{n}
u^{1}(\mathbf{x},\omega)-\mu^{0}(\omega)\partial_{n}
u^{0}(\mathbf{x},\omega)~;~\mathbf{x} \in \Gamma_{h}.
\end{equation}
The uniqueness of the solution to the forward-scattering problem
is assured by the radiation condition in the substratum:
\begin{equation}\label{boundrad.7}
u^{0}(\mathbf{x},\omega)-u^{i}(\mathbf{x},\omega)\sim ~ \text{
outgoing waves}~~;~~ \|\mathbf{x}\|\rightarrow \infty,~~x_{2}>h~.
\end{equation}
\subsection{Recovery of the  space-frequency displacements from the space-time
displacements}\label{xomegaboundvalpb}
The spectra of the displacements are obtained from the time
records of the displacements by Fourier inversion, i.e.,
\begin{equation}\label{xomegaboundvalpb.1}
 u^{m}(\mathbf{x},\omega)= \frac{1}{2\pi}\int_{-\infty}^{\infty}u^{m}(\mathbf{x},t)
 e^{\mbox{i}\omega t}
  dt~;~j=0,1,2j,~j\in\mathbb{B}~.
\end{equation}
\section{Field representations in the space-frequency framework
for $N_{b}=\infty$}

Since the formulation for the case  of $N_{b}<\infty$ identical
(or non-identical) blocks was given in the companion paper, it
will not be repeated here.

The new feature in the present investigation is the possibility of
the existence of an infinite number (i.e., $N_{b}=\infty$) of
identical (in shape, dimensions and composition) blocks, separated
(horizontally) one from the other by the constant distance $d$
(called the period).

 At present, the blocks are identified by indices in the set
$\mathbb{B}=\{...,-1,$ $0,1,...\}=\mathbb{Z}$, with the
understanding that the center of the segment $\Gamma_{bs}^{0}$ is
at the origin.

Owing to the plane wave nature of the incident wave, the periodic
nature of $\Gamma_{G}$, and the fact that the blocks are assumed
to be identical in height, width, and composition, one can show
that the field is quasi-periodic (this constituting the so-called
{\it Floquet relation}, i.e.,
\begin{equation}\label{w2.5.31}
  u^{j}(x_{1}+nd,x_{2})=
   u^{j}(x_{1},x_{2})e^{\mbox{i}k_{1}^{i}nd}~;~\forall
  \mathbf{x}\in\Omega_{j}~;~\forall n\in\mathbb{Z}~;~j=0,1~.
\end{equation}
wherein $k_{1}^{i}:=k^{0}s^{i}$.
\subsection{Field in $\Omega_{0}$}\label{wss2.7}
By referring to the companion paper, and after use of  the Green's
second identity, the field in $\Omega_0$ can be shown to take the
form:
\begin{multline}\label{fieldrepomega0.3}
u^{0}(\mathbf{x},\omega)=
u^{i}(\mathbf{x},\omega)+
\int_{-\infty}^{\infty}B^{0}(k_{1},\omega)
\exp\left\{\mbox{i}\left[k_{1}x_{1}+k_{2}^{0}(x_{2}-h)\right]\right\}
  dk_{1}~;
  \\
  ~\mathbf{x}\in \Omega_{0}~,
\end{multline}
wherein:
\begin{multline}\label{fieldrepomega0.4}
B^{0}(k_{1},\omega)=\frac{i}{4\pi
k_{2}^{0}}\int_{-\infty}^{\infty}
\left\{\partial_{y_{1}}u^{0}(y_{1},h,\omega)+
ik_{2}^{0}u^{0}(y_{1},h,\omega)\right\} \times
\\
\exp\left( -i
k_{1}y_{1}\right) dy_{1}~,
\end{multline}
By a suitable change of variables and use of the Floquet relation
$B^{0}(k_{1})$ can be cast in the form:
\begin{multline}\label{w2.5.32}
B^{0}(k_{1})=\frac{\mbox{i}}{4\pi k_{2}^{0}}\int_{-d/2}^{d/2}
\left\{\partial_{y_{1}}u^{0}(y_{1},h,\omega)+
\mbox{i}k_{2}^{0}u^{0}(y_{1},h,\omega)\right\}\times
\\
\exp\left[-\mbox{i}k_{1}y_{1}\right]dy_{1}
\sum_{n=-\infty}^{\infty}\exp\left[\mbox{i}xnd\left(k_{1}^{i}-
k_{1}\right)\right]~.
\end{multline}
With the help of the Poisson summation formula \cite{MF53} it follows that:
\begin{equation}\label{w2.5.33}
\sum_{n=-\infty}^{\infty}\exp\left[ind\left(k_{1}^{i}-
k_{1}\right)\right]=\frac{2\pi}{d}\sum_{n=-\infty}^{\infty}
\delta\left(k_{1n}-k_{1}\right)~,
\end{equation}
wherein $\delta(~)$ is the Dirac delta distribution and
\begin{equation}\label{w2.5.34}
k_{1n}=k_{1}^{i}+\frac{2n\pi}{d}~.
\end{equation}
On the other hand, $F(x)\delta(x-y)=F(y)\delta(x-y)$ in the sense
of distributions, so that
\begin{equation}\label{w2.5.35}
B^{0}(k_{1})=\sum_{n=-\infty}^{\infty}B^{0}_{n}
\delta\left(k_{1n}-k_{1}\right)~,
\end{equation}
where
\begin{multline}\label{w2.5.36}
B^{0}_{n}=\frac{\mbox{i}}{2k_{2n}^{0}d}\int_{-d/2}^{d/2}
\left\{\partial_{y_{1}}u^{0}(y_{1},h,\omega)+
\mbox{i}k_{2n}^{0}u^{0}(y_{1},h,\omega)\right\}\times
\\
\exp\left( -\mbox{i}k_{1n}y_{1}\right) dy_{1}~,
\end{multline}
and
\begin{equation}\label{w2.6.37}
k_{2n}^{j}=\sqrt{\left ( k^{j}\right ) ^{2}- k_{1n}^{2}}~~;~~\Re
(k_{2n}^{j})\geq 0~~,~~\Im(k_{2n}^{0})\geq 0~~,~~\omega\geq
0~;~j=0,1~.
\end{equation}
The introduction of (\ref{w2.5.35}) into (\ref{fieldrepomega0.3})
and the use of the sifting property of the Dirac delta
distribution result in:
\begin{equation}\label{w2.5.38}
u^{0}(\mathbf{x},\omega)=
u^{i}(\mathbf{x},\omega)+\sum_{n=-\infty}^{\infty}B_{n}^{0}
\exp\left\{\mbox{i}\left[k_{1n}x_{1}+k_{2n}^{0}(x_{2}-h)\right]\right\}
  ~;~\mathbf{x}\in \Omega_{0}~.
\end{equation}
However, we can write
\begin{equation}\label{w2.5.39}
u^{i}(\mathbf{x},\omega)= \sum_{n=-\infty}^{\infty}A_{n}^{i}
\exp\left\{\mbox{i}\left[k_{1n}x_{1}-k_{2n}^{0}x_{2}\right]\right\}
  ~;~\mathbf{x}\in \mathbb{R}^{2}~,
\end{equation}
wherein
\begin{equation}\label{w2.5.40}
A_{n}^{i}=S(\omega)\delta_{n0}~,
\end{equation}
$\delta_{nm}$ is the Kronecker symbol, $k_{10}=k_{1}^{i}$, and
$k_{20}^{j}=\sqrt{\left( k^{j}\right) ^{2}-\left( k_{1}^{i}\right)
^{2}}:=k_{2}^{ji}$, so that
\begin{multline}\label{w2.5.41}
u^{0}(\mathbf{x},\omega)=\sum_{n=-\infty}^{\infty}A_{n}^{i}
\exp\left\{\mbox{i}\left[k_{1n}x_{1}-k_{2n}^{0}x_{2}\right]\right\}+
\\
\sum_{n=-\infty}^{\infty}B_{n}^{0}
\exp\left\{\mbox{i}\left[k_{1n}x_{1}+k_{2n}^{0}(x_{2}-h)\right]\right\}
  ~;~\mathbf{x}\in \Omega_{0}~,
\end{multline}
with the understanding that
$\mathbf{a}^{0}:=\{A_{n}^{0}~;~n\in\mathbb{Z}\}$ is a known vector
and $\mathbf{b}^{0}:=\{B_{n}^{0}~;~n\in\mathbb{Z}\}$ an unknown
vector.
\subsection{Field in $\Omega_{1}$}\label{wss2.8}
By proceeding in the same manner as previously, we find:
\begin{multline}\label{w2.5.42}
u^{1}(\mathbf{x},\omega)=\sum_{n=-\infty}^{\infty}A_{n}^{1}
\exp\left\{\mbox{i}\left[k_{1n}x_{1}-k_{2n}^{1}x_{2}\right]\right\}+
\\
\sum_{n=-\infty}^{\infty}B_{n}^{1}
\exp\left\{\mbox{i}\left[k_{1n}x_{1}+k_{2n}^{1}x_{2}\right]\right\}
  ~;~\mathbf{x}\in \Omega_{1}~,
\end{multline}
with the understanding that both
$\mathbf{a}^{1}:=\{A_{n}^{1}~;~n\in\mathbb{Z}\}$ and
$\mathbf{b}^{1}:=\{B_{n}^{1}~;~n\in\mathbb{Z}\}$ are unknown
vectors.
\subsection{Field in $\Omega_{2}$}\label{wss2.8}
The Floquet relation actually extends to $\Omega_{2}$ wherein it
takes the form
\begin{equation}\label{w2.5.42a}
u^{2j}(x_{1},x_{2},\omega)=u^{20}(x_{1},x_{2},\omega)e^{ik_{1}^{i}jd}
~;~j\in\mathbb{Z}~.
\end{equation}
By referring to the companion paper, the displacement field in the
zeroth-th block takes the form:
\begin{multline}\label{w2.5.47a}
u^{20}(\mathbf{x},\omega):=u^{2}(\mathbf{x},\omega)=
\sum_{m=0}^{\infty}B_{m}^{2}\cos\left[ k^{2}_{1m}\left(
x_{1}+\frac{w}{2}\right) \right] \cos\left[ k^{2}_{2m}\left(
x_{2}+b\right) \right]
\\
~~;~~ \mathbf{x}\in\Omega_{2}^{(0)}~,
\end{multline}
with:
\begin{equation}\label{w2.5.47b}
k_{1m}^{2}=\frac{m\pi}{w}~~,~~k_{2m}^{2}=\sqrt{(k^{2})^{2}-(k_{1m}^{2})^{2}}~~,~~\Re
k_{2m}^{2}\geq 0~~,~~\Im k_{2m}^{2}\geq 0~~,~~\omega\geq 0~,
\end{equation}
it being understood that $\mathbf{b}^{2}:=\{B_{m}^{2}~~;~~n\in
\mathbb{N}\}$ is an unknown vector.
\section{Determination of the various unknown coefficients by
application of the boundary and continuity conditions on
$\Gamma_{G}$ and  $\Gamma_{h }$}
\subsection{Application of the boundary and continuity conditions
concerning the traction on $\Gamma_{G}$}
From (\ref{boundrad.1}) and (\ref{boundrad.4}) we obtain
\begin{multline}\label{w7.1.1}
\mu^{1}\int_{-d/2}^{d/2}\partial_{x_{2}}u^{1}(x_{1},0,\omega)
\exp(-\mbox{i}k_{1l}x_{1})dx_{1}-
\\
\mu^{2}\int_{-w/2}^{w/2}
\partial_{x_{2}}u^{2}(x_{1},0,\omega)
\exp(-\mbox{i}k_{1l}x_{1})dx_{1}=0 ~~;~~ \forall l\in\mathbb{Z}~.
\end{multline}
Introducing the appropriate field representations therein and
making use of the orthogonality relation
\begin{equation}\label{w7.1.2}
\int_{-d/2}^{d/2}\exp[\mbox{i}(k_{1n}-k_{1l})x_{1}]dx_{1}= d\delta_{nl}
~~;~~ \forall n,l\in\mathbb{Z}~,
\end{equation}
gives rise to
\begin{equation}\label{w7.1.3}
A_{n}^{1}-B_{n}^{1}=\frac{w}{\mbox{i}d}
e^{\mbox{i}k_{1n}w/2}\sum_{m=0}^{\infty}B_{m}^{2}
\frac{\mu^{2}k_{2m}^{2}}{\mu^{1}k_{2n}^{1}}
I_{mn}^{-}\sin(k_{2m}^{2}b)
 ~~;~~ \forall
n\in\mathbb{Z}~,
\end{equation}
wherein
\begin{multline}\label{w7.1.4}
I_{mn}^{\pm}=\int_{0}^{1}\exp(\pm
ik_{1n}w\eta)\cos(k_{1m}^{2}w\eta)d\eta=
\\
\frac{i^{m}}{2}e^{\pm\mbox{i}k_{1n}\frac{w}{2}} \left\{
\mbox{sinc}\left[ (k_{1m}^{2}\pm k_{1n})\frac{w}{2}\right]
+(-1)^{m}\mbox{sinc}\left[ (-k_{1m}^{2}\pm
k_{1n})\frac{w}{2}\right] \right\}~.
\end{multline}
\subsection{Application of the continuity condition
concerning the displacement on $\Gamma_{G}$}
From (\ref{boundrad.1})  we obtain
\begin{multline}\label{w7.2.1}
\int_{-\frac{w}{2}}^{\frac{w}{2}}u^{1}(x_{1},0,\omega)\cos\left[
k_{1n}^{2}(x_{1}+w/2)\right] dx_{1}-
\\
\int_{-\frac{w}{2}}^{\frac{w}{2}}u^{2}(x_{1},0,\omega)\cos\left[
k_{1n}^{2}(x_{1}+w/2)\right] dx_{1}=0 ~~;~~ l=0,1,2,...~.
\end{multline}
Introducing the appropriate field representations therein, and
making use of the orthogonality relation
\begin{multline}\label{w7.2.2}
\int_{-\frac{w}{2}}^{\frac{w}{2}}\cos\left[
k_{1m}^{2}(x_{1}+w/2)\right] \cos\left[
k_{1n}^{2}(x_{1}+w/2)\right]
dx_{1}=\frac{w}{\epsilon_{m}}\delta_{mn} \\
 ~~;~~ \forall
m~,~n=0,1,2,....~,
\end{multline}
gives rise to
\begin{equation}\label{w7.2.3}
B_{m}^{2}=\frac{\epsilon_{m}}{\cos(k_{2m}^{2}b)}\sum_{n\in\mathbb{Z}}
\left[ A_{n}^{1}+B_{n}^{1}\right] I_{mn}e^{-\mbox{i}k_{1n}w/2}
 ~~;~~
\forall m=0,1,2,....~.
\end{equation}
\subsection{Application of the  continuity conditions
concerning the traction on $\Gamma_{h}$}
From (\ref{boundrad.6})  we obtain
\begin{multline}\label{w7.3.1}
\mu^{0}\int_{-d/2}^{d/2}\partial_{x_{2}}u^{0}(x_{1},h,\omega)
\exp(-\mbox{i}k_{1l}x_{1})dx_{1}-
\\
\mu^{1}\int_{-d/2}^{d/2}\partial_{x_{2}}u^{1}(x_{1},h,\omega)
\exp(-\mbox{i}k_{1l}x_{1})dx_{1}=0 ~~;~~ \forall l\in\mathbb{Z}~.
\end{multline}
Introducing the appropriate field representations therein, and
making use of the orthogonality relation (\ref{w7.1.2}), gives
rise to
\begin{multline}\label{w7.3.2}
-\mu^{0}k_{2n}^{0}A_{n}^{0}e^{-\mbox{i}k_{2n}^{0}h}+
\mu^{0}k_{2n}^{0}B_{n}^{0}+
\mu^{1}k_{2n}^{1}A_{n}^{1}e^{-\mbox{i}k_{2n}^{1}h}-
\mu^{1}k_{2n}^{1}B_{n}^{1}e^{\mbox{i}k_{2n}^{1}h}=0 ~~;
\\~~
\forall n\in\mathbb{Z}~.
\end{multline}
\subsection{Application of  the continuity condition
concerning the displacement on $\Gamma_{h}$}
From (\ref{boundrad.5})  we obtain
\begin{multline}\label{w7.4.1}
\int_{-d/2}^{d/2}u^{0}(x_{1},h,\omega)\exp(-\mbox{i}k_{1l}x_{1})dx_{1}-
\\
\int_{-d/2}^{d/2}u^{1}(x_{1},h,\omega)\exp(-\mbox{i}k_{1l}x_{1})dx_{1}=0
~~;~~ \forall l\in\mathbb{Z}~.
\end{multline}
Introducing the appropriate field representations therein, and
making use of the orthogonality relation (\ref{w7.1.2}), gives
rise to
\begin{equation}\label{w7.4.2}
A_{n}^{0}e^{-\mbox{i}k_{2n}^{0}h}+B_{n}^{0}-
A_{n}^{1}e^{-\mbox{i}k_{2n}^{1}h}-B_{n}^{1}e^{\mbox{i}k_{2n}^{1}h}=
0 ~~;~~
\forall n\in\mathbb{Z}~.
\end{equation}
\subsection{Determination of the various unknowns}\label{det}
\subsubsection{Elimination of $B_{m}^{2}$ to obtain a linear
system of equations
for $B_{n}^{0}$.}\label{b0}
After a series of substitutions, the following matrix equation  is
obtained  for $B_{n}^{0}$:
\begin{equation}\label{bOeq}
C_{n}^{10}B_{n}^{0}-\sum_{m=-\infty}^{\infty}D_{nm}B_{m}^{0}=F_{n}~~;~~\forall
n \in \mathbb{Z}~,
\end{equation}
wherein:
\begin{equation}
C_{n}^{jl}:=\cos\left(k_{2n}^{1} h \right)-\mbox{i}
\frac{\mu^{j}k_{2n}^{j}}{\mu^{l}k_{2n}^{l}}\sin\left(k_{2n}^{1} h
\right)~~; ~~\forall n \in \mathbb{Z}~,~j,l=0,1~,
\end{equation}
\begin{equation}
D_{nm}:=\sum_{l=0}^{\infty}\frac{\mbox{i}\epsilon_{l}\mu_2
k_{2l}^{2}w}{\mu^{0} k_{2n}^{0}d} I_{ln}^{-}I_{lm}^{+}
\tan\left(k_{2l}^{2} b\right)C_{m}^{01}
e^{\mbox{i}\left(k_{1n}-k_{1m} \right)w/2}~,
\end{equation}
and
\begin{multline}
F_{n}=S(\omega)e^{-\mbox{i}
k_{2n}^{0}h}C_{n}^{10}\delta_{n0}+
\\
\sum_{l=0}^{\infty}\frac{\mbox{i}S(\omega)e^{-\mbox{i}k_{20}^{0}h}
w \epsilon_l \mu^{2} k_{2l}^2 \sin\left(k_{2l}^2 b \right)
I_{ln}^{-} I_{l0}^{+}}{\mu^{0} k_{2n}^{0} d}C_{0}^{01}
e^{\mbox{i}\left(k_{1n}-k_1^{i} \right)w/2}~.
\end{multline}
Eq. (\ref{bOeq}) is a matrix equation, which, in principle,
enables
 the determination of the vector
$\mathbf{b}^{0}:=\{B_{n}^{0}~;~n=0,\pm 1,\pm 2,...\}$.
\subsubsection{Elimination of $B_{n}^{0}$ to obtain a linear system
of equations for $B_{m}^{2}$}
The procedure is again to make a series of substitutions which now
leads to the linear system for $B_{l}^{2}~;~ \forall l\in
\mathbb{N}$:
\begin{equation}\label{bm2eq1}
B_{l}^{2}=P_{l}+\sum_{m=0}^{\infty}Q_{ml}B_{m}^{2}~~;~~\forall l
\in \mathbb{N}~,
\end{equation}
wherein
\begin{equation}
Q_{ml}=\sum_{n=-\infty}^{\infty}\frac{\mbox{i}\epsilon_{l}w
k_{2m}^{2} \mu^{2} \sin\left(k_{2m}^{2}b \right)
I_{mn}^{-}I_{ln}^{+}C_{n}^{01}} {\mu^{0}k_{n2}^{0}d
\cos\left(k_{2l}^{2}b \right) C_{n}^{10}}~,
 \label{Qml}
\end{equation}
and
\begin{equation}
\begin{array}{ll}
\displaystyle P_{l}&\displaystyle
=\sum_{n=-\infty}^{\infty}A_{n}^{i}\frac{2\epsilon_{l}
\exp\left(-\mbox{i}k_{n2}^{0}h\right)\exp\left(-\mbox{i}k_{n1}
\frac{w}{2}\right)I_{ln}^{+} } {\cos\left(k_{2l}^{2}b
\right)C_{n}^{10}}\\[12pt] \displaystyle &\displaystyle
=S(\omega)\frac{2\epsilon_{l}\exp\left(-\mbox{i}k_{2}^{i}h\right)
\exp\left(-\mbox{i}k_{1}^{i}\frac{w}{2}\right)I_{l0}^{+}
}{\cos\left(k_{2l}^{2}b \right) C_{0}^{10}}~.\\[8pt]
\end{array}
\label{Pl}
\end{equation}
Eq. (\ref{bm2eq1}) is a matrix equation, which, in principle,
enables  the determination of the vector
$\mathbf{b}^{2}:=\{B_{m}^{2}~;~m=0,1,2,...\}$.
\section{Modal Analysis}
\subsection{General considerations}
At this point, it is important to recall that the ultimate goal of
this investigation is to predict the response of an urban site to
a seismic wave. This response takes the form of the displacement
field at various locations on the ground as a function of time.
Thus, the field quantities of interest are the {\it space-time}
expressions of $u^{j}~;~j=0,1,2$.

On account of what was written above (see (\ref{w33.3}) and
(\ref{fieldrepomega0.3})), the space-time framework diffracted
field in $\Omega_{0}$ can be written as
\begin{equation}\label{fieldrepomega0.3a}
u^{d}(\mathbf{x},\omega)= \int_{-\infty}^{\infty}dk_{1}
\int_{-\infty}^{\infty}d\omega ~ B^{0}(k_{1},\omega)
\exp\left\{\mbox{i}\left[k_{1}x_{1}+k_{2}^{0}(x_{2}-h)-\omega t
\right]\right\}~,
\end{equation}
with similar types of expressions for the fields in $\Omega_{1}$
and $\Omega_{2}$, as well as for the (incident) excitation field.
These expressions possess at least four important features.

The first  feature, underlined in \cite{grobyetwirgin2005} and
\cite{grobyetwirgin2005II}, is that the time framework fields are
expressed as integrals over the horizontal wavenumber $k_{1}$ and
angular frequency $\omega$ of functions that represent, for each
$k_{1}$ and $\omega$ (and assuming there is no material
attenuation in the media), either a propagating or evanescent
plane wave, the amplitude of which is a function such as
$B^{0}(k_{1},\omega)$.

A second important feature, underlined in
\cite{grobyetwirgin2005}, \cite{grobyetwirgin2005II} and in the
companion paper, is that the amplitude functions, such as
$B^{0}(k_{1},\omega)$, exhibit resonant behavior (i.e., can become
large in the presence of material losses, or even larger (and
sometimes, infinite) in the absence of material losses), in the
neighborhood of certain values, $k_{1}^{\bigstar}$ of $k_{1}$ and
$\omega^{\bigstar}$ of $\omega$, which are characteristic of the
{\it modes} of the structure giving rise to these fields.

The third feature, brought out in \cite{grobyetwirgin2005},
\cite{grobyetwirgin2005II}, is that resonances can be provoked by
the solicitation when: a) the frequency $\omega$ of one of the
spectral components of the latter  is equal to one of the natural
frequencies $\omega^{\bigstar}$ and b) the horizontal wavenumber
of one of the component plane waves of the excitation field is
equal to $k_{1}^{\bigstar}$.

The fourth feature is that a mode (such as the well-known Love
mode) corresponds to $\|k_{1}^{\bigstar}\|$  larger than
$\|k^{0}\|$, which means that the plane wave associated with an
excited mode is necessarily evanescent in $\Omega_{0}$.
Consequently,
 to bring $\|k_{1}\|$  (which is a sort of
momentum) up to the required level, requires a momentum boost,
which is provided either by the incident field (i.e., the latter
should contain evanescent wave components with horizontal
wavenumbers $\|k_{1}\|>|k^{0}\|$) and/or by the scattering
structure (site) itself.

In \cite{grobyetwirgin2005}, \cite{grobyetwirgin2005II}, the site
had horizontal, flat boundaries and interfaces and all the media
were homogeneous, so that it could not provide the required
momentum boost. Moreover, it was shown in
\cite{grobyetwirgin2005}, \cite{grobyetwirgin2005II} that if the
incident field takes the form of a propagating (bulk) plane wave,
the (Love) modes of the site cannot be excited, this being
possible only if the incident field contains the required
evanescent wave component,  as  is the situation in which  the
wave is radiated by a line source.

Herein, the incident field takes the form of a propagating plane
wave and the momentum boost is provided by the periodic uneveness
of the surface (in quanta of $\frac{2\pi}{d}$), as manifested by
the presence of evanescent waves in the field representations, so
that we can expect the configuration to exhibit resonant behavior
corresponding to the excitation of some sort of modes.

The remainder of this section is devoted to the characterization
of these modes and to the methods for finding the
$(k_{1}^{\bigstar},\omega^{\bigstar})$ with which they are
associated.
\subsection{The emergence of the quasi-Love modes of the configuration from
the iterative solution of the matrix equation for
$B_n^0$}\label{quasiLove}
Adding and substracting the same term on the left side of the
matrix equation (\ref{bOeq}) gives
\begin{equation}
\left(C_{n}^{10}
-D_{nn}\right)B_{n}^{0}=\sum_{m=-\infty}^{\infty}D_{nm}B_{m}^{0}\left(
1-\delta_{mn}\right) +F_{n}~~;~~\forall n \in \mathbb{Z}~,
\end{equation}
from which we obtain
\begin{equation}
B_{n}^{0}=\frac{\sum_{m=-\infty}^{\infty}D_{nm}B_{m}^{0}\left(
1-\delta_{mn}\right) +F_{n}}{C_{n}^{10} -D_{nn}}~~;~~\forall n \in
\mathbb{Z}~.
\end{equation}
An iterative approach for solving this matrix system consists in
computing successively:
\begin{equation}\label{bOeq1}
B_{n}^{0(0)}=\frac{F_{n}}{C_{n}^{10} -D_{nn}}~~;~~\forall n \in
\mathbb{Z}~,
\end{equation}
\begin{equation}\label{bOeq2}
B_{n}^{0(1)}=B_{n}^{0(0)}+\frac{\sum_{m=-\infty}^{\infty}D_{nm}B_{m}^{0(0)}\left(
1-\delta_{mn}\right) }{C_{n}^{10} -D_{nn}}~~;~~\forall n \in
\mathbb{Z}~,
\end{equation}
and so forth.

The $l$-th order iterative approximation of the solution  is thus
of the form
\begin{equation}\label{qL1}
B_{n}^{0(l)}=\frac{\mathcal{N}_n^{(l)}}{C_{n}^{10}-D_{nn}}:=
\frac{\mathcal{N}_n^{(l)}}{\mathcal{D}_n}~,
\end{equation}
wherein
\begin{equation}\label{qL2}
\mathcal{N}_n^{(0)}=F_{n}~,
\end{equation}
\begin{equation}\label{qL3}
\mathcal{N}_n^{(l>0)}=F_{n}+\sum_{m=-\infty}^{\infty}D_{nm}B_{m}^{0(l-1)}\left(
1-\delta_{mn}\right)  ~,
\end{equation}
from which it becomes apparent that the solution $B_{n}^{0(l)}$,
\textit{to any order $l$ of approximation}, is expressed as a
fraction, the denominator of which (not depending on the order of
approximation), can become small for certain values of $k_{1n}$
and $\omega$ so as to make $B_{n}^{0(l)}(\omega)$, and (possibly)
the field in the substratum, large at these values.

When this happens, a \textit{natural mode of the configuration},
comprising the blocks, the soft layer and the hard half
substratum, is excited, this taking the form of a
\textit{resonance} with respect to $B_n^{0(l)}$, i.e., with
respect to the field in the substratum. As $B_n^{0}$ is related to
$A_n^{1}$ and $B_n^{1}$ via (\ref{w7.3.2})-(\ref{w7.4.2}), the
structural resonance also manifests  itself in the layer for the
same $k_{1n}$ and $\omega$.
\newline
\newline
\textit{Remark} \newline The matrix equation (\ref{bOeq}) can be
written as
\begin{equation}\label{b0eqt}
\left(\mathbf{C} - \mathbf{D}\right)\mathbf{b}^0=\mathbf{f}
\end{equation}
wherein:  the matrix  $\mathbf{C}^{10}$ has components
$C_{nm}^{10}=C_{n}^{10} \delta_{nm}$,  $\mathbf{D}$  has
components $D_{nm}$, and the vectors $\mathbf{b}^0$ and
$\mathbf{f}$  have components $B_{n}^0$ and $F_n$ respectively.
The \textit{modes} of the configuration are obtained by turning
off the excitation \cite{wiba96}, embodied in the vector
$\mathbf{f}$. Thus, the non-trivial solution of the homogeneous
matrix equation (\ref{b0eqt}) is the solution of (the {\it
dispersion relation})
\begin{equation}\label{b0eqt1}
\mbox{det}\left(\mathbf{C}^{10} - \mathbf{D}\right)=0
\end{equation}
wherein $\mbox{det} (\mathbf{M})$ signifies the determinant of the
matrix $\mathbf{M}$.
\newline
\newline
The equations
\begin{equation}\label{b0eqt1a}
\mathcal{D}_{n}=C_{n}^{10}-D_{nn}=0~~;~~n=0,\pm 1,.... ~,
\end{equation}
associated with a singularity in the iterative procedure
(\ref{qL1}), also correspond to an approximation of the dispersion
equation of the modes (\ref{b0eqt1}) when the off-diagonal
elements of the matrix $\mathbf{C}^{10}-\mathbf{D}$ are small
compared to the diagonal elements. Such a situation does not
necessarily prevail, but it is nevertheless useful  to obtain a
first idea of the natural frequencies of the modes from the simple
relations (\ref{b0eqt1a}) rather than from the much more
complicated relation (\ref{b0eqt1}).

As shown in the companion paper (section 6.2), and  in
\cite{grobyetwirgin2005,grobyetwirgin2005II}, \textit{it is
impossible to excite a Love mode in a configuration without blocks
consisting of a soft layer overlying a hard halfspace when the
incident wave is a plane bulk wave}. This case corresponds to
 $b\rightarrow 0$.

Let us return to the denominator of the expression of $B_{n}^{0}$,
which takes the form:
\begin{equation}\label{qL4}
\mathcal{D}_{n}=C_{n}^{10}-D_{nn} =C_{n}^{10}-
\\ \sum_{l=0}^{\infty}\frac{\mbox{i}\epsilon_{l}\mu_2
k_{2l}^{2}w}{\mu^{0} k_{2n}^{0}d} I_{ln}^{-}I_{ln}^{+}
\tan\left(k_{2l}^{2} b\right)C_{n}^{01} ~.
\end{equation}
\newline
{\it Remark}
\begin{equation}\label{qL4a}
C_{n}^{10}=0 ~,
\end{equation}
is the dispersion relation  for ordinary \textit{Love modes}.
\newline
\newline
{\it Remark}
\begin{equation}\label{qL4c}
\mathcal{D}_{n}=0~,
\end{equation}
 is
then  the dispersion relation of what we term  \textit{quasi-Love
modes} which are generally different from ordinary Love modes.
\newline
\newline
\textit{Remark}
\newline
When $b\rightarrow 0$, the  dispersion relation for quasi-Love
modes becomes the dispersion relation
$\mathcal{D}_{n}=C_{n}^{10}=0$ for ordinary Love modes.
\newline
\newline
\textit{Remark}
\newline
For small $\displaystyle\frac{\mu^2 k_{2m}^{2}w}{\mu^1
k_{2n}^{1}d}$,  the quasi-Love modes are a small perturbation of
ordinary Love modes.

\subsection{The emergence of the quasi displacement-free
base block  modes and quasi-Cutler modes of the configuration from
iterative solutions of the linear system of equations for
$B_{m}^{2(l)}$ }\label{quasidisf}
%
\subsubsection{Approximate dispersion relations arising from
the first type of iterative scheme}
Let us consider  (\ref{bm2eq1}), which can be re-written as:
\begin{equation}\label{it0}
B_{l}^{2}\left(1 -Q_{ll}\right)=P_{l}+
\sum_{m=0}^{\infty}Q_{ml}\left(1-\delta_{ml} \right)B_{m}^{2}~~;
~~\forall l \in \mathbb{N}~.
\end{equation}
A (first type of) iterative procedure for solving this linear set
of equations leads to:
\begin{equation}\label{it1}
B_{l}^{2(0)}=\frac{P_{l}}{1 -Q_{ll}}~~;~~\forall l \in
\mathbb{N}~,
\end{equation}
\begin{equation}\label{it2}
B_{l}^{2(1)}(\omega)=B_{l}^{2(0)}(\omega)+
\frac{\sum_{m=0}^{\infty}Q_{ml}\left(1-\delta_{ml}
\right)B_{m}^{2(0)}}{1 -Q_{ll}}~~;~~l=1,2,...~.
\end{equation}
This procedure signifies that $B_{l}^{2(p)}$ becomes large when $1
-Q_{ll}$ is small, and that this occurs at all orders $p$ of
approximation. The fact that $B_{l}^{2(p)}$ becomes inordinately
large is associated with the excitation of a natural mode of the
configuration. The equations $1 -Q_{ll}=0 ~;~l\in\mathbb{N}$ are
the approximate dispersion relations of the $l$-th natural modes
($l\in\mathbb{N}$) of the configuration.
\newline
\newline
{\it Remark}
\newline
We say that {\it these dispersion relations are approximate in
nature} because they are obtained by neglecting the off-diagonal
terms in the matrix equation (\ref{it0}). This may not be
legitimate, but it is nevertheless useful to get an idea of the
true natural frequencies by examination of the solutions of the
approximate dispersion relations $1 -Q_{mm}=0~;~m\in\mathbb{N}$.

Let us therefore examine the latter in more detail:
\begin{equation}\label{quadisfre}
\mathcal{F}_{m}(k_{1},\omega):=\cot\left(k_{2m}^{2}
b\right)-\sum_{n=-\infty}^{\infty} \mbox{i}\epsilon_{m}\frac{w}{d}
\frac{k_{2m}^{2}\mu^{2}
}{k_{n2}^{0}\mu^{0}}\frac{C_{n}^{01}}{C_{n}^{10}}
I_{mn}^{-}I_{ln}^{+}=0~;~m\in\mathbb{N}~,
\end{equation}
which shows that the modes of the configuration result from the
\textit{interaction} of the fields in two substructures: the {\it
superstructure} (i.e., the blocks above the ground), associated
with the term
\begin{equation}
\mathcal{F}_{1m}(k_{1},\omega)=
\mathcal{F}_{1m}=\cot\left(k_{2m}^{2} b\right) ~,
\end{equation}
and the {\it substructure} (i.e., the soft layer plus the  hard
half space below the ground) associated with the term
\begin{equation}
\mathcal{F}_{2m}(k_{1},\omega)=
\mathcal{F}_{2m}=\sum_{n=-\infty}^{\infty}
\mbox{i}\epsilon_{m}\frac{w}{d} \frac{k_{2m}^{2}\mu^{2}
}{k_{n2}^{0}\mu^{0}}\frac{C_{n}^{01}}{C_{n}^{10}}
I_{mn}^{-}I_{ln}^{+}~.
\end{equation}
Each of these two substructures possesses its own modes, i.e.,
arising from $\mathcal{F}_{1m}=0$, for the superstructure, and
$\mathcal{F}_{2m}=0$, for the substructure, but the modes of the
complete structure  are neither the modes of the superstructure
nor those of the substructure, since they are defined by
\begin{equation}
\mathcal{F}_{1m}(k_{1},\omega)-
\mathcal{F}_{2m}(k_{1},\omega)=0~;~m\in\mathbb{N}~,
\end{equation}
which again emphasizes the fact that the modes of the complete
structure result from the {\it interaction} of the modes of the
component structures.

In order to obtain the natural  frequencies of the complete
structure, we first analyze the natural frequencies of each
substructure, and assume that all the media are non-dissipative
(i.e. elastic).

The solutions of the (approximate) dispersion relations for the
superstructure are:
\begin{multline}
\cot\left(k_{2m}^{2} b\right)=0~\Leftrightarrow~ k_{2m}^{2}
b=\frac{(2n+1)\pi}{2}~\Leftrightarrow~
\\
\omega=\omega_{mn}=c^{2}\sqrt{\left(\frac{(2n+1)\pi}{2b}\right)^{2}+
\left(\frac{m\pi}{w}\right)}~~;~~n,~m=0,1,2,...~,
\end{multline}
which are the natural frequencies of vibration of a block with
\textit{displacement-free base} (i.e., at these natural
frequencies, $\left.u^{2}\right|_{x_2=0}$ vanishes on the base
segment of the block).

Next consider the dispersion relations for the geophysical
(sub)structure
$\mathcal{F}_{2m}(k_{1},\omega)=0~;~m\in\mathbb{N}$. As  pointed
out in the companion paper, the sum in this relation can be split
into three parts corresponding to: i) propagative waves in both
the substratum and the layer, ii) evanescent waves in the
substratum and propagative waves the layer,  iii) evanescent waves
in both the substratum and  layer. Only the second part can lead
to a vanishing  denominator, and also to the satisfaction of the
dispersion relation of  Love modes.
\newline
\newline
\textit{Remark}
\newline
For small $\frac{\mu^{2}}{\mu^{0}}$, the quasi displacement-free
base block modes are a small perturbation of the displacement-free
base block modes.
\newline
\newline
\textit{Remark}
\newline
For small $\frac{w}{d}$, the quasi displacement-free base block
modes are a small perturbation of the displacement-free base block
modes.
\newline
\newline
\textit{Remark}
\newline
We notice that the approximate dispersion relations for the
configuration involving an infinite set of equispaced identical
blocks are similar to the dispersion relations obtained (in the
companion paper) for a small number of blocks, in that they betray
the existence of a combination of quasi-Love and quasi
displacement-free base block modes, which are
 small perturbations of Love and displacement-free
base block modes respectively for small $\displaystyle
\frac{\mu^{2}}{\mu^{0}}$ and/or small $ \frac{w}{d}$.
\subsubsection{Approximate dispersion relations resulting
from a second type of iterative scheme}
Let  $\mathbf{Q}$ denote the matrix of  components $Q_{ml}$,
$\mathbf{I}$ the identity matrix, and $\mathbf{b}$ and
$\mathbf{p}$ the vectors of  components $B_{l}^{2}$ and $P_{l}$
respectively.

The system of linear equations (\ref{bm2eq1}) can be written as
the matrix equation (for the determination of the unknown vector
$\mathbf{b}$)
\begin{equation}\label{matr1}
\left(\mathbf{I}-\mathbf{Q}\right)\mathbf{b}=\mathbf{p}~.
\end{equation}
The \textit{modes} of the configuration are obtained  by turning
off the excitation \cite{wi95}, embodied in $\mathbf{p}$. The
non-trivial solutions of (\ref{matr1}) are then obtained from
\begin{equation}\label{matr2}
\mbox{det}\left(\mathbf{I}-\mathbf{Q}\right)=0~.
\end{equation}
An iterative (partition) procedure  for solving this equation,
which is different from the preceding one, and is particularly
appropriate if the off-diagonal elements of the matrix are small
(but non neglected) compared to the diagonal elements, is first to
consider the matrix to have one row and one column, i.e.,
\begin{equation}\label{matr3}
1-Q_{00}=0~,
\end{equation}
then to consider it to have two rows and two columns,
\begin{equation}\label{matr4}
\mbox{det}
\left(
\begin{array}{ll}
\displaystyle 1-Q_{00}&\displaystyle -Q_{01}\\
\displaystyle Q_{10}&\displaystyle 1-Q_{11}
\end{array}
\right)=
\left(1-Q_{00}\right)\left(1-Q_{11}\right)-Q_{01}Q_{10}=0~,
\end{equation}
and so forth.

We shall not go into the details of these various approximate
expressions of the dispersion relation because they become
extremely involved beyond (\ref{matr3}). However, it is of
historical and didactic interest to examine (\ref{matr3}) in
detail. This is done in the next section.
\subsubsection{Solution of the zeroth-order dispersion relation arising
in the two types of iterative schemes: the Cutler
mode}\label{cutler}
We rewrite the lowest-order approximation of the dispersion
relation (\ref{matr1}) (equivalent to (\ref{qL4}) for $n=0$) in
the form first given in \cite{wi88}:
\begin{equation}\label{quadisfre2}
\mathcal{D}_{0}=1-\sum_{n=-\infty}^{\infty}\frac{\mbox{i}
k_{2}^{2} \mu^{2}} {k_{2n}^{0}\mu^{0}}\frac{w}{d}\tan\left(k^{2}
b\right) I_{0n}^{-}I_{0n}^{+}\frac{C_{n}^{01}}{C_{n}^{10}}=0~.
\end{equation}
However:
\begin{equation}\label{quadisfre21}
I_{0n}^{-}I_{0n}^{+}=\mbox{sinc}^{2}\left(
k_{1n}\frac{w}{2}\right) ~,
\end{equation}
so that
\begin{equation}\label{quadisfre22}
\mathcal{D}_{0}=1-\mbox{i}\frac{w}{d}\frac{k_{2}^{2}
\mu^{2}}{k^{0}\mu^{0}}\tan\left( k^{2} b\right)
\sum_{n\in\mathbb{Z}} \frac{k^{0}}
 {k_{2n}^{0}} \mbox{sinc}^{2}\left(
k_{1n}\frac{w}{2}\right) \frac{C_{n}^{01}}{C_{n}^{10}}=0~.
\end{equation}
We now consider the cases (first studied in
\cite{cu44,ro51,hu54,bopa58,hoha58,auga76,gupl78,wi88}) in which
the layer is filled with the same material $M^{0}$ as that of the
substratum (actually, in most of the cited publications, $M^{2}$
was also taken equal to $M^{0}$, but this is not done here, for
the moment at least). Thus, $\mu^{1}=\mu^{0}$ and $k^{1}=k^{0}$.
In addition, we recall that $M^{0}$ is non-dissipative, so that
$\mu^{0}$ and $k^{0}$ are real. We shall also suppose, to simplify
matters, that $\mu^{2}$ and $k^{2}$ are real (i.e., $M^{2}$ is
lossless).

Then
\begin{equation}\label{quadisfre23}
C_{n}^{01}=C_{n}^{10}=\exp(ik_{2n}^{1}h)=\exp(ik_{2n}^{0}h)~
\Rightarrow~\frac{C_{n}^{01}}{C_{n}^{10}}=1~,
\end{equation}
so that
\begin{equation}\label{quadisfre24}
\mathcal{D}_{0}=1-\mbox{i}\frac{w}{d}\frac{k_{2}^{2}
\mu^{2}}{k^{0}\mu^{0}}\tan\left( k^{2} b\right)
\sum_{n\in\mathbb{Z}} \frac{k^{0}}
 {k_{2n}^{0}} \mbox{sinc}^{2}\left(
k_{1n}\frac{w}{2}\right)=0~.
\end{equation}
If, in addition, we assume that $M^{2}=M^{0}$, then
\begin{equation}\label{quadisfre25}
\mathcal{D}_{0}=1-\mbox{i}\frac{w}{d}\tan\left( k^{0} b\right)
\sum_{n\in\mathbb{Z}} \frac{k^{0}}
 {k_{2n}^{0}} \mbox{sinc}^{\eta}\left(
k_{1n}\frac{w}{2}\right)=0~,
\end{equation}
with $\eta=2$. This dispersion relation is identical to the one
obtained in \cite{auga76} (wherein $\eta$ is not given a definite
value).

Let us stick to (\ref{quadisfre24}) for the moment and make the
substitutions
\begin{equation}\label{quadisfre25}
\delta:=\frac{w}{d}~~,~~\alpha:=\frac{k^{2}}{k^{0}},~~\kappa:=\frac{k_{2}^{2}
\mu^{2}}{k^{0}\mu^{0}}=\alpha\frac{\mu^{2}}{\mu^{0}}~~,~~\beta=\frac{b}{d}~,
\end{equation}
therein, so that
\begin{equation}\label{quadisfre26}
\mathcal{D}_{0}=1-\mbox{i}\delta\kappa\tan\left( k^{2} b\right)
\sum_{n\in\mathbb{Z}} \frac{k^{0}}
 {k_{2n}^{0}} \mbox{sinc}^{2}\left(
k_{1n}\frac{w}{2}\right) =0~.
\end{equation}
Suppose that there exists an integer $N$ for which one of the
terms in the series is very large compared to the rest of the
series. We thus write
\begin{equation}\label{quadisfre27}
\mathcal{D}_{0}=1-\mbox{i}\delta\kappa\tan\left( k^{2}
b\right)\left[ \frac{k^{0}}
 {k_{2N}^{0}} \mbox{sinc}^{2}\left(
k_{1N}\frac{w}{2}\right) + R_{0}\right] =0~,
\end{equation}
with
\begin{equation}\label{quadisfre28}
R_{0}:= \sum_{n\in(\mathbb{Z}-\{N\})} \frac{k^{0}}
 {k_{2n}^{0}} \mbox{sinc}^{2}\left(
k_{1n}\frac{w}{2}\right) ~.
\end{equation}
Let $k_{1}:=k_{1N}$ and $k_{2}:=k_{2N}^{0}$. It follows that
\begin{equation}\label{quadisfre29}
k_{1n}=k_{1}+2(n-N)\frac{\pi}{d}~~\Rightarrow~k_{1n}\frac{w}{2}=
k_{1}\frac{w}{2}+(n-N)\pi\frac{w}{d}=k_{1}\frac{w}{2}+(n-N)\pi\delta~.
\end{equation}
Now, for the $n=N$ term to be large, at the very least
\begin{equation}\label{quadisfre30}
\left \|k_{1}\frac{w}{2}\right \|<<1~.
\end{equation}
and
\begin{equation}\label{quadisfre31}
k_{2N}^{0}\approx 0~.
\end{equation}
The first of these two conditions is synonymous with
\begin{equation}\label{quadisfre32}
\mbox{sinc}\left( k_{1}\frac{w}{2}\right) \approx 1~.
\end{equation}
A corollary is that
\begin{multline}\label{quadisfre33}
\sin\left( k_{1n}\frac{w}{2}\right)=\sin\left(
k_{1}\frac{w}{2}\right) \cos\left( (n-N)\pi\delta\right)
+
\\
\cos\left( k_{1}\frac{w}{2}\right) \sin\left(
(n-N)\pi\delta\right) \approx
\sin\left( (n-N)\pi\delta\right)~.
\end{multline}
We shall also assume that
\begin{equation}\label{quadisfre34}
\delta\approx 1~,
\end{equation}
so that
\begin{equation}\label{quadisfre35}
\sin\left( k_{1n}\frac{w}{2}\right)\approx \sin\left(
(n-N)\pi\right)\approx 0~~\Rightarrow R_{0}\approx 0~,
\end{equation}
whence
\begin{equation}\label{quadisfre36}
1-\mbox{i}\delta\kappa\tan\left( k^{2} b\right) \frac{k^{0}}
 {k_{2}} \mbox{sinc}^{2}\left(
k_{1}\frac{w}{2}\right)=0~,
\end{equation}
which is the dispersion relation obtained by Rotman \cite{ro51}.

It is more precise to take account of (\ref{quadisfre32}) so that
\begin{equation}\label{quadisfre37}
1-\mbox{i}\delta\kappa\tan\left( k^{2} b\right) \frac{k^{0}}
 {k_{2}}=0~,
\end{equation}
which, when $\kappa=1$ (i.e., the case  $M^{2}=M^{0}$),
constitutes the dispersion relation first obtained by Cutler
\cite{cu44}. We shall now examine this relation in detail.

Since all the parameters in the Cutler dispersion relation are
real, the latter has no solution unless $k_{2}=\sqrt{(k^{0})^{2}-
(k_{1})^{2}}$ is imaginary, i.e.,
$k_{2}=\mbox{i}\|\sqrt{(k^{1})^{2}- (k_{0})^{2}}\|$, which occurs
if $|k_{1}|>k^{0}$. If we recall that imaginary $k_{2}$
corresponds to an evanescent (i.e., surface) wave, then we can say
that the {\it Cutler mode} is associated with the excitation of a
surface wave.

We now inquire as to the conditions under which  the Cutler mode
can be excited. The first step is to find the values of
$(k_{1},\omega)$ which are solutions of
\begin{equation}\label{quadisfre38}
1-\delta\kappa\tan\left( k^{2} b\right) \frac{k^{0}}
 {\|k_{2}\|}=0~~;~~|k_{1}|>k^{0}~.
\end{equation}
Another point of view is to consider $k_{1}$ to be the wavenumber
(of a surface wave) to which is associated the phase velocity
$c_{1}$ such that
\begin{equation}\label{quadisfre39}
k_{1}=\frac{\omega}{c_{1}}~.
\end{equation}
Then it is easy to obtain (from (\ref{quadisfre38}))
\begin{equation}\label{quadisfre40}
c_{1}=c^{0}\sqrt{ \frac{1}{1+\left[ \delta\kappa \tan\left(\frac{
\omega b}{c^{2}}\right) \right] ^{2}}}~.
\end{equation}
\newline
{\it Remark}
\newline
This relation, first published in \cite{gupl78}, shows that even
if $M^{2}$ is non-dispersive (recall that it was assumed, from the
start, that $M^{0}$ is non-dispersive), then the phase velocity of
the Cutler mode is dispersive, i.e., $c_{1}=c_{1}(\omega)$ (which,
of course, is the reason why one speaks of a dispersion relation
in connection with a (e.g., Cutler) mode).
\newline
\newline
{\it Remark}
\newline
The Cutler mode corresponds to a {\it slow (surface) wave} with
repect to the bulk plane waves in $\Omega_{0}$, since $c_{1}\leq
c^{0}$.
\newline
\newline
{\it Remark}
\newline
$c_{1}(\omega)$ is a periodic function of $\omega$, since
$c_{1}(\omega+\frac{l\pi c^{2}}{b})=c_{1}(\omega)~;~\forall
l\in\mathbb{N}$.
\newline
\newline
{\it Remark}
\newline
$c_{1}(\omega)=c^{0}$ for $\frac{\omega b}{c^{2}}=l\pi~;~\forall
l\in\mathbb{N}$.
\newline
\newline
{\it Remark}
\newline
$c_{1}(\omega)=0$ for $\frac{ \omega
b}{c^{2}}=(2l+1)\frac{\pi}{2}~;~\forall l\in\mathbb{N}$.
\newline
\newline
{\it Remark}
\newline
The phase velocity of the Cutler mode is all the closer to the
phase velocity of bulk waves in $M^{0}$ for all $\omega$, the
smaller is $\kappa$. On the contrary, for a Cutler mode with phase
velocity very different from that of bulk  waves in $M^{0}$, we
must have a large contrast between the material properties of
$M^{0}$ and $M^{2}$.
\newline
\newline
Let us now inquire as to the means of actually exciting a Cutler
mode with an incident  plane bulk wave. At first, this seems
impossible (for the same reason it is not possible to excite a
Love mode with an incident plane bulk wave). But we must not
forget that the field in $\Omega_{0}$ is composed not only of
diffracted plane bulk waves, but also of diffracted evanescent
waves, the possibility of these waves to exist being due to the
uneven (at present, periodically uneven) geometry of the
stress-free surface at the site.

The discussion concerning the Cutler mode began with the
assumptions: i) the term in the expression of the diffracted field
in $\Omega^{0}$ corresponding to the $N$-th order diffracted plane
wave dominates all the other terms, and ii) this diffracted wave
is an evanescent wave, i.e., $k_{2N}^{0}$ is imaginary. In the
dispersion equation context, $k_{1}$ is a variable that has no
particular connection with the solicitation. When  the site is
solicited by a plane bulk wave, then
$k_{1N}=k_{1}^{i}+\frac{2\pi}{d}N$, with
$k_{1}^{i}=k^{0}\sin\theta^{i}$ the factor directly related to the
solicitation ($\theta^{i}$ the angle of incidence). Thus, for the
$N$-th order evanescent diffracted wave to be excited, we must
have
\begin{equation}\label{quadisfre41}
k_{1}=k_{1N}~~\Rightarrow k_{1}=k_{1}^{i}+\frac{2\pi}{d}N~,
\end{equation}
with $N$ such that
$k_{2N}^{0}=i\|\sqrt{(k_{1N})^{2}-(k^{0})^{2}}\|$. This so-called
{\it coupling relation}, i.e.,(\ref{quadisfre41}), translates the
fact that the periodic topography adds the momentum
$\frac{2\pi}{d}N$ necessary to convert the incident bulk wave into
an evanescent wave (whose phase velocity is smaller than that of
the bulk wave, and whose horizontal wavenumber is therefore larger
than the wavenumber $k^{0}$ of the incident bulk wave).

In order for a {\it resonance} to occur in the $N$-th order mode,
the frequency of one of the components of the spectrum of the
excitation must be equal to a natural frequency of the mode.
Although this is a necessary condition, it is not a sufficient
condition, because we must also have
$\|k_{1}\|=\|k_{1N}\|>\|k^{0}\|$.

A  remark is in order concerning what happens when $R_{0}$ is not
neglected in the expression of $\mathcal{D}_{0}$. The subset, in
this remainder term, involving the  $n$ for which $k_{2n}$ is
imaginary (i.e., corresponding to evanescent waves), will modify
somewhat the (real) solutions of what formerly constituted the
Cutler dispersion relation, whereas the subset involving the $n$
for which $k_{2n}$ is real (i.e., corresponding to propagative
waves) adds an imaginary part to the (real) solutions of what
formerly constituted the Cutler dispersion relation. Thus, the
evanescent Cutler wave becomes a leaky wave, i.e., a wave with
complex $k_{1}$.
\subsubsection{Solution of the zeroth-order dispersion
relation when $M^{1}\neq
M^{0}$: the quasi-Cutler mode}
The dispersion relation  (\ref{quadisfre2}) has been studied in
\cite{wi88} and is a subset of the dispersion relations analyzed
in sect. \ref{quasiLove}. Not much more, other than what is
revealed by a numerical analysis, can be added to the text in
sects. \ref{quasiLove} and \ref{cutler}, due to the complexity of
(\ref{quadisfre2}).

To make a long story short, one finds that the presence of the
layer transforms the Cutler mode into a {\it quasi Cutler mode}
which is all the closer to a Cutler mode the smaller is the layer
thickness $h$ and/or the closer the material parameters of the
layer are to those of the substratum. If, on the other hand, $h$
is not very small, and/or the material parameters of the layer are
very different from those of the substratum, then the quasi-Cutler
mode becomes an entity entirely different from that of the Cutler
mode; in fact, it ressembles a Love mode, so that it is better to
represent the phenomena in terms of quasi-Love modes (as in sect.
\ref{quasiLove}) than in terms of quasi-Cutler modes.
\section{Computation of the fields $u^{0}$, $u^{1}$
and $u^{2}$}
%
\subsection{Computation of  $u^{0}$}
The quasi-modal  coefficients $B_{m}^{2}(\omega)$, $\forall m \in
\mathbb{N}$ are obtained by employing the partition procedure
(i.e., reducing the infinite-order matrix in (\ref{matr1}) to a
$M\times M$ matrix and the vectors to $M$-tuple vectors, solving
the finite-order matrix equation so obtained, and increasing $M$
until convergence is obtained of the successive $M$-th order
approximate solutions). Once the $B_{m}^{2}(\omega)$, $\forall m
\in \mathbb{N}$ are computed in this manner, the field in the
block domain $\Omega_{2}$ is obtained via (\ref{w2.5.47a}). This
field vanishes on the ground at the frequencies of occurrence of
the \textit{displacement-free base} modes of the block.
\subsection{Computation of  $u^{1}$}
Let us next consider the field in the layer. Combining
(\ref{w7.1.3}), (\ref{w7.2.3}), (\ref{w7.3.2}) and (\ref{w7.4.2})
leads, via (\ref{w2.5.41}), to:
\begin{multline}
 u^{1}(\mathbf{x},\omega)=\frac{2 S(\omega) \exp\left(\mbox{i}
 \left(k_{1}^{i}x_{1}-k_{02}^{2} h \right) \right) \cos\left(k_{02}^{1}x_{2}
 \right)} {C_{0}^{10}}+
 \\
\sum_{n=-\infty}^{\infty}
\sum_{m=0}^{\infty}B_{m}^{2}\mbox{i}\frac{\mu^{2}k_{2m}^{2} w }
{\mu^{0} k_{n2}^{0}d} \frac{I_{mn}^{-}}{C_{n}^{10}}
\sin\left(k_{2m}^{2} b\right) \exp\left(\mbox{i}k_{n1}\left(
\frac{w}{2}+x_{1}\right) \right) \times
\\
\left( \cos
\left(k_{n2}^{1}\left(x_{2}-h\right) \right)+\mbox{i}
\frac{\mu^{0}k_{n2}^{0}}{\mu^{1}k_{n2}^{1}}
\sin\left(k_{n2}^{1}\left(x_{2}-h\right) \right) \right) ~.
 \label{u1}
\end{multline}
\subsection{Computation of  $u^{2}$}
Let us finally consider the field in the substratum. Combining
(\ref{w7.1.3}), (\ref{w7.2.3}), (\ref{w7.3.2}) and (\ref{w7.4.2})
leads, via (\ref{w2.5.42}), to:
\begin{multline}
 u^{0}(\mathbf{x},\omega)= u^{i}(\mathbf{x},\omega)+
 \\
 S(\omega)\exp\left(\mbox{i}\left(k_{1}^{i}x_{1}+
 k_{02}^{0}\left(x_{2}-2h \right) \right) \right)\frac{\cos\left(k_{02}^{1}h \right)+
 \mbox{i}\frac{\mu^{1}k_{02}^{1}}{\mu^{0}k_{02}^{0}}\sin\left(k_{02}^{1}h \right) }
 {C_{0}^{10}}+
 \\
\sum_{n=-\infty}^{\infty}\mbox{i}
\exp\left(\mbox{i}\left(k_{n1}x_{1}+ k_{n2}^{0}\left(x_{2}-h
\right) \right) \right) \times
\\
\sum_{m=0}^{\infty} B_{m}^{2}
\frac{\mu_{2}k_{2m}^{2}w}{\mu^{0}k_{n2}^{0}d}\frac{I_{mn}^{-}}{C_{0}^{10}}
\sin\left(k_{2m}^{2}b \right)\exp\left(\mbox{i}k_{n1}\frac{w}{2}
\right)  ~. \label{u0}
\end{multline}
\subsection{Comments on the fields  $u^{1}$ and  $u^{2}$}
Expressions (\ref{u1})~ and ~(\ref{u0}) ~indicate ~that  ~both~
displacement fields  ~$u^{1}(\mathbf{x},\omega)$ ~and~
$u^{2}(\mathbf{x},\omega)$ are composed of: i) the field obtained
in the absence of the blocks and induced in the layer or
substratum by the incident plane wave, ii) the field induced by
the presence of the blocks, which appears as a field radiated by
an infinite number of identical  source distributions (each one
related to a given block). In particular, each of these induced
sources takes the form of a ribbon source of width $w$ located at
the base segment of a block when it is related to the zeroth-order
quasi-mode (see the companion paper).

When a mode is excited (i.e., at a resonance frequency), one or
several of the $B_{m}^{2}$ can become large, in which case it is
possible for the fields to become large at resonance. This will be
demonstrated in the numerical examples which follow.
\section{Numerical results for the seismic response in two idealized cities}
%
\subsection{Preliminaries}
The numerical results are obtained in two manners: i) by the
Mode-Matching (MM) method (described in the previous section) as
it applies to  configurations consisting of an infinite number of
equally-spaced, equally-sized rectangular blocks, and ii) by the
Finite-Element (FE) method (briefly described in the companion
paper, and in more detail in \cite{grobyettsogka,grts05}) as it
applies to configurations with a large (but finite) number
 of equally-spaced, equally-sized rectangular blocks.

The discussions concerning the numerical aspects of the dispersion
relations will be based on material stemming from the MM. For the
purpose of the analysis, and to allow for an easier comparison
with the results exposed in the companion paper, we re-write  (the
lowest-order approximation of the dispersion relation)
(\ref{quadisfre2}) in the form
\begin{multline}\label{quadisfreprelim}
\cos\left(k^{2}b\right)-\sum_{n=-\infty}^{\infty} \frac{\mbox{i}w
k^{2} \mu^{2}}{\mu^{0}k_{n2}^{0}d}
I_{0n}^{-}I_{0n}^{+}\frac{C_{n}^{01}}{C_{n}^{10}}\sin\left(k^{2}
b\right)=\mathcal{F}_{1}-\mathcal{F}_{2}=0~,
\end{multline}
wherein $\mathcal{F}_{1}=\cos\left(k_{2m}^{2} b\right)$.
\subsection{Ten and an infinite number of blocks in a Nice-like site}
\subsubsection{Parameters}
The sources of earthquakes in the city of Nice (France) are
usually deep and located almost vertically below the Nice site (a
few kilometers laterally from the city). Modeling the solicitation
of the site by a normally-incident  (i.e., $\theta^i=0$, which
leads to $k_{1}^{i}=0$) plane incident wave is thus realistic
\cite{segu03}. The central frequency of the Ricker pulse
associated with the solicitation is chosen to be $\nu_0=2Hz$.

With $N_{b}$ the number of blocks, we consider here both the cases
$N_{b}=10$ and $N_{b}=\infty$.

 The parameters of the underground of our idealized
Nice urban site are: $\rho^{0}=2200$ Kg/m$^{3}$, $c^{0}$=1000 m/s,
$Q^{0}=\infty$, $\rho^{1}=1800$ Kg/m$^{3}$, $c^{1}_{ref}$=200 m/s,
$Q^{1}=25$, with the soft layer thickness being $h=50$ m.

The Haskell eigenfrequencies, which are usually close to the
quasi-Love mode frequencies, are then
$\nu_{m}^{HASK}=\frac{2m+1}{2}\frac{c^{1}}{2h}=$1Hz, 3Hz, 5Hz,....

The blocks are chosen to be $b=$30m high, $w=$10m wide, and their
center-to-center spacing is $d=$50m. Their material constants are
chosen to be: $\rho^{2}=250$ Kg/m$^{3}$, $c^{2}_{ref}$=240 m/s,
$Q^{2}=10$.

Thus, the displacement-free based block eigenfrequencies
(solutions of $\mathcal{F}_{2}=0$), $\nu_{m}^{DF}={(2m+1)c^{2}}
{4b}$, are $\nu_{m}^{DF}=2Hz, 6Hz$... and the quasi-Cutler mode
natural frequencies (which are specific to the periodic nature of
the site, with characteristic dimension $d$), are obtained from
$\mathcal{F}=\mathcal{F}_{1}-\mathcal{F}_{2}=0$.
\subsubsection{Infinite number of blocks}
\begin{figure}[ptb]
\begin{center}
\includegraphics[width=6cm] {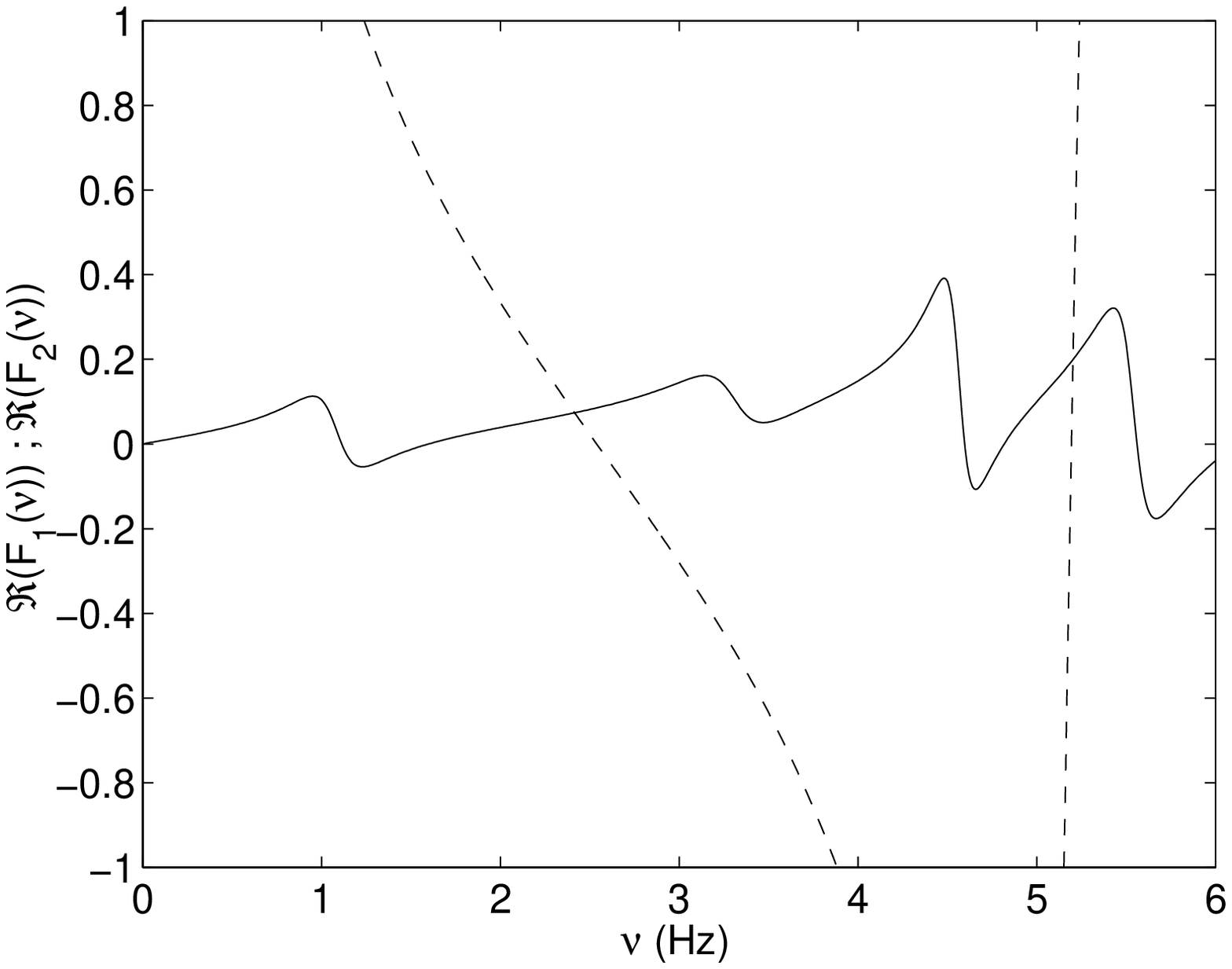}
\includegraphics[width=6cm] {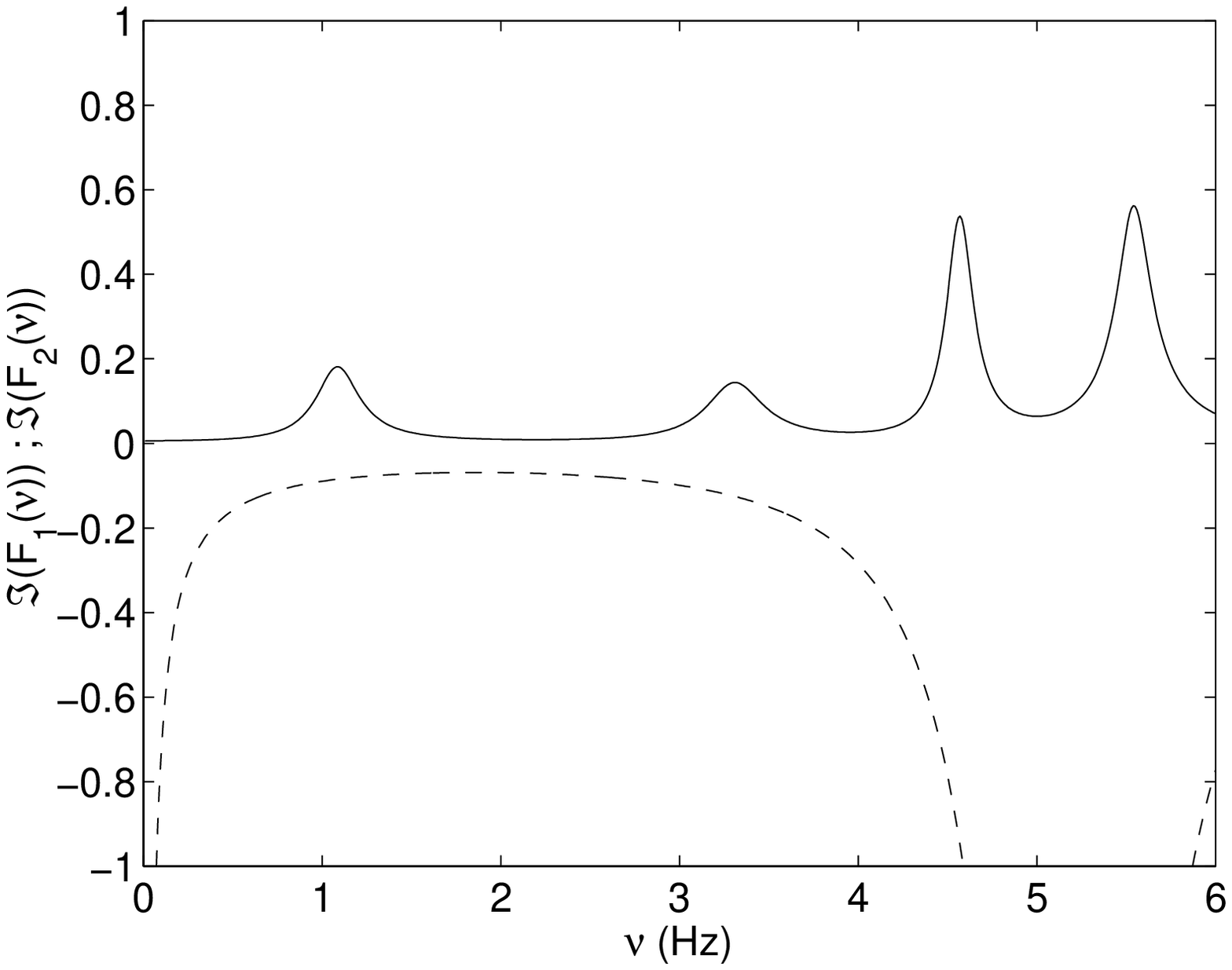}
\caption{Indications concerning the solution of the dispersion
relation $\mathcal{F}_{1}-\mathcal{F}_{2}=0$ for the Nice-like
city. In the left panel:
 the solid  and dashed curves depict $\Re(\mathcal{F}_1)$
and $\Re(\mathcal{F}_2)$ respectively versus  frequency ($\nu$ in
$Hz$) . In the right panel:   the solid and dashed  curves depict
$\Im(\mathcal{F}_1)$ and $\Im(\mathcal{F}_2)$ respectively versus
frequency.} \label{modeNiceInf}
\end{center}
\end{figure}
Fig.  \ref{modeNiceInf} gives an indication of the natural
frequencies of the modes of the global configuration. Recall that
such a natural frequency is a solution of
$\mathcal{F}_{1}-\mathcal{F}_{2}=0$, requiring that
$\Re(\mathcal{F}_1)=\Re(\mathcal{F}_2)$, at the least. This occurs
at $\nu\approx$ 2.5Hz, 5.1Hz,.... in the frequency range of the
figure. The attenuation associated with a particular mode (at a
frequency $\nu^{\star}$) is related to
$\Im(\mathcal{F}_1(\nu^{\star}))-\Im(\mathcal{F}_2(\nu^{\star}))$.

One observes in the figure that the quasi-Love modes, which are
expected to occur near 1Hz, 3Hz, 5Hz,... are either not excited or
are strongly attenuated, while what appears to be the
quasi-displacement free base block mode at $\nu^{QDFB}\approx 2.5$
is excited with a relatively-low attenuation. On the contrary,
what appears to be the quasi stress-free base block mode at
$\nu^{QDFB}\approx 5.1Hz$, is associated with a large attenuation
and should therefore have little effect on the global response of
the  site.

The notable features of this response are that it is dominated by
the quasi-displacement-free block modes, and that the influence of
the periodic nature of the distribution of blocks, which manifests
itself by the quasi-Cutler modes, probably appears at frequencies
higher than those in the range of the figure.
\subsubsection{Comparison of responses for $N_{b}=10$ and
$N_{b}=\infty$}
\begin{figure}[ptb]
\begin{center}
\includegraphics[width=6cm] {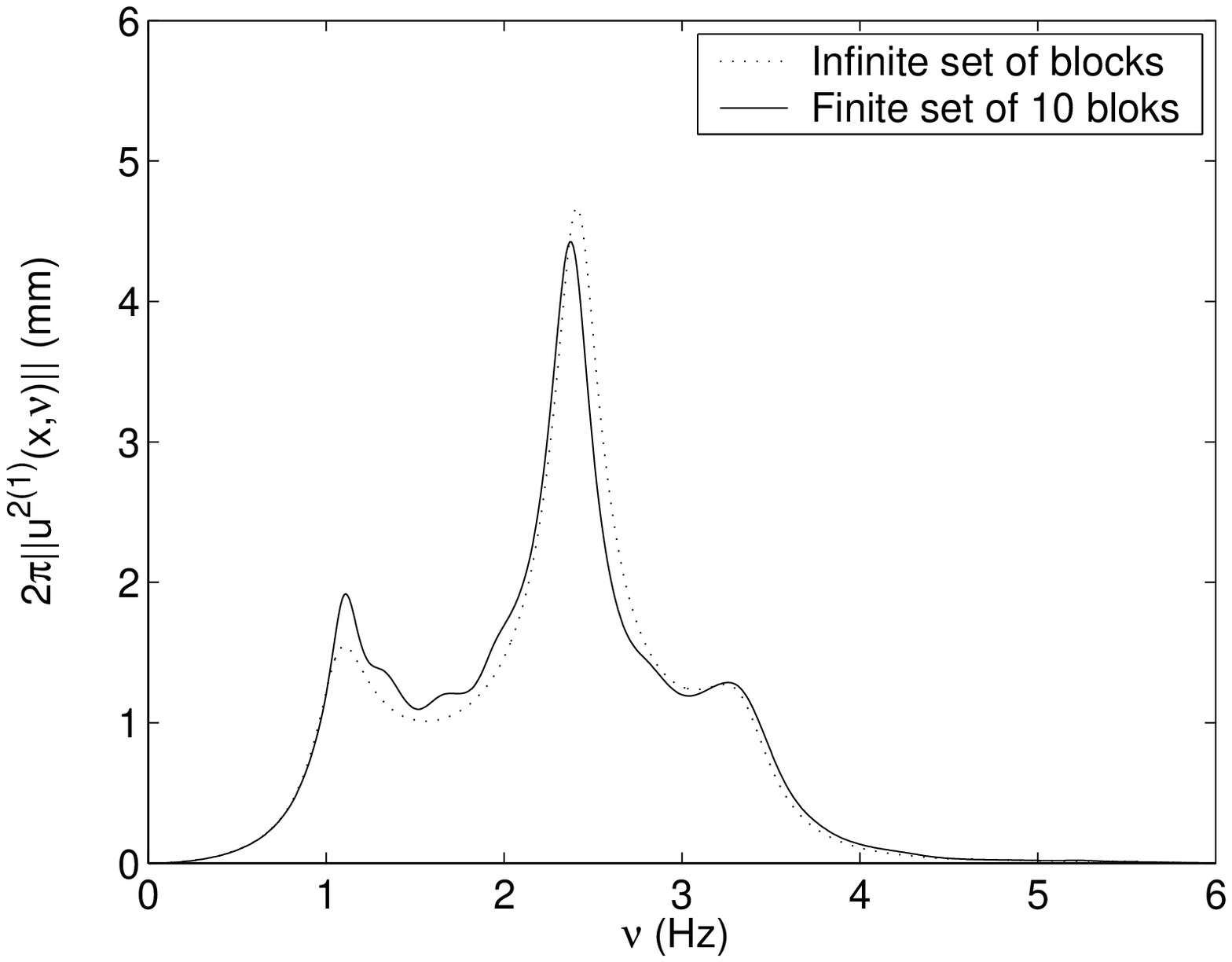}
\includegraphics[width=6cm] {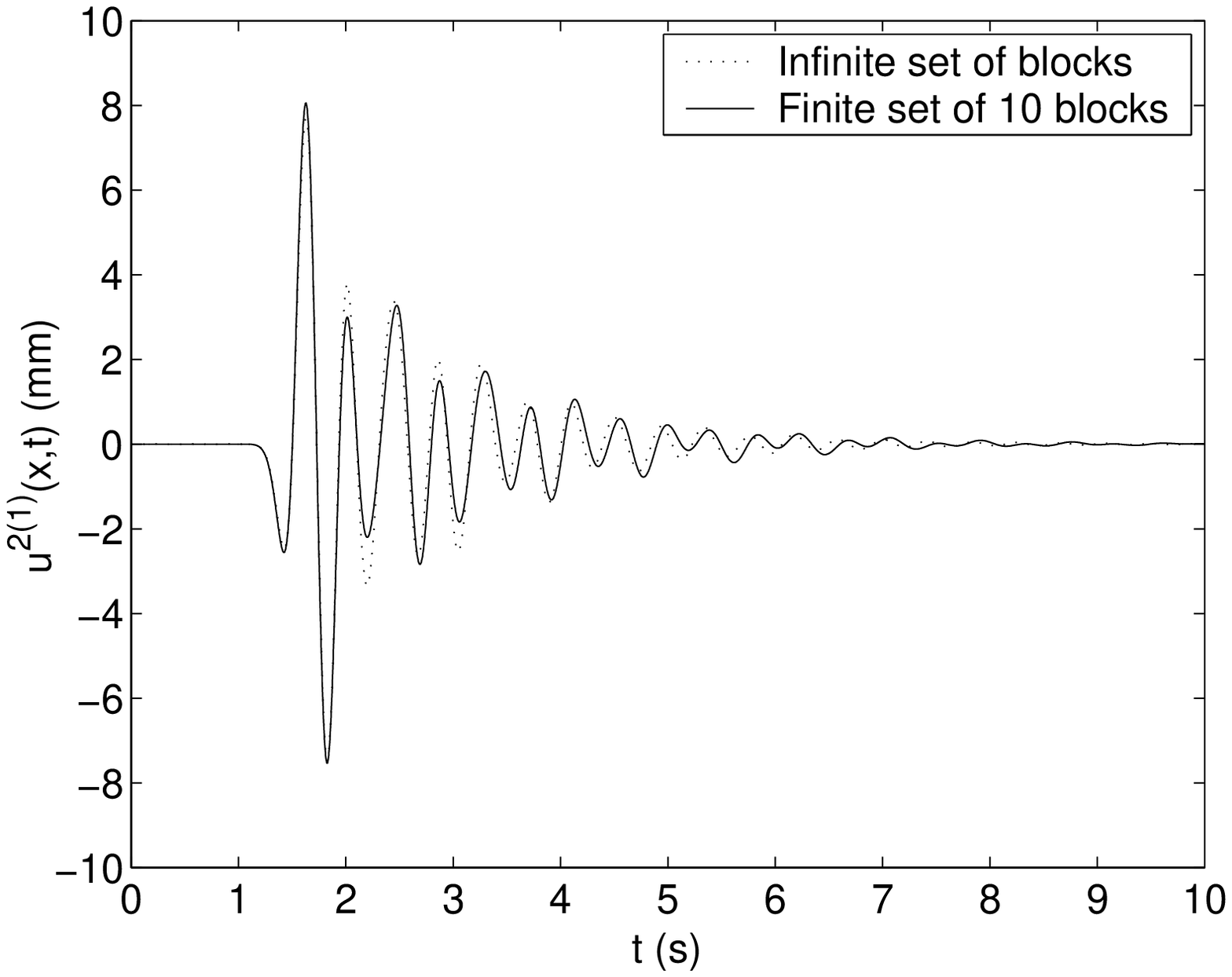}
\caption{$2\pi$ times the  spectrum (left panel) and time history
(right panel) of the total displacement at the center of the top
segment of a block, which is the {\it leftmost one} of a set of 10
blocks (solid curves, obtained by the FE method), or any one of an
infinite set of blocks (dashed curves, obtained by the MM method )
 in a Nice-like site. } \label{SpecttimesumNice10InfB1}
\end{center}
\end{figure}

\begin{figure}[ptb]
\begin{center}
\includegraphics[width=6cm] {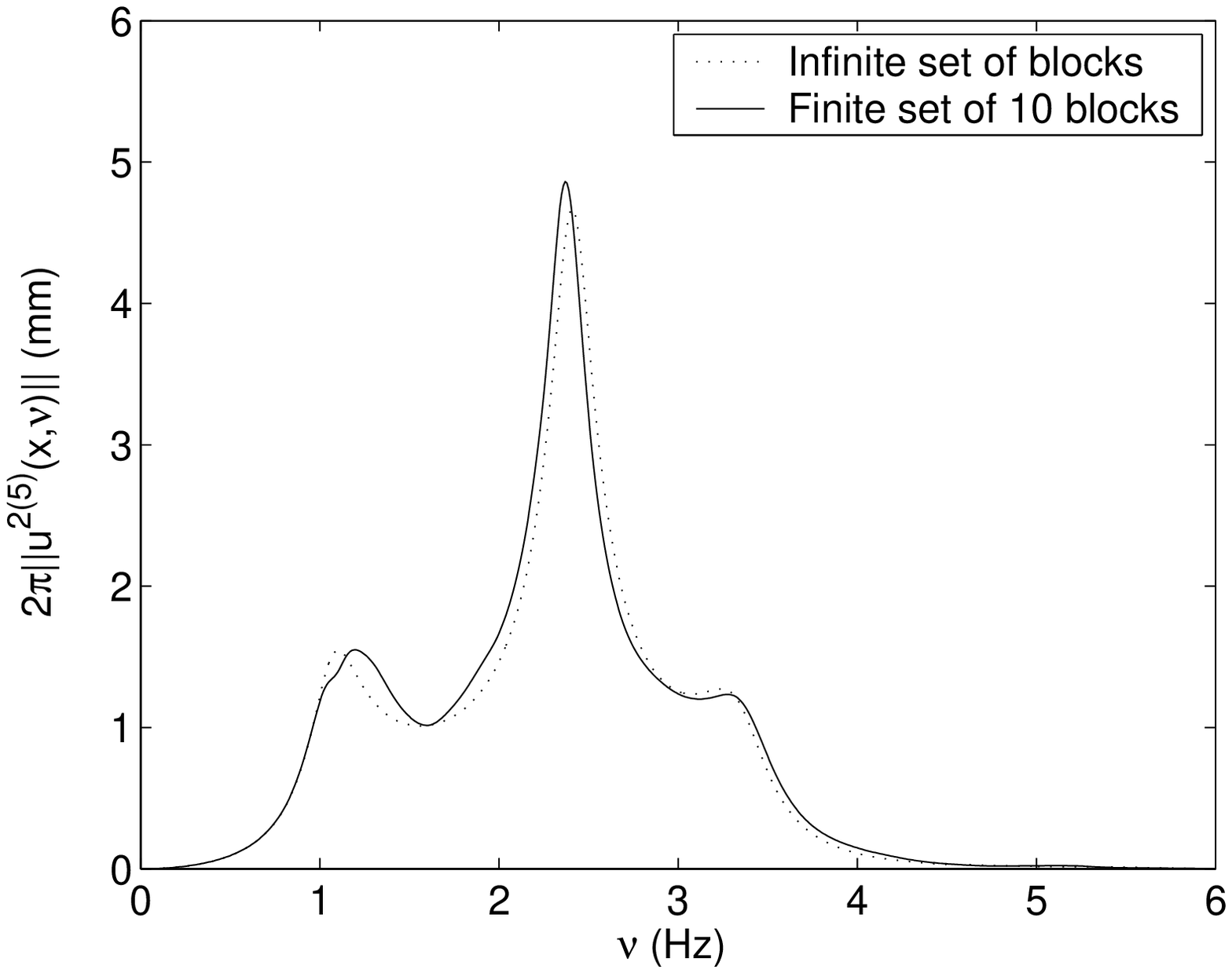}
\includegraphics[width=6cm] {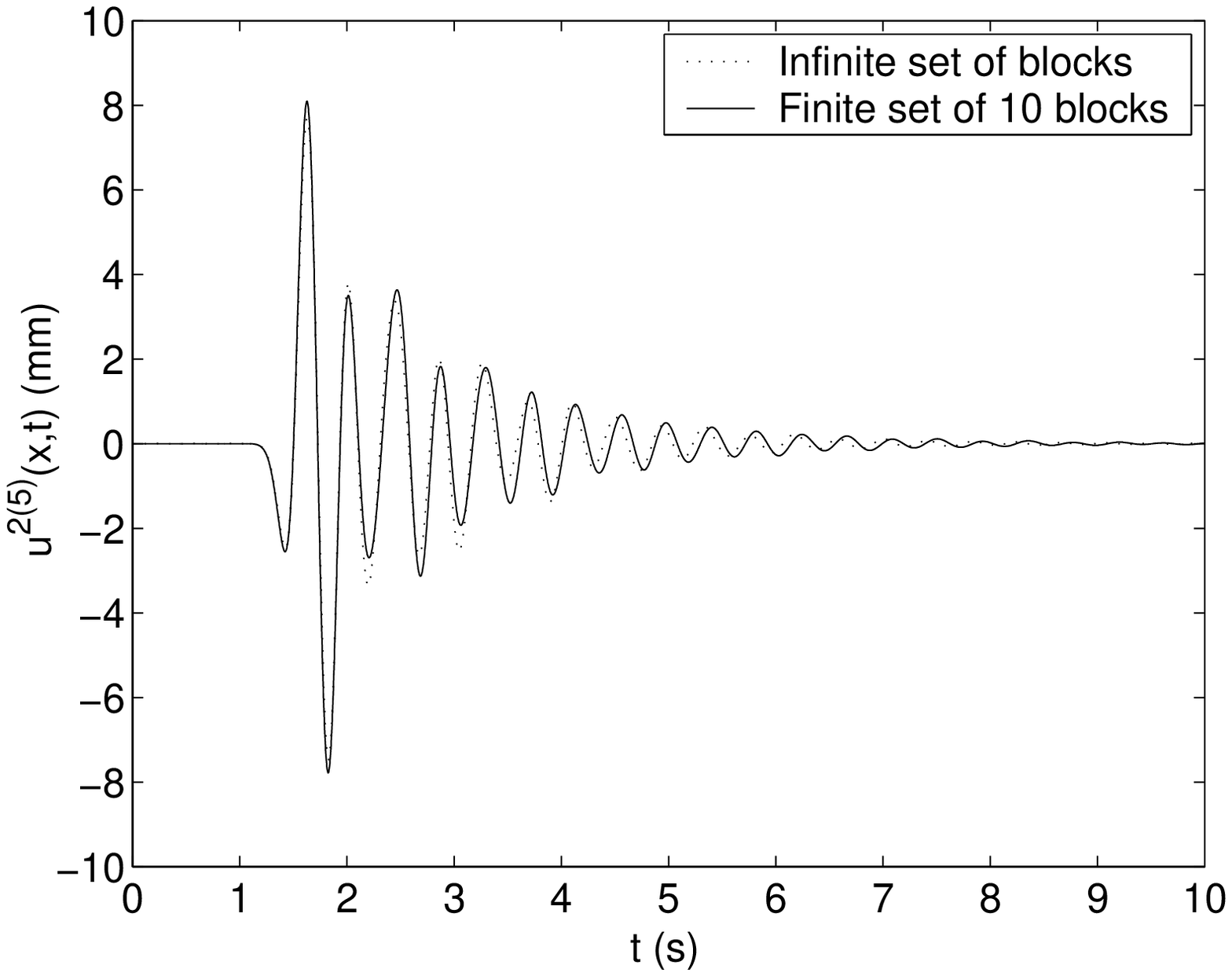}
\caption{$2\pi$ times the  spectrum (left panel) and time history
(right panel) of the total displacement at the center of the top
segment of a block, which is the {\it fifth one} of a set of 10
blocks (solid curves, obtained by the FE method), or any one of an
infinite set of blocks (dashed curves, obtained by the MM method )
 in a Nice-like site. } \label{SpecttimesumNice10InfB5}
\end{center}
\end{figure}

 We now compare both the spectra and time
histories for an infinite number of identical blocks with those of
a finite number (i.e., 10) of the same blocks. The point of
observation is at the center of the top segment of a block. The
block under examination  is the leftmost one in fig.
\ref{SpecttimesumNice10InfB1}, and the fifth one in fig.
\ref{SpecttimesumNice10InfB5}. Since the angle of incidence is 0,
both the spectra and time histories of response do not change from
one block to another in the  $N_{b}=\infty$ structure.
We note the close similarity of both the frequency and time domain
responses relative to the $N_{b}=10$ and $N_{b}=\infty$
structures, even for the leftmost block  of the finite
configuration. This means that for a Nice-like site, the model
involving an infinite number of identical blocks can used to model
and/or to analyze the response phenomena in such a city.

Other than this, as concerns the response spectra, the left-most
peak at $\approx 1.1$ Hz and the rightmost peak at $\approx 3.2$
Hz seem to be related to the excitation of the Haskell
pseudo-modes (which is another way of saying that the
corresponding quasi-Love modes are weakly excited, since these
peaks would exist even if the buildings had nearly zero height).
The large peak in the response spectra at $\approx 2.5$Hz is
associated with the excitation of the quasi-displacement free base
block mode, as expected from examination of the solutions of the
dispersion relation (see fig. \ref{modeNiceInf}). The
 lengthening of the duration ($\approx$ 8 sec) of
the response, with respect to the duration of the incident pulse
of  $\approx$ 1 sec, is mainly due to the excitation of this mode,
but since the quality factor of the resonance of this mode is
relatively small, the lengthening of the duration is
relatively-modest. Thus, it seems that both the amplification of
the amplitude, and the lengthening of duration, of seismic motion
in a periodic or quasi-periodic portion of  Nice,  cannot take
catastrophic proportions.  Insofar as this conclusion concerns the
amplitude of ground motion, it is in agreement with the
conclusions in \cite{segu03}.
\subsection{Finite and infinite number of blocks in a Mexico City-like site}
\subsubsection{Parameters}
The sources of the major earthquakes in Mexico City have usually
been shallow and located in the subduction zone off the Pacific
coast, approximately $350km$ from Mexico City), so that modeling
the solicitation of this city by a plane incident bulk wave is not
realistic \cite{grobyetwirgin2005,grobyetwirgin2005II}.
Nevertheless, we assume such a (normally-incident) plane bulk wave
solicitation, notably to enable an easy  quantitative comparison
between the finite and infinite city responses and qualitative
comparison with previous studies
\cite{wiba96,clau01,segu03,boro04,boro06}. The central frequency
of the Ricker pulse associated with the solicitation is chosen to
be $\nu_0=0.5Hz$.

The underground  of the city is characterized by: $\rho^{0}=2000$
kg/m$^{3}$, $c^{0}$=600 m/s, $Q^{0}=\infty$, $\rho^{1}=1300$
kg/m$^{3}$, $c^{1}_{ref}$=60 m/s, $Q^{1}=30$, with the soft layer
thickness  being $h=50$ m. The Haskell frequencies are
 0.3Hz, 0.9Hz, 1.5Hz,....

The blocks are 50m high, 30m in width, and their center-to-center
spacing are successively chosen to be 65m, 150m and 300m. The
material constants of the blocks are: $\rho^{2}=325$ Kg/m$^{3}$,
$c^{2}_{ref}$=100m/s, $Q^{2}=100$. The natural frequencies of the
displacement-free base block modes are $\nu_{m}^{SFB}=$ 0.5Hz,
1.5Hz,..., and the quasi-Cutler mode natural frequencies (which
are specific to the periodic nature of the site, with
characteristic dimension $d$), are obtained from $\mathcal{F}=0$.
\subsubsection{Dispersion characteristics of the modes for
an infinite number of blocks}
\begin{figure}[ptb]
\begin{center}
\includegraphics[width=6.0cm] {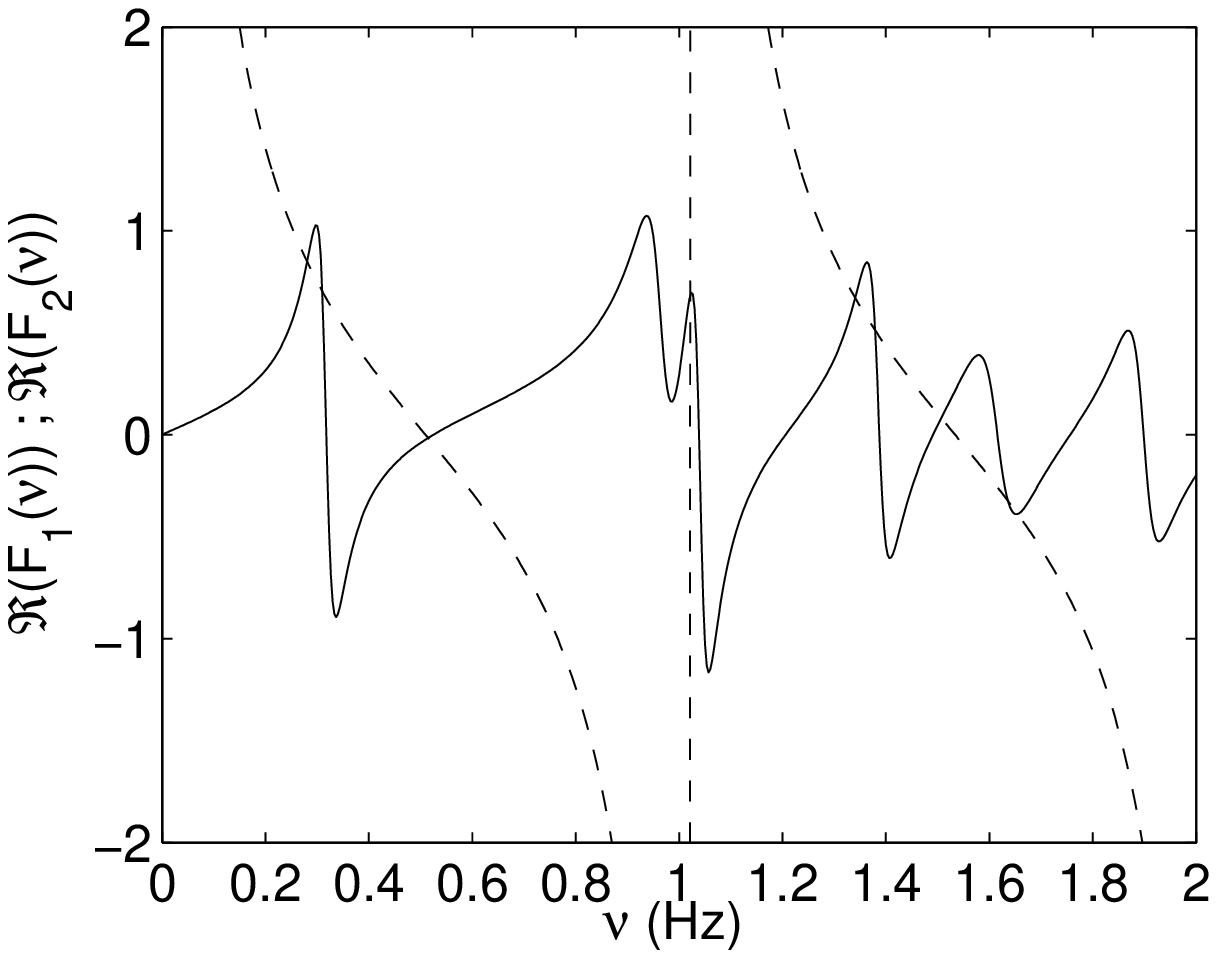}
\includegraphics[width=6.0cm] {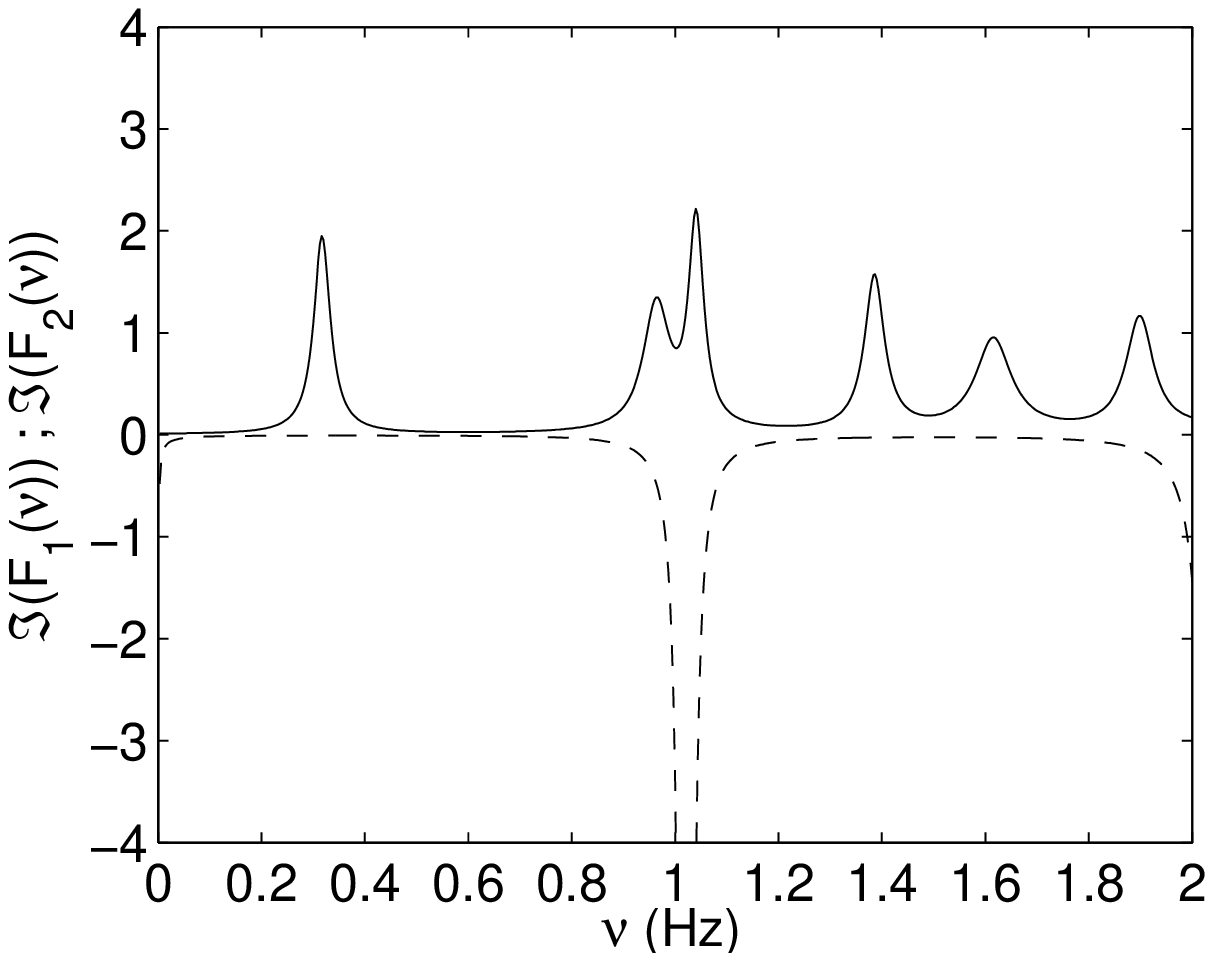}
\includegraphics[width=6.0cm] {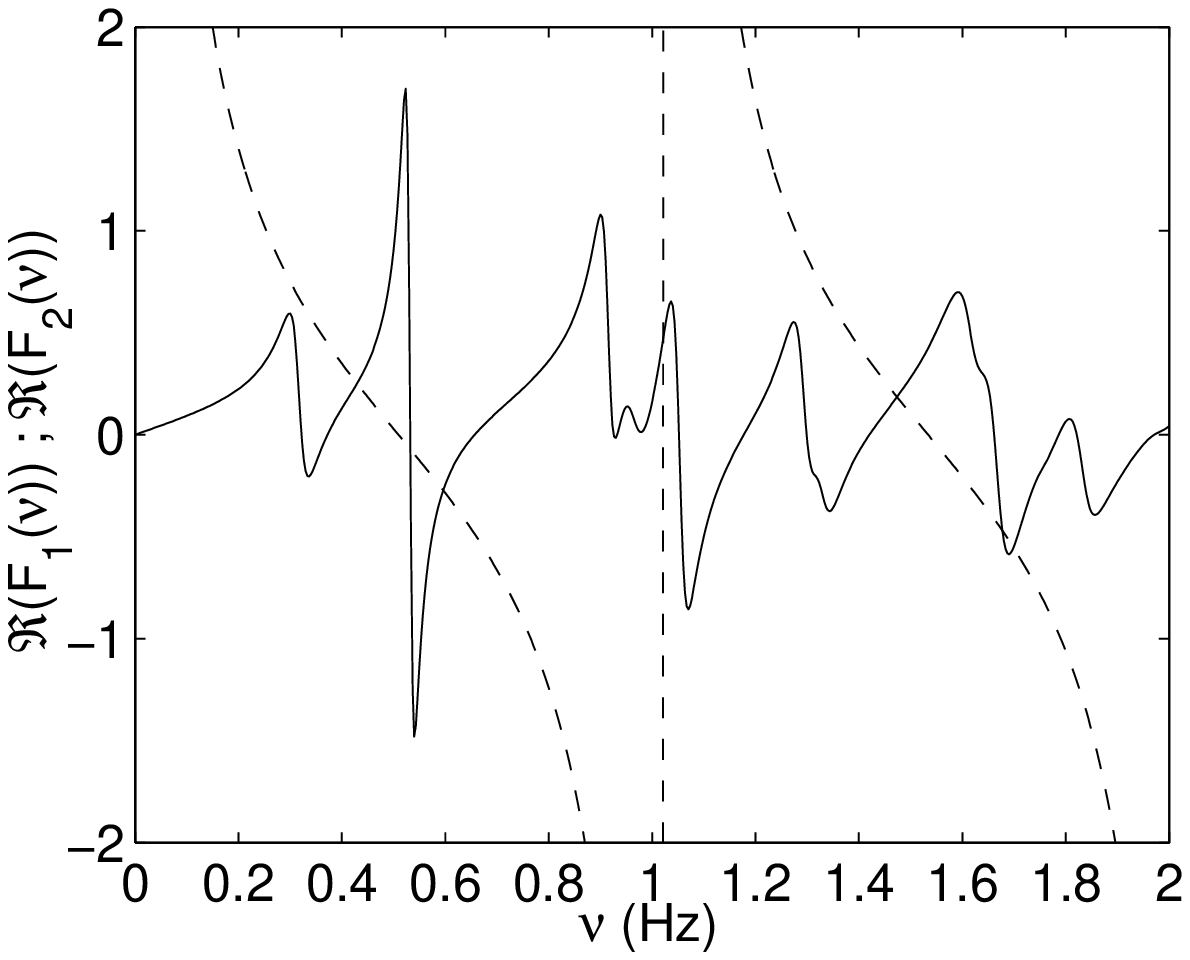}
\includegraphics[width=6.0cm] {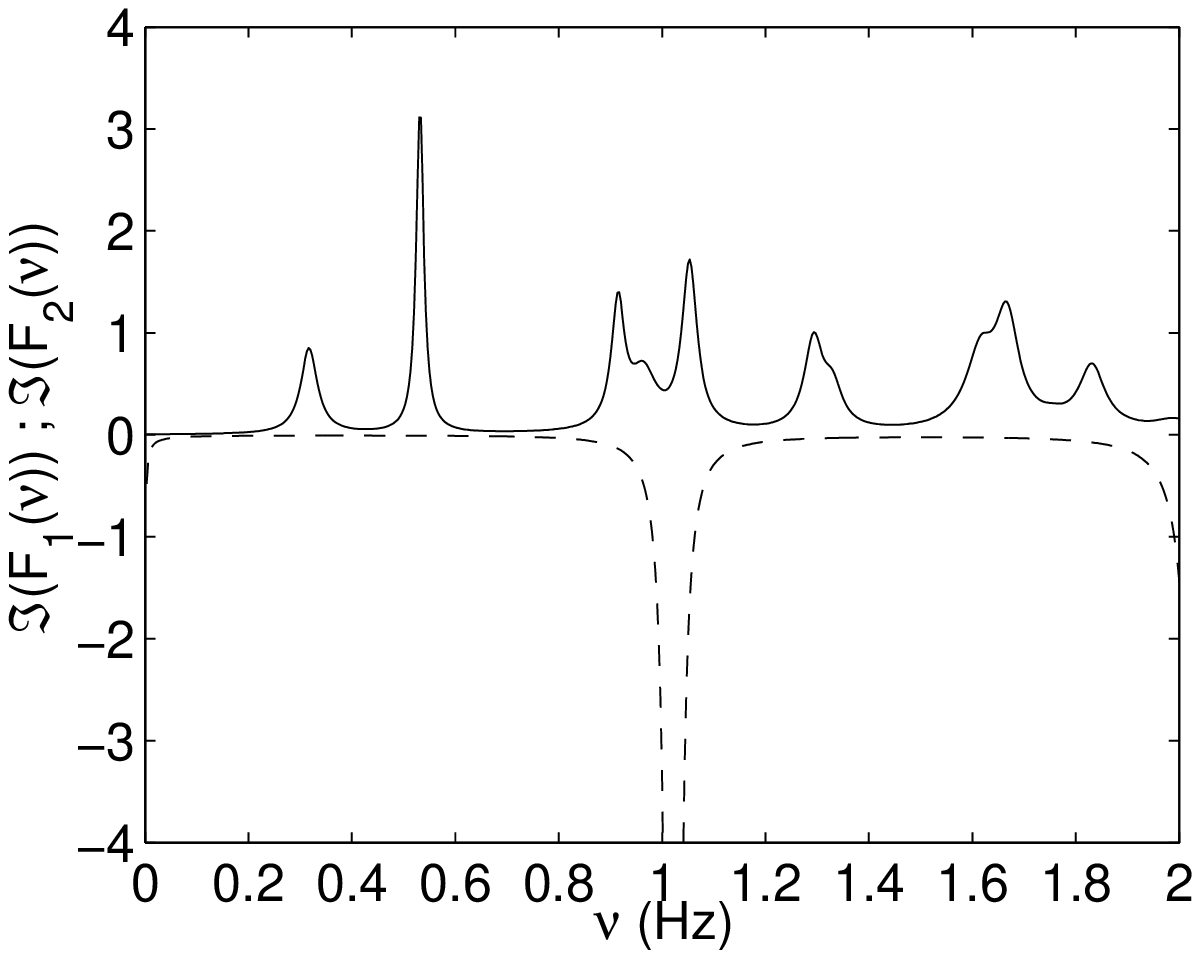}
\includegraphics[width=6.0cm] {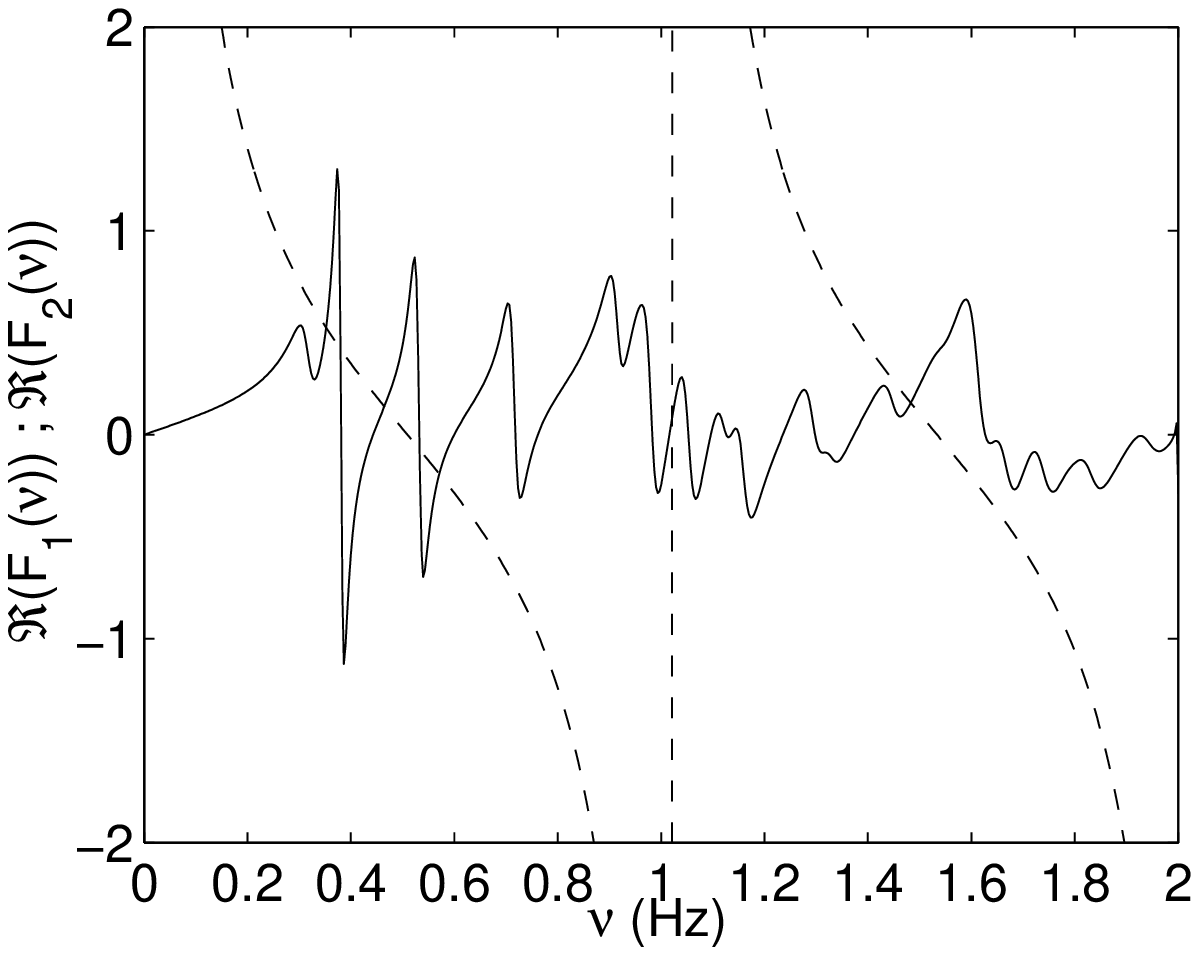}
\includegraphics[width=6.0cm] {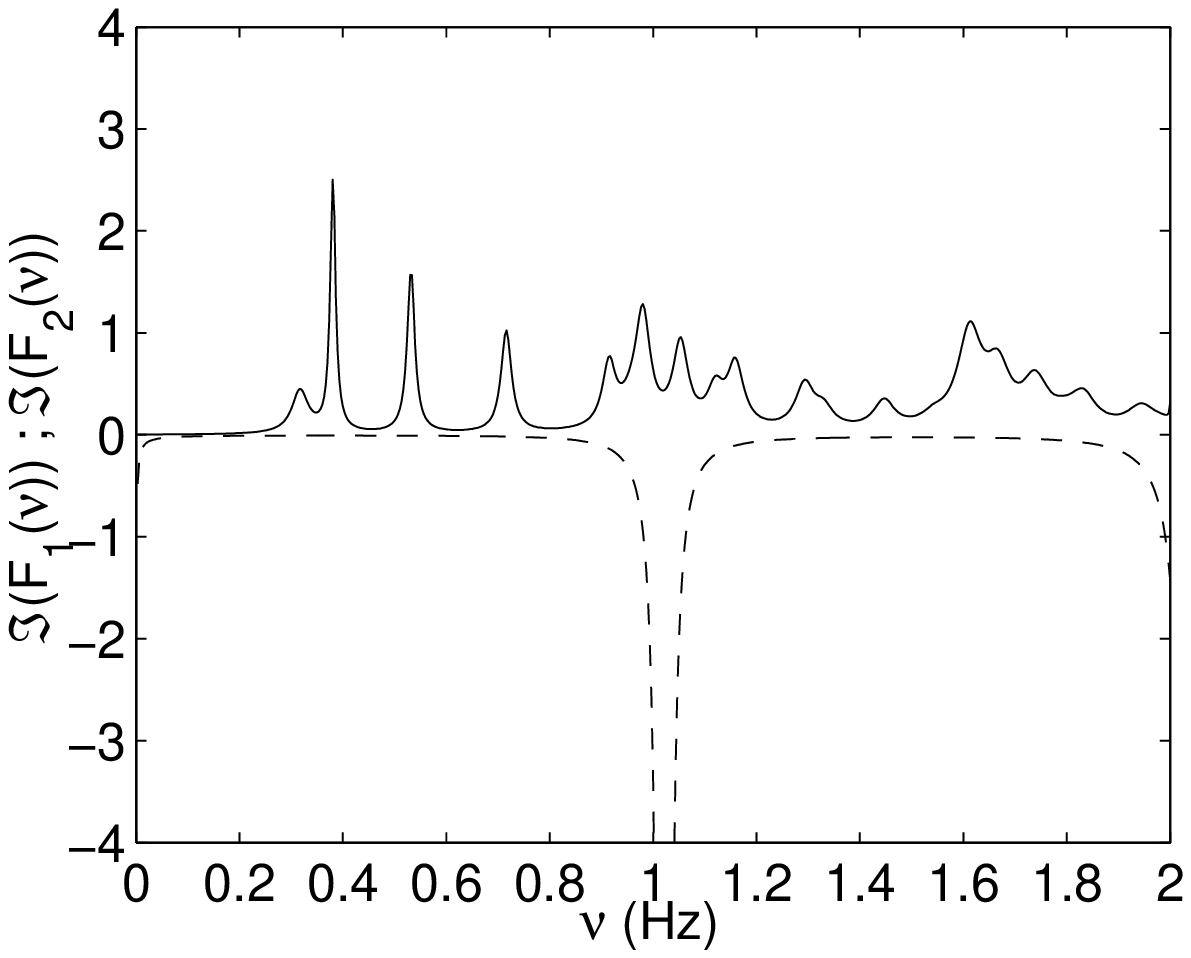}
\caption{Indications concerning the solution of the dispersion
relation $\mathcal{F}_{1}-\mathcal{F}_{2}=0$ for the Mexico
City-like urban site. In the left panels:
 the solid  and dashed curves depict $\Re(\mathcal{F}_1)$
and $\Re(\mathcal{F}_2)$ respectively versus  frequency ($\nu$ in
$Hz$) . In the right panels:   the solid and dashed  curves depict
$\Im(\mathcal{F}_1)$ and $\Im(\mathcal{F}_2)$ respectively versus
frequency. The top, middle and bottom panels are related to
$d=65$m, $d=150$m and $d=300$m respectively.}
\label{modeMexicoInf}
\end{center}
\end{figure}

Fig. \ref{modeMexicoInf} gives  indications concerning  the
solutions of the dispersion relation of the global configuration
(i.e., blocks plus underground) for center-to-center spacings
$d=65$m, $d=150$m and $d=300$m. For a center-to-center spacing
$d=65m$, the influence of the periodic nature of the city
(embodied in the parameter $d$)  appears essentially at a  high
frequency outside the spectral bandwidth of the solicitation,
while for $d=150m$ and $d=300m$, this influence can make itself
felt in the spectral range of the solicitation through the
quasi-Cutler modes.
Quasi-Love and quasi-displacement free base block modes should be
excited at $\approx 0.3Hz$ and $\approx 0.5Hz$ respectively, both
associated with a low attenuation. The stress-free base block mode
is excited at $\approx$ 1Hz, but is highly-attenuated for all
three $d$.

Quasi-Cutler modes are clearly excited with a low attenuation. In
particular, for a relatively large center-to-center spacing (i.e.
$d=300$), the spectral density of this type of mode is large. The
fundamental quasi-Love natural frequency is probably  very close
to the fundamental quasi-Cutler natural frequency.
\subsubsection{Seismic response for an infinite number of blocks
with period $d=65$m compared to that of no blocks}
\begin{figure}[ptb]
\begin{center}
\includegraphics[width=6.0cm] {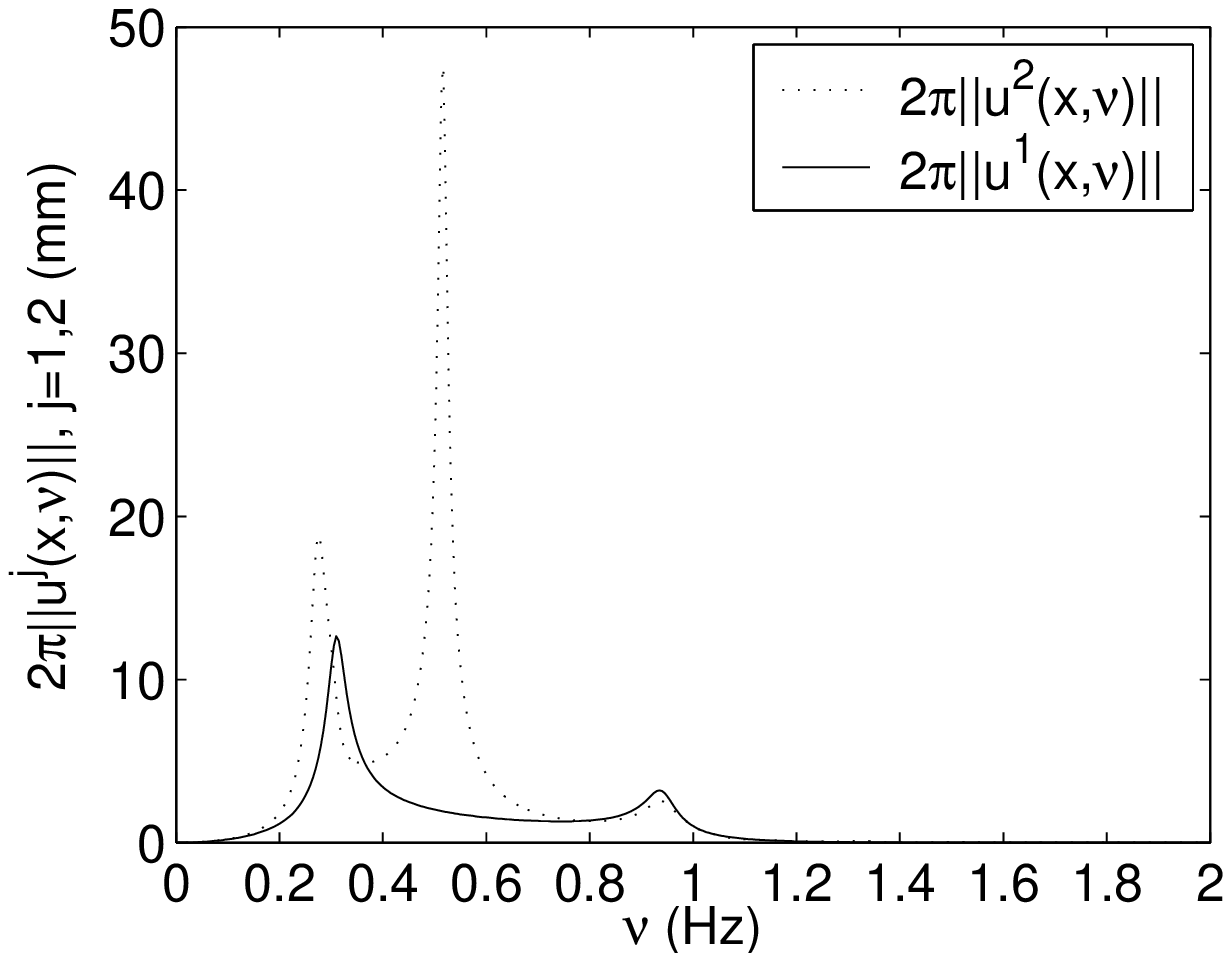}
\includegraphics[width=6.0cm] {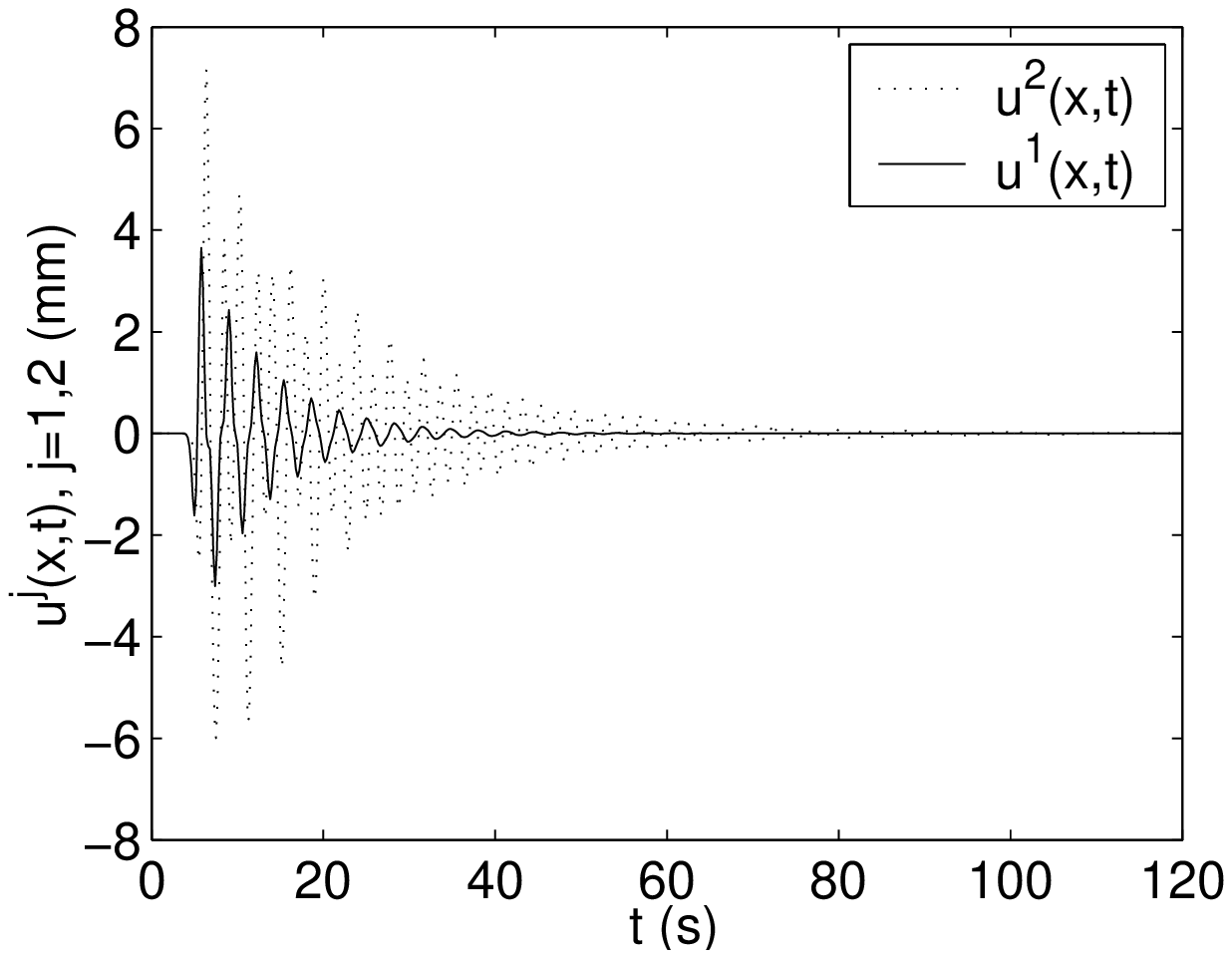}
\includegraphics[width=6.0cm] {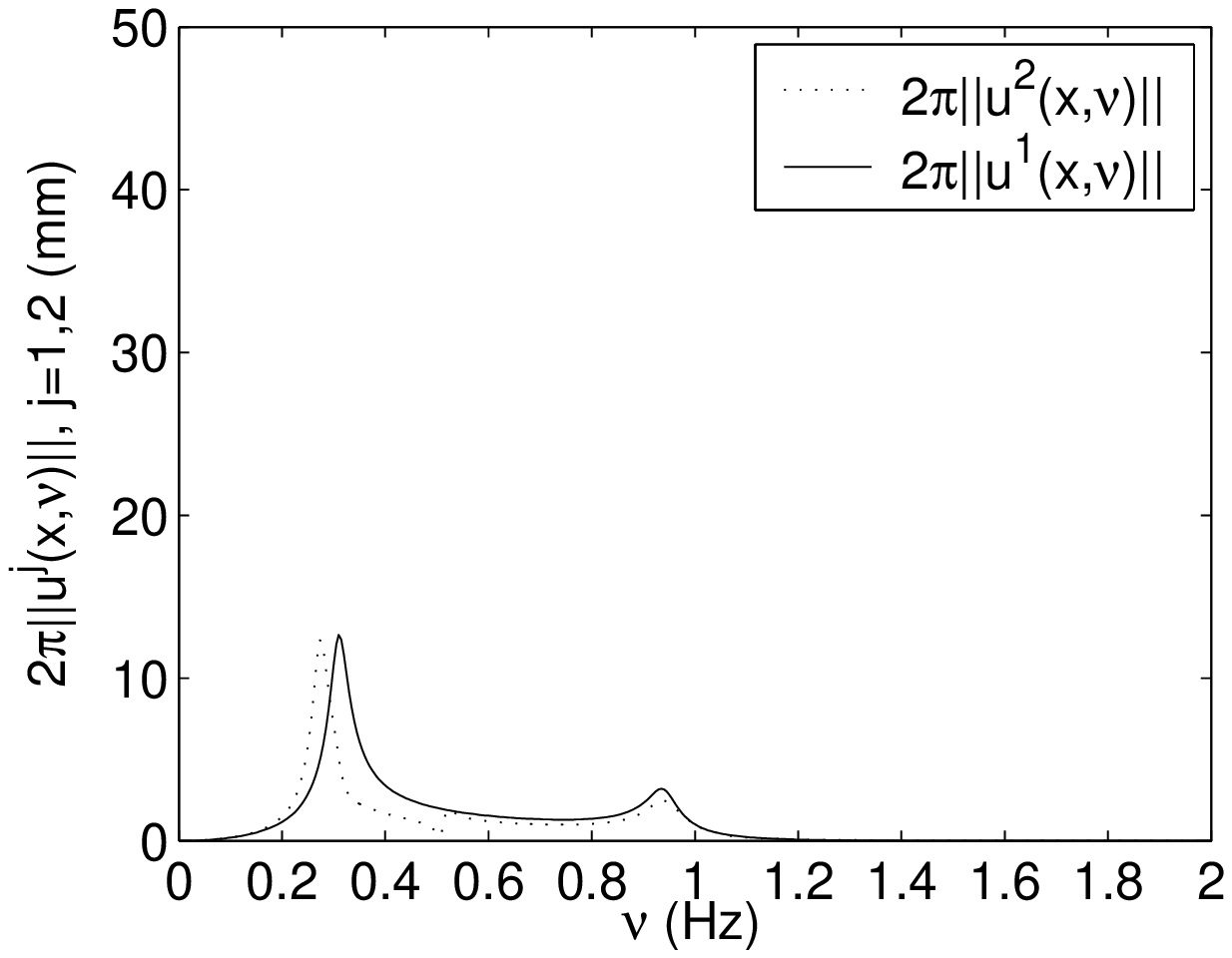}
\includegraphics[width=6.0cm] {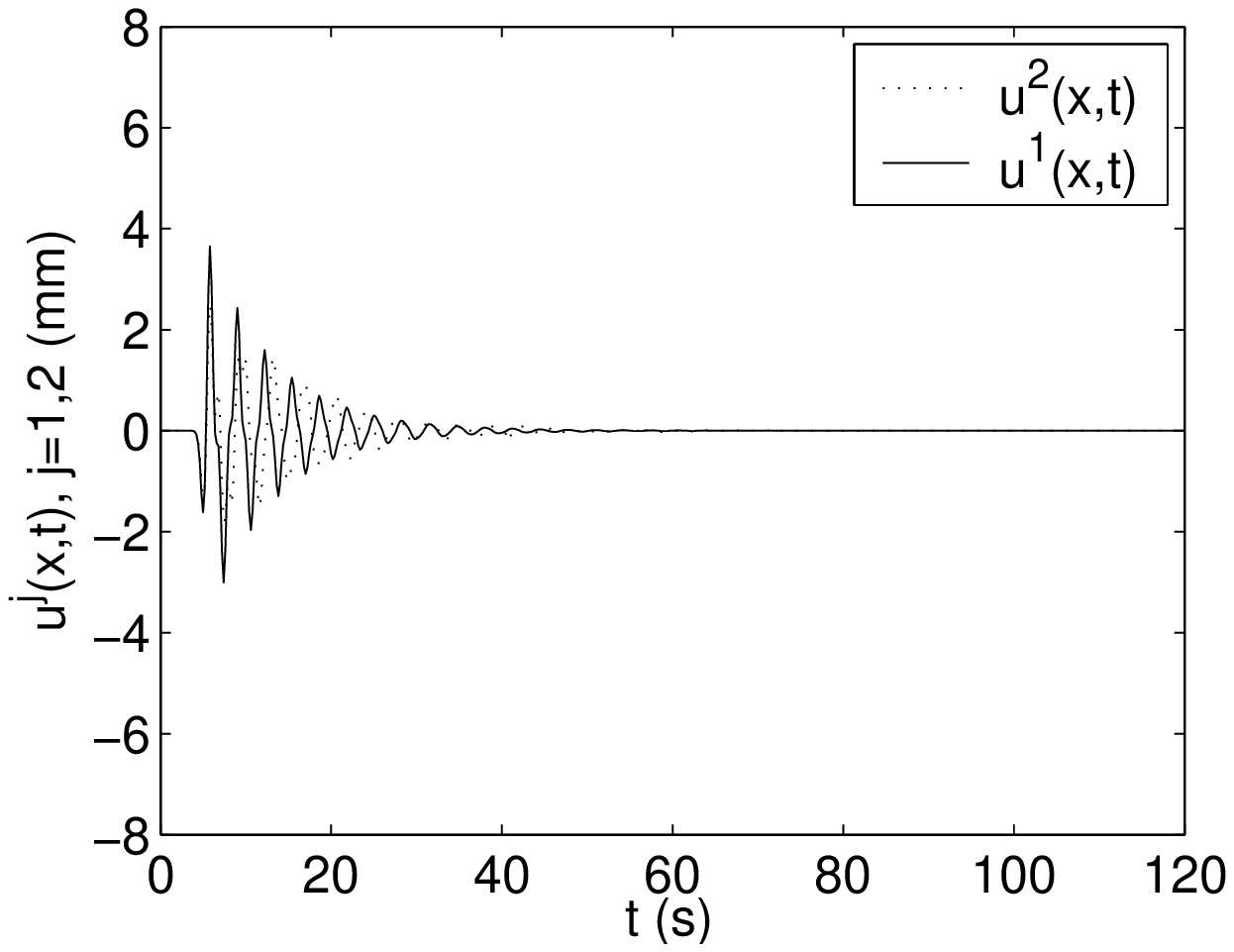}
\includegraphics[width=6.0cm] {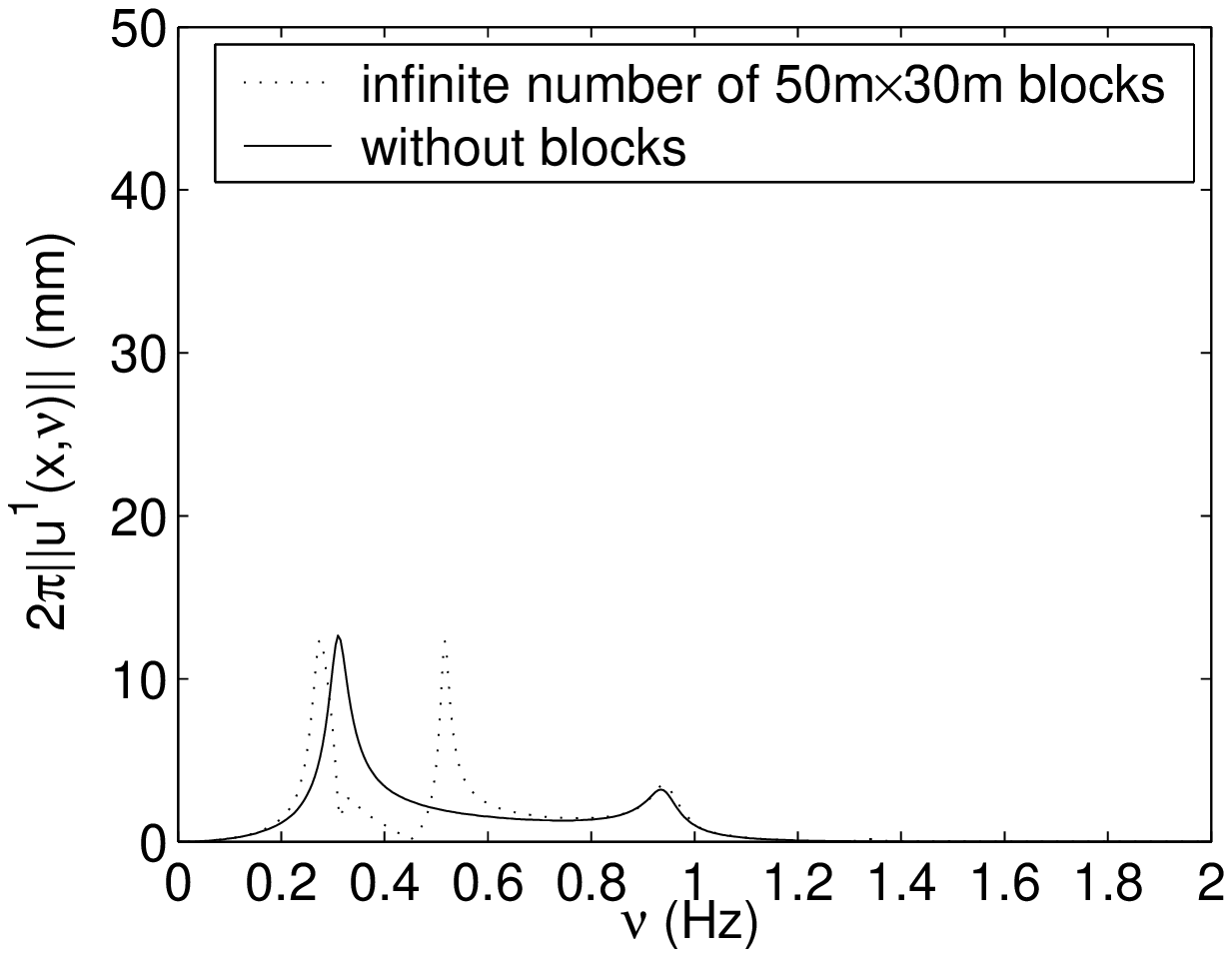}
\includegraphics[width=6.0cm] {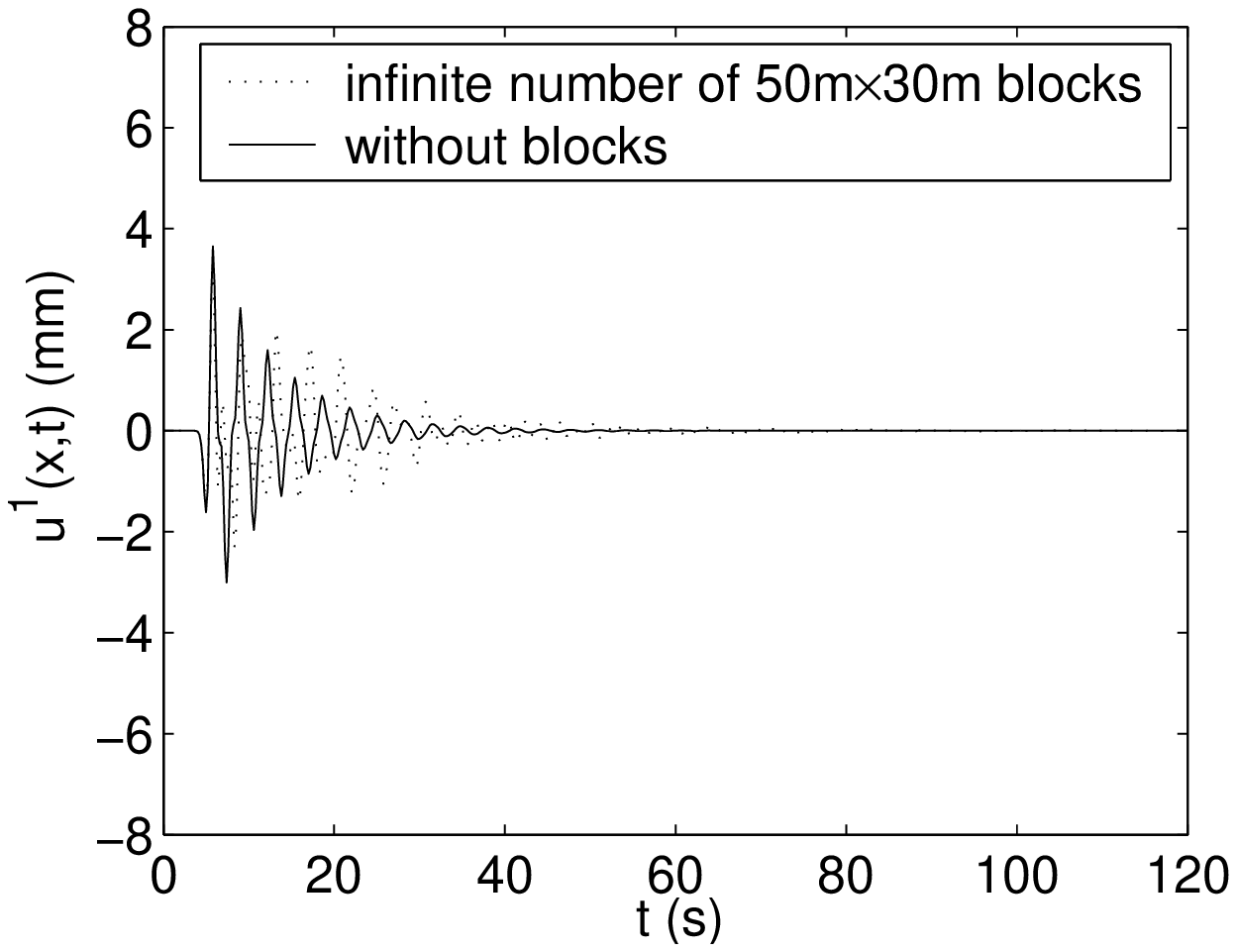}
\caption{Comparison of $2\pi$ times the  spectrum (left panels)
and time history (right panels) of the total displacement on the
ground in absence of blocks (solid curves) with  the displacement
at three locations  in the presence of blocks (dashed curves): i)
on the ground, at a location halfway between two adjacent blocks
(bottom panels), ii) at the center of the base segment of a block
(middle panels), iii) at the center of the summit segment of a
block (top panels).} \label{CompMexicod65}
\end{center}
\end{figure}
In fig.  \ref{CompMexicod65} we compare  the spectra and time
histories in the presence and absence of the $d=65$ set of blocks.

The first resonance peak in the spectra is shifted to a lower
frequency and is of higher amplitude when the blocks are present.
This is the indication of the excitation of the fundamental
quasi-Love mode responsible for what was termed the
{\textit{soil-structure interaction} in the companion paper. The
peak relative to the excitation of the perturbed displacement-free
base block mode has a relatively high amplitude, due to both the
spectrum of the incident wave and to the presence of the other
blocks. Effectively, the resonance frequency does not correspond
to that of the displacement-free base block mode  (at which
frequency the response is nil at the center of the base segment of
the block), and the particular shape of the spectrum at the center
of the base segment of the block is characteristic of the
excitation of a multi-displacement-free base block mode and not to
the excitation of a quasi displacement-free base block mode (as
shown in the companion paper). This coupled mode also takes into
account  the so-called structure-soil-structure interaction as
(see the companion paper) was suggested in sect. \ref{quasidisf}.

A splitting of the first peak appears in the ground (between the
blocks) response. The second of the split peaks has a low quality
factor, which fact means the quasi-absence of beatings in the time
history of response.

The temporal displacements are of higher amplitude and duration in
the presence of blocks than in the absence of blocks, particularly
at the center of the top segment of the block. These results are
evocative of those obtained in the companion paper for two blocks.
\clearpage
\subsubsection{Comparison of the seismic responses for one, two
and an infinite number of blocks
with center-to-center separations $d=65$m}
In fig. \ref{CompMexicod65hautonetwoinf}, we compare the spectra
and time histories of response for  one-block, two-block, and
infinite-block configurations at the center of the {\it top
segment} of one of the blocks.
\begin{figure}[ptb]
\begin{center}
\includegraphics[width=6.0cm] {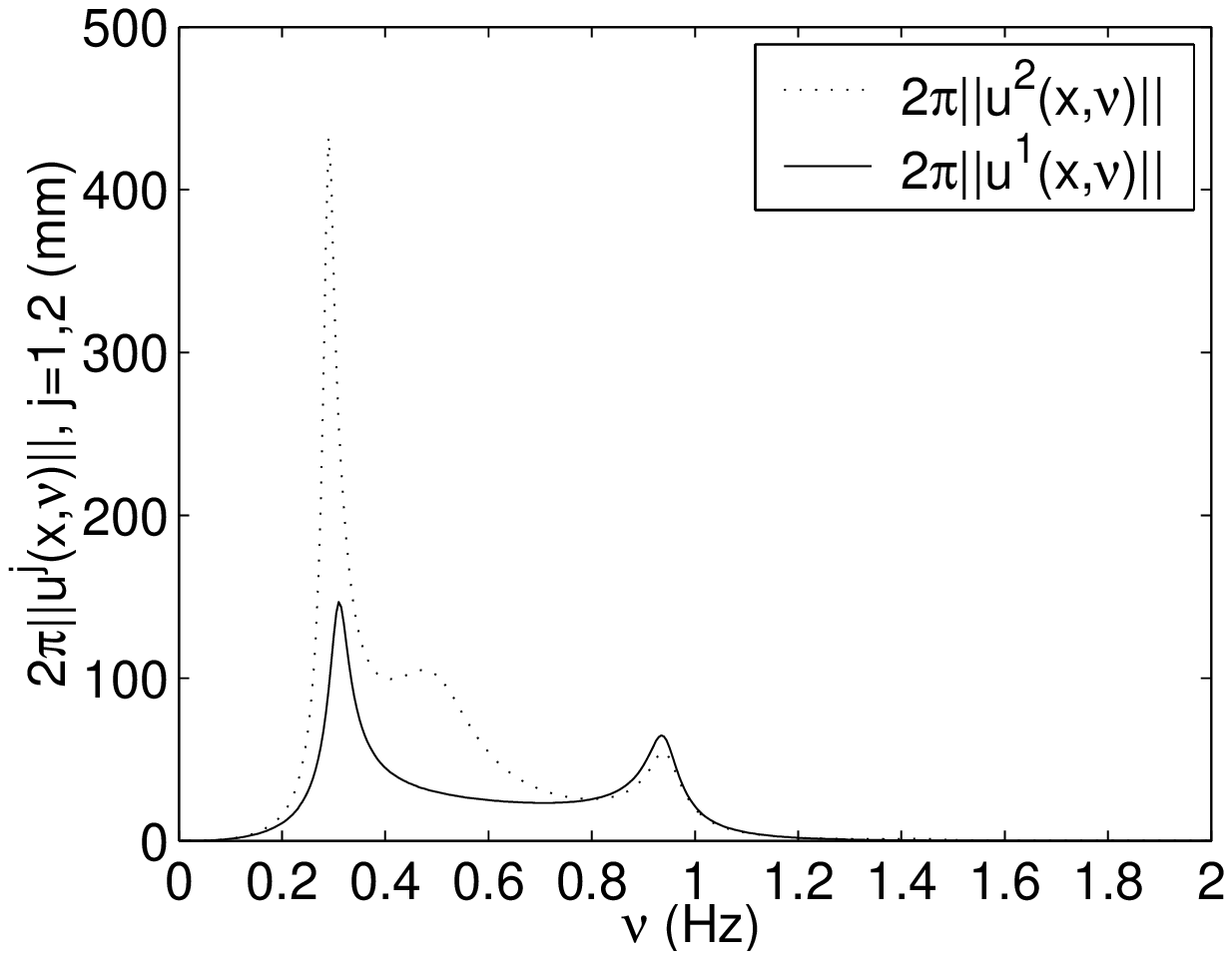}
\includegraphics[width=6.0cm] {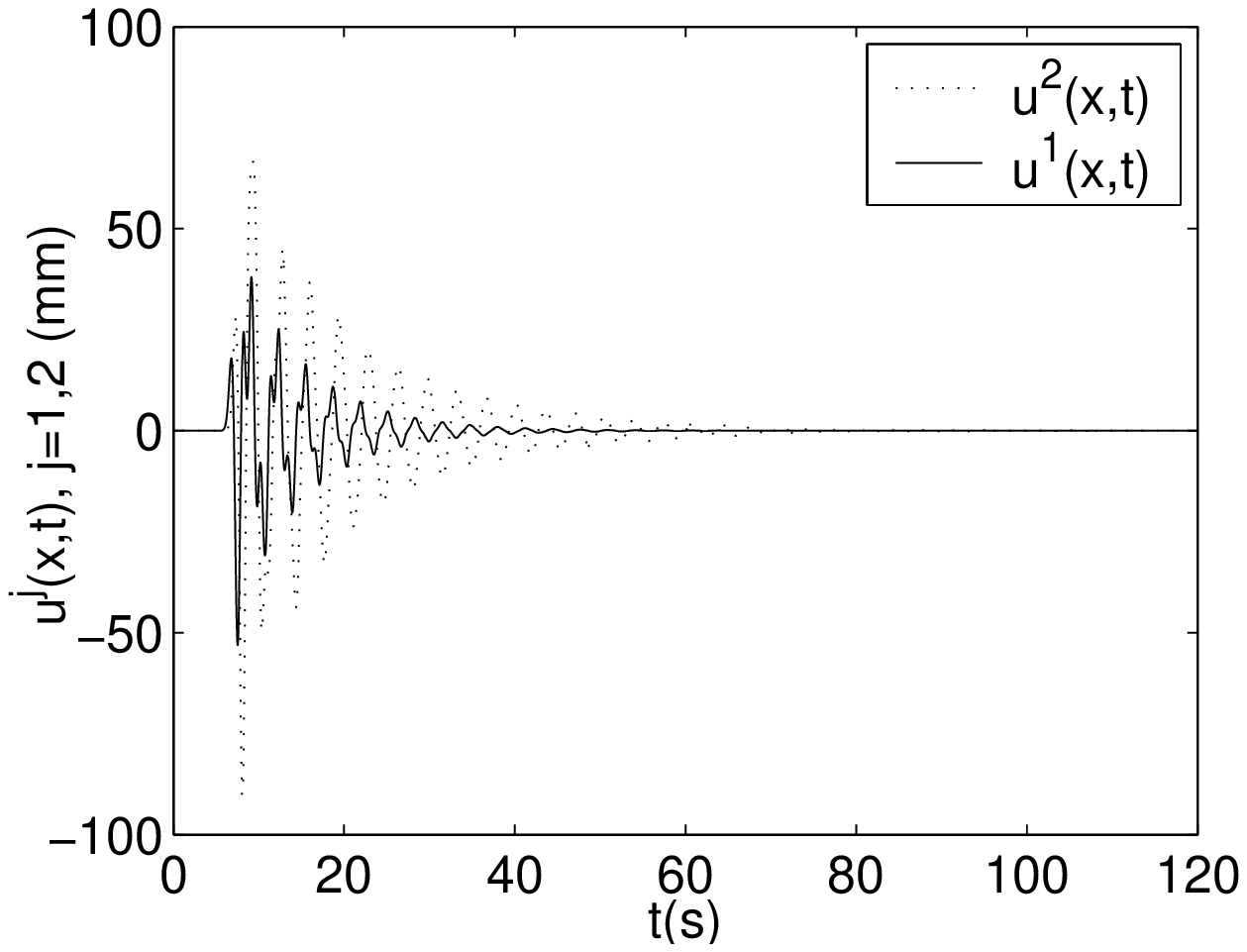}
\includegraphics[width=6.0cm] {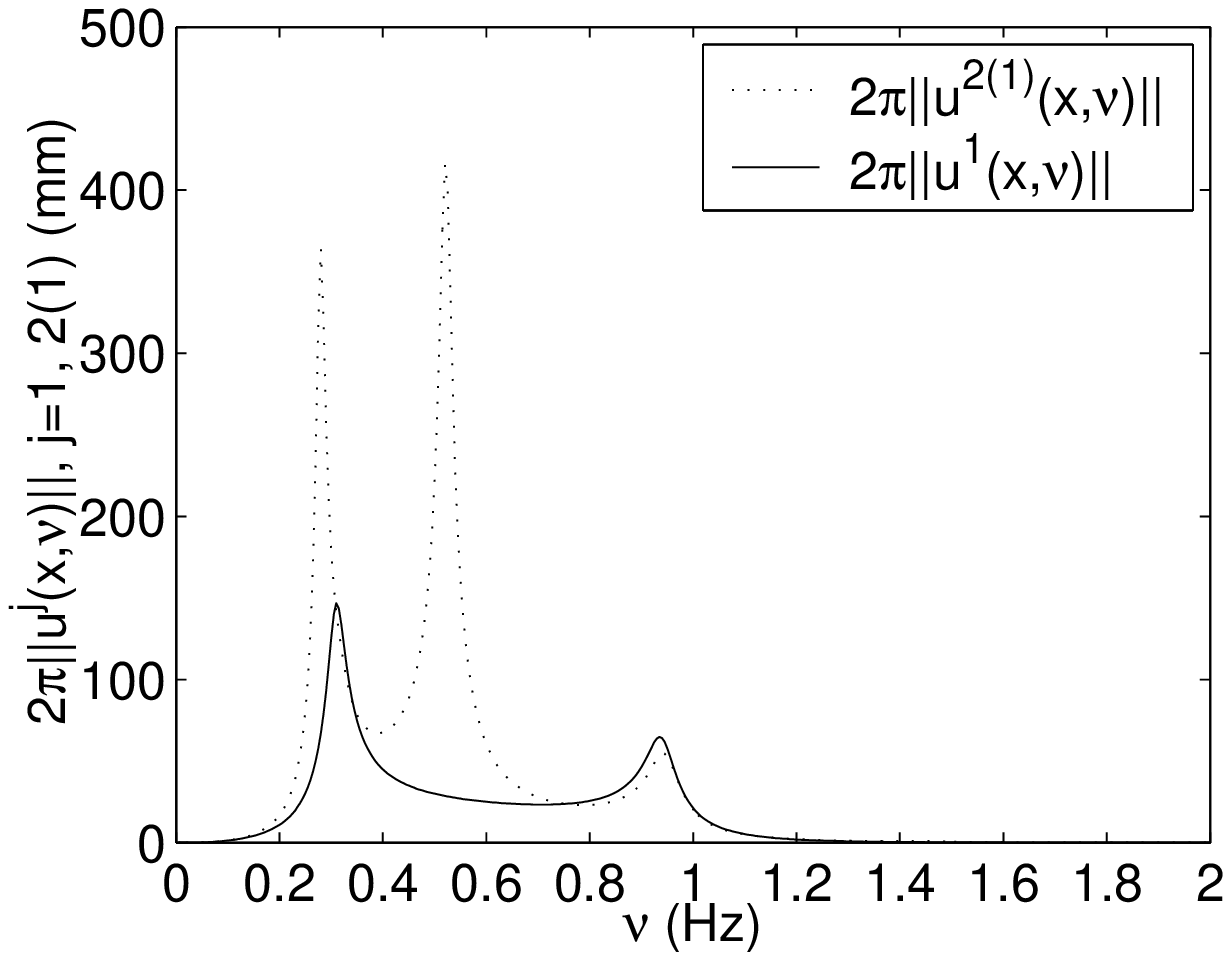}
\includegraphics[width=6.0cm] {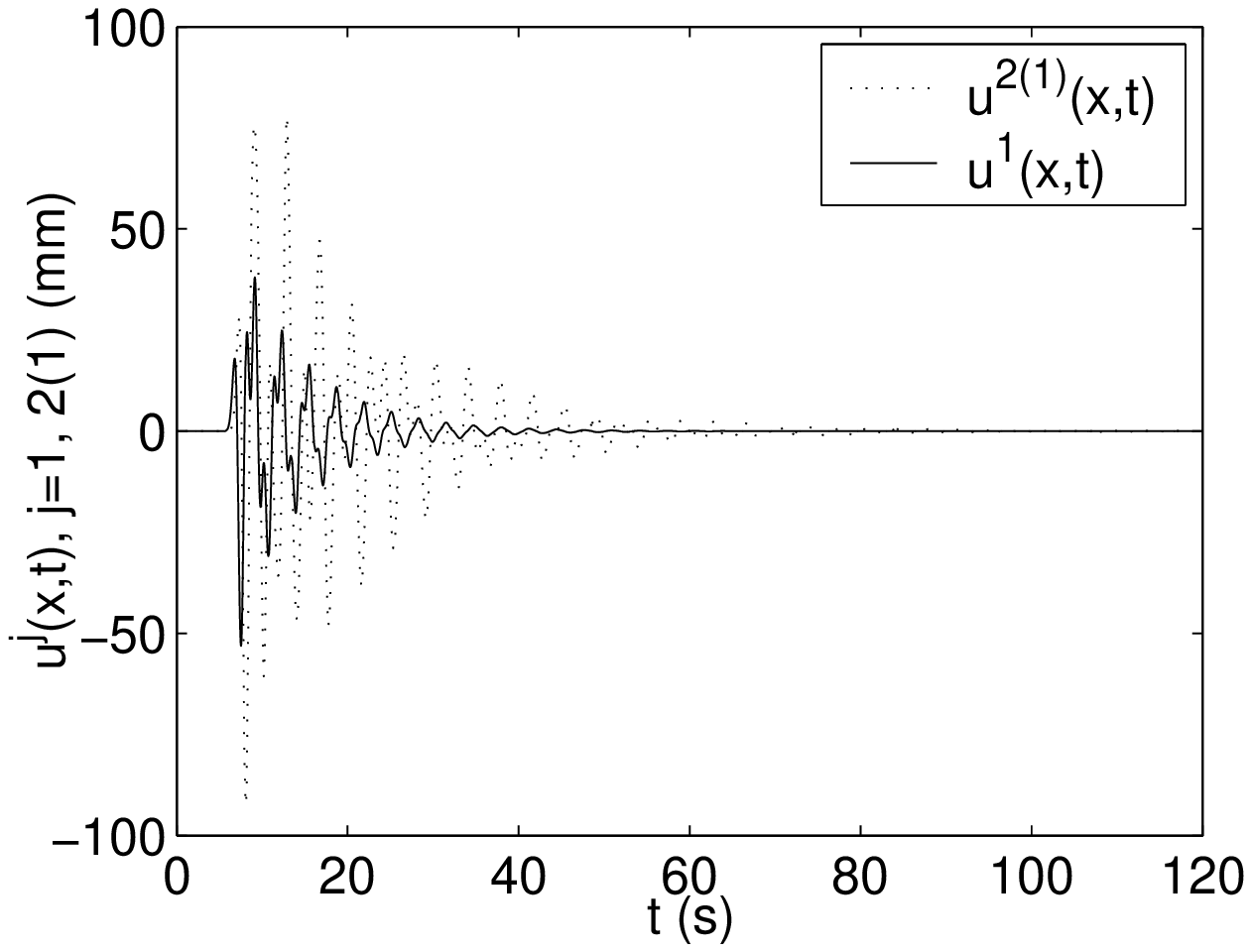}
\includegraphics[width=6.0cm] {COMPSPECTRE5030HAUTd65.eps}
\includegraphics[width=6.0cm] {COMPTEMP5030HAUTd65.eps}
\caption{Comparison of $2\pi$ times the  spectrum (left panels)
and time history (right panels) of the total displacement on the
ground in absence of blocks (solid curves) with  the displacement
 at  the center of the {\it top segment} of a $50m\times
30m$ block (dashed curves) for: i) one such blocks (top panels),
ii) two such blocks with center-to-center spacing $d=65m$ (middle
panels), and iii) an infinite number of such blocks with
center-to-center spacing $d=65m$ (bottom panels). The one- and
two-block results are taken from the companion paper and apply to
the response to the wave emitted by a deep line source situated at
(0m,3000m) whose amplitude, at ground level is different from that
of the plane wave soliciting the infinite block structure.}
\label{CompMexicod65hautonetwoinf}
\end{center}
\end{figure}
This figure shows that the effect of increasing the number of
blocks is essentially to increase the height of the second (i.e.,
higher-frequency) resonance peak at the expense of the first
resonance peak, and thus to introduce stronger high-frequency
oscillations in the temporal response. As the quality factor of
the second resonance peak increases with $N_{b}$, the durations
also increase with $N_{b}$. The combined effect of the increased
quality factors and higher frequency oscillations is increased
cumulative motion of the block.

In fig. \ref{CompMexicod65basonetwoinf}, we compare the spectra
and time histories of seismic response for one-block, two-block,
and infinite-block configurations at the center of the {\it bottom
segment} of one of the blocks.
\begin{figure}[ptb]
\begin{center}
\includegraphics[width=6.0cm] {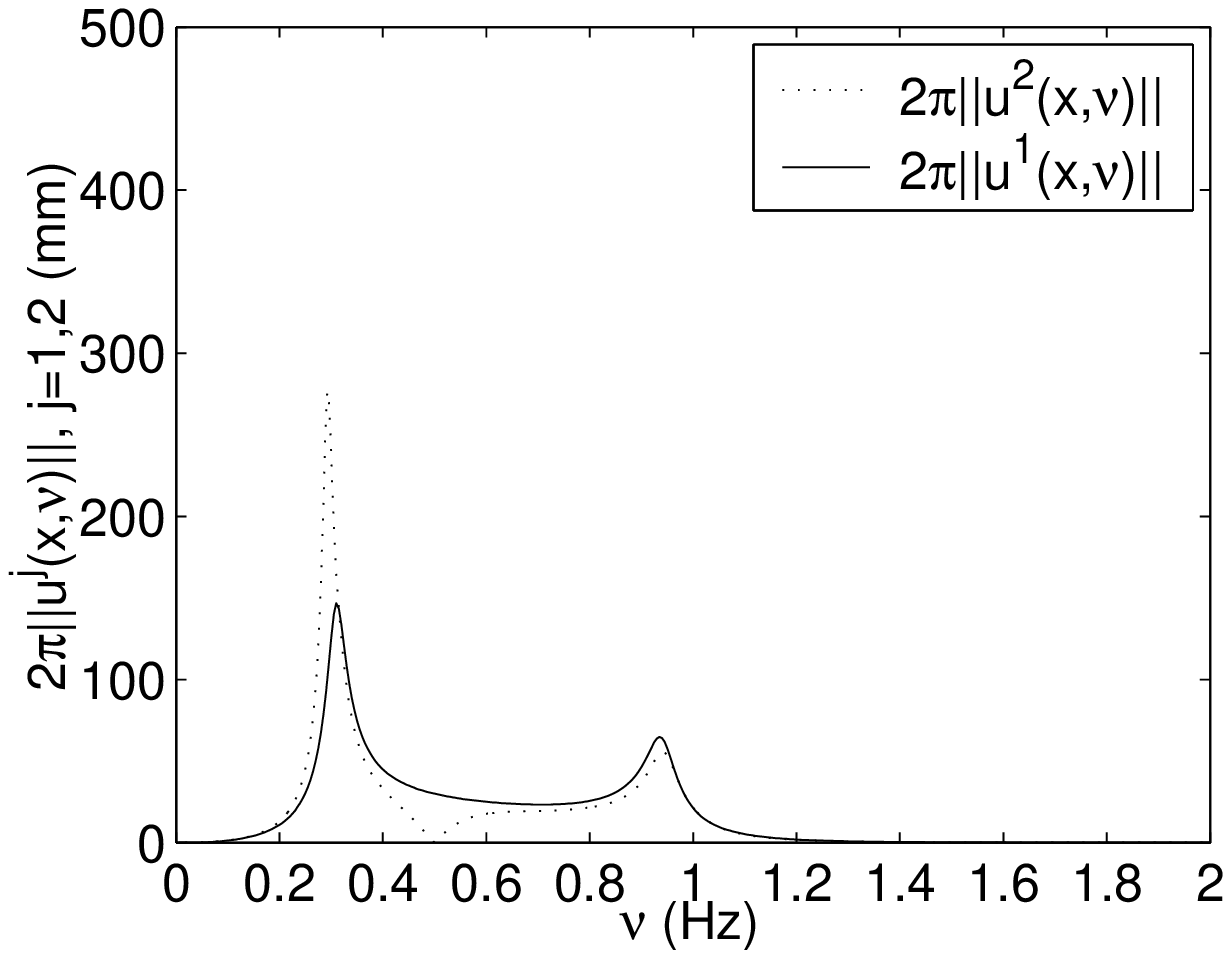}
\includegraphics[width=6.0cm] {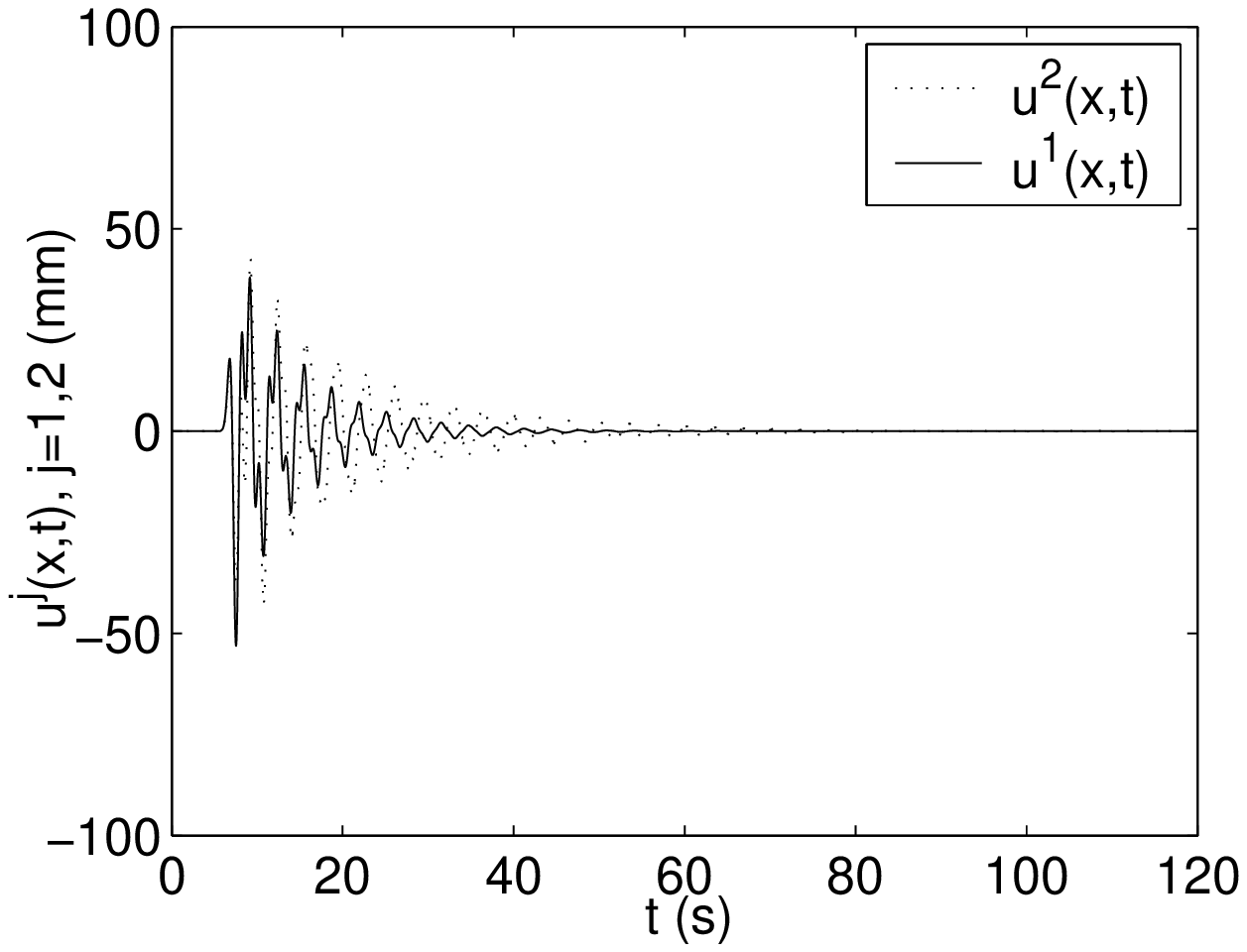}
\includegraphics[width=6.0cm] {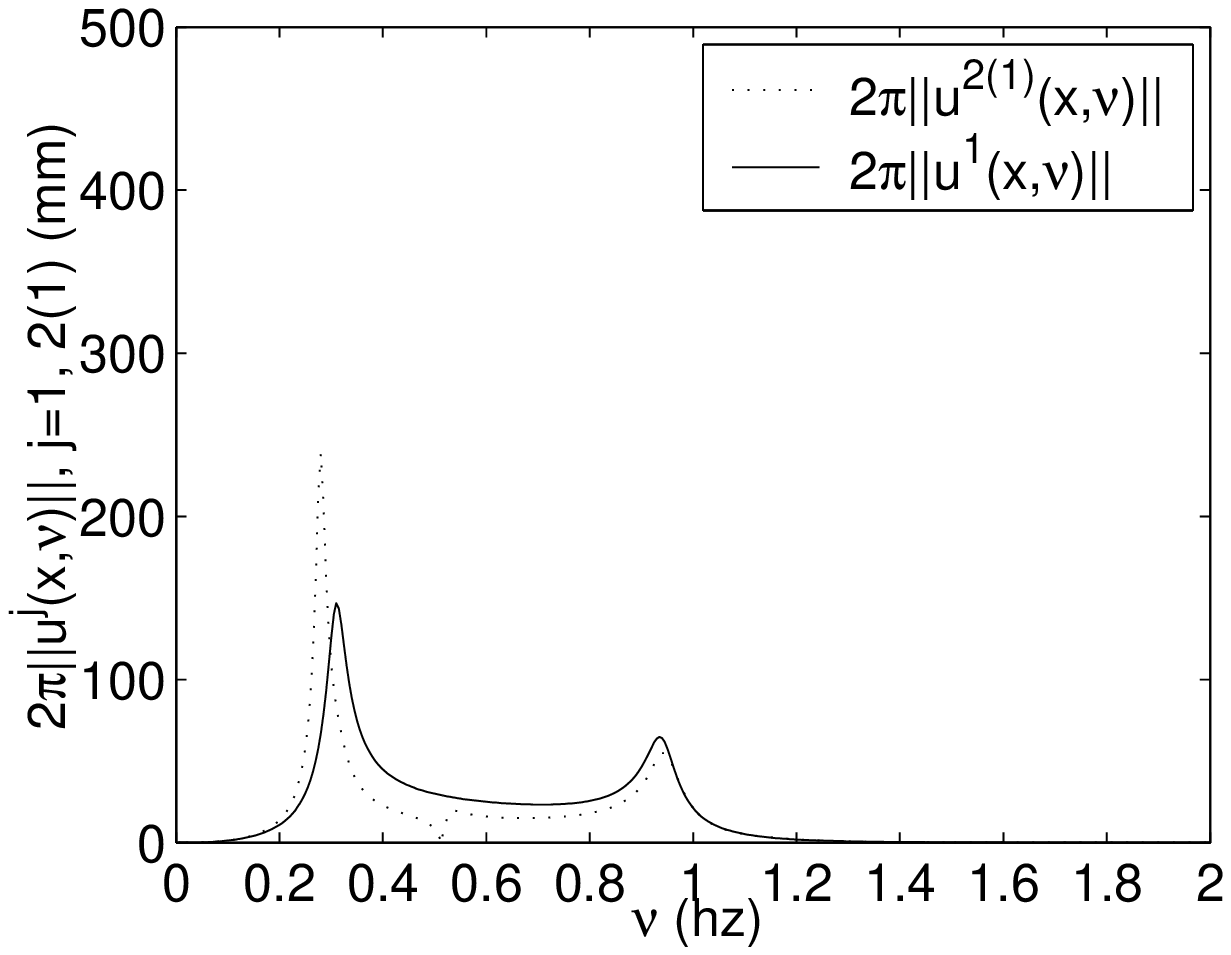}
\includegraphics[width=6.0cm] {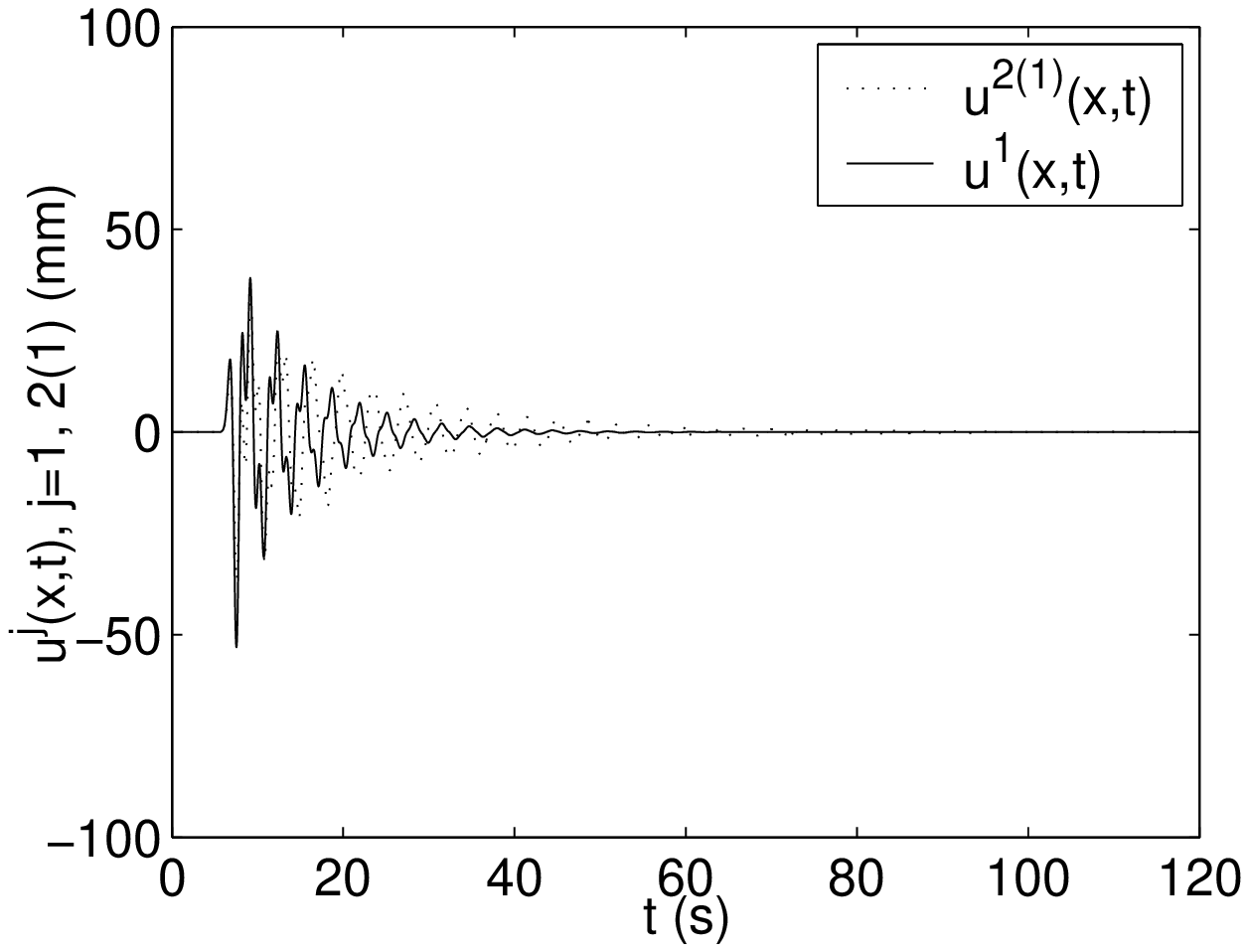}
\includegraphics[width=6.0cm] {COMPSPECTRE5030BASd65.eps}
\includegraphics[width=6.0cm] {COMPTEMP5030BASd65.eps}
\caption{Comparison of $2\pi$ times the  spectrum (left panels)
and time history (right panels) of the total displacement on the
ground in absence of blocks (solid curves) with  the displacement
at  the center of the {\it bottom segment} of a $50m\times 30m$
block (dashed curves) for: i) one such blocks (top panels), ii)
two such blocks with center-to-center spacing $d=65m$ (middle
panels), and iii) an infinite number of such blocks with
center-to-center spacing $d=65m$ (bottom panels). The one- and
two-block results are taken from the companion paper and apply to
the response to the wave emitted by a deep line source situated at
(0m,3000m) whose amplitude, at ground level is different from that
of the plane wave soliciting the infinite block structure.}
\label{CompMexicod65basonetwoinf}
\end{center}
\end{figure}
Since the second resonance now is synonymous with a minimum of
response, the aforementioned effects are not produced at the base
of the block. In fact, with increasing $N_{b}$, we actually
observe a decrease of the height of the first resonance peak,
whose effect is to decrease the amplitude and duration of motion
in the time histories.
\clearpage
\subsubsection{Comparison of the seismic responses for ten,
twenty, fourty block configurations with that of a configuration
having an infinite number of blocks for center-to-center
separations $d=65$m}
\begin{figure}[ptb]
\begin{center}
\includegraphics[width=6.0cm] {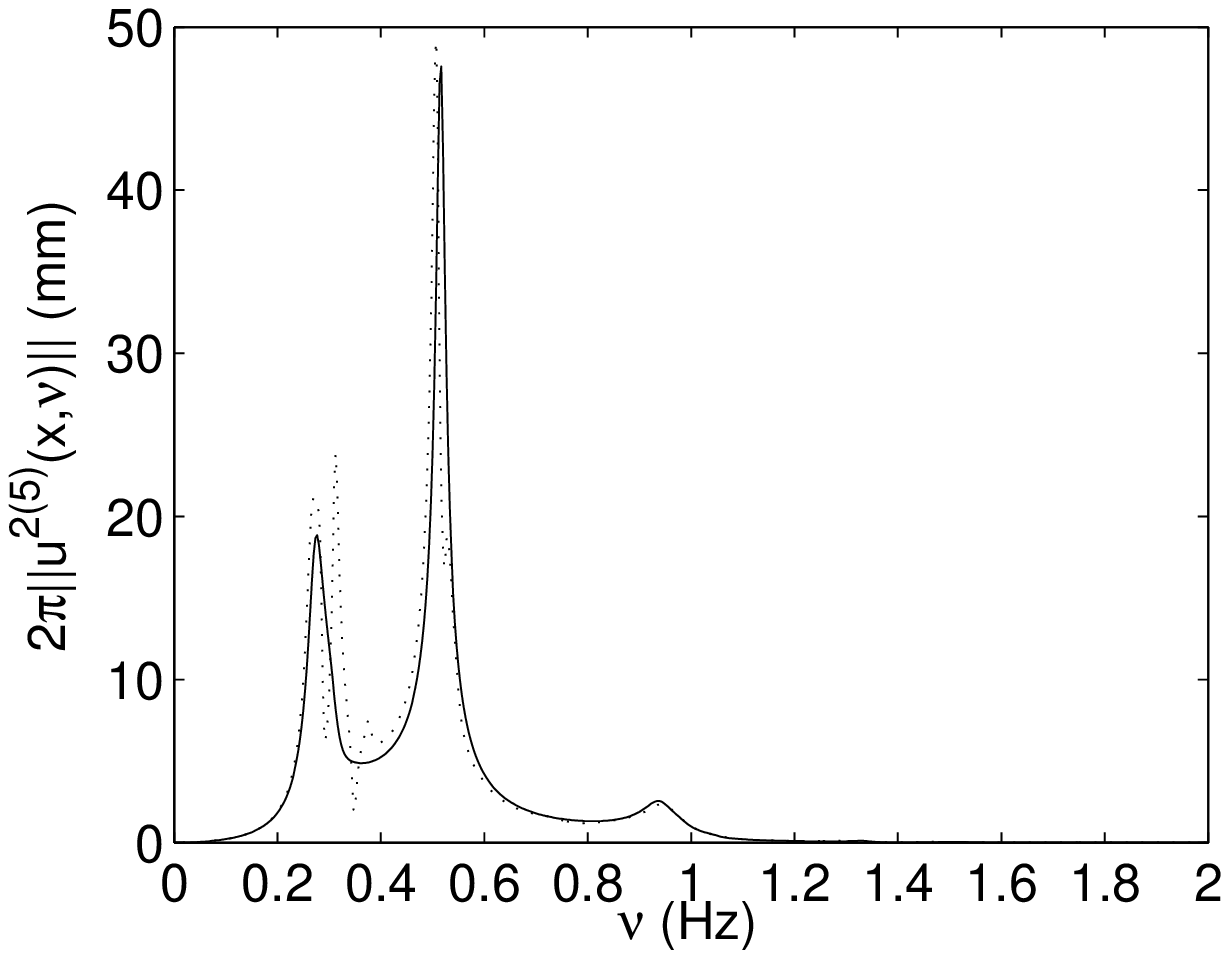}
\includegraphics[width=6.0cm] {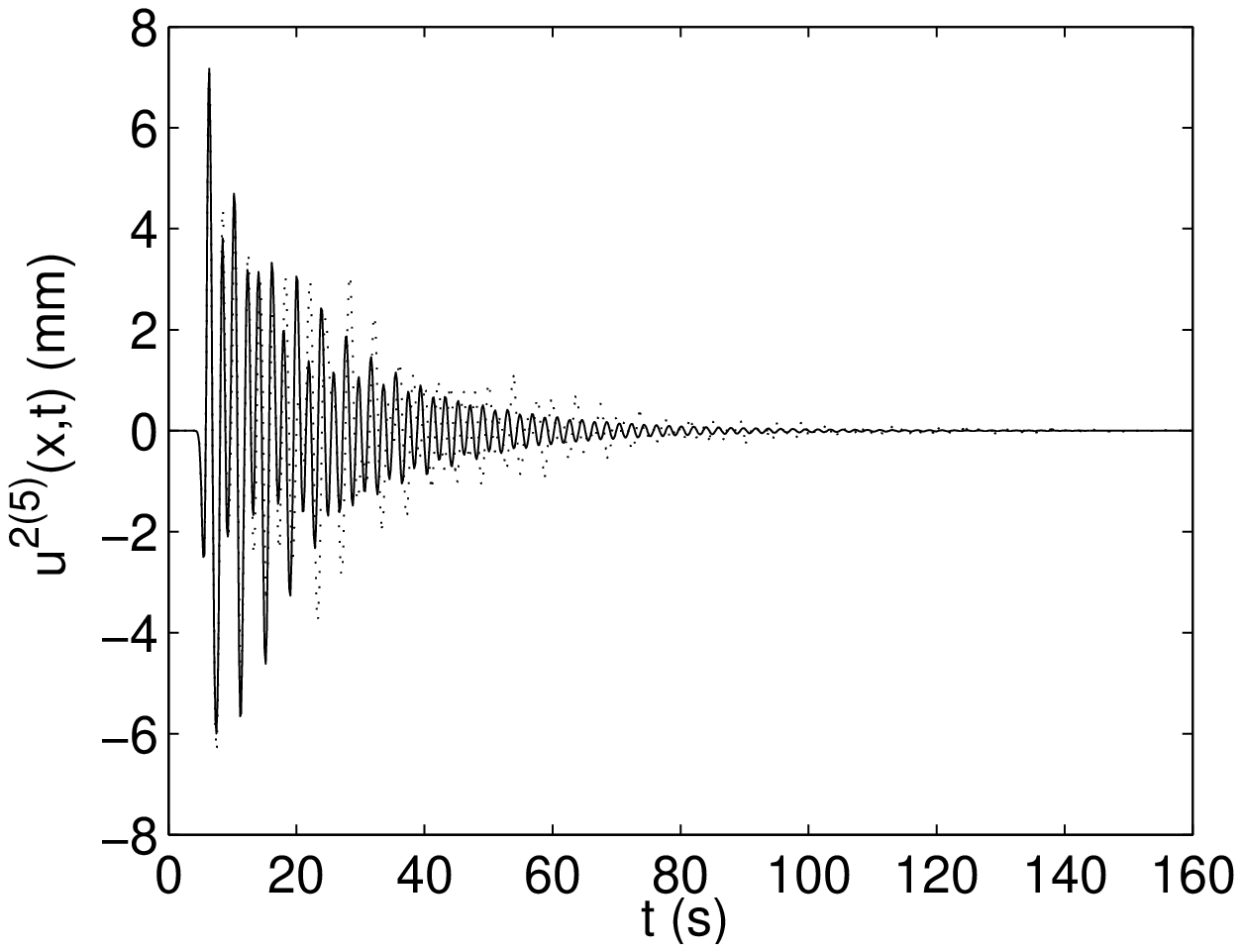}
\includegraphics[width=6.0cm] {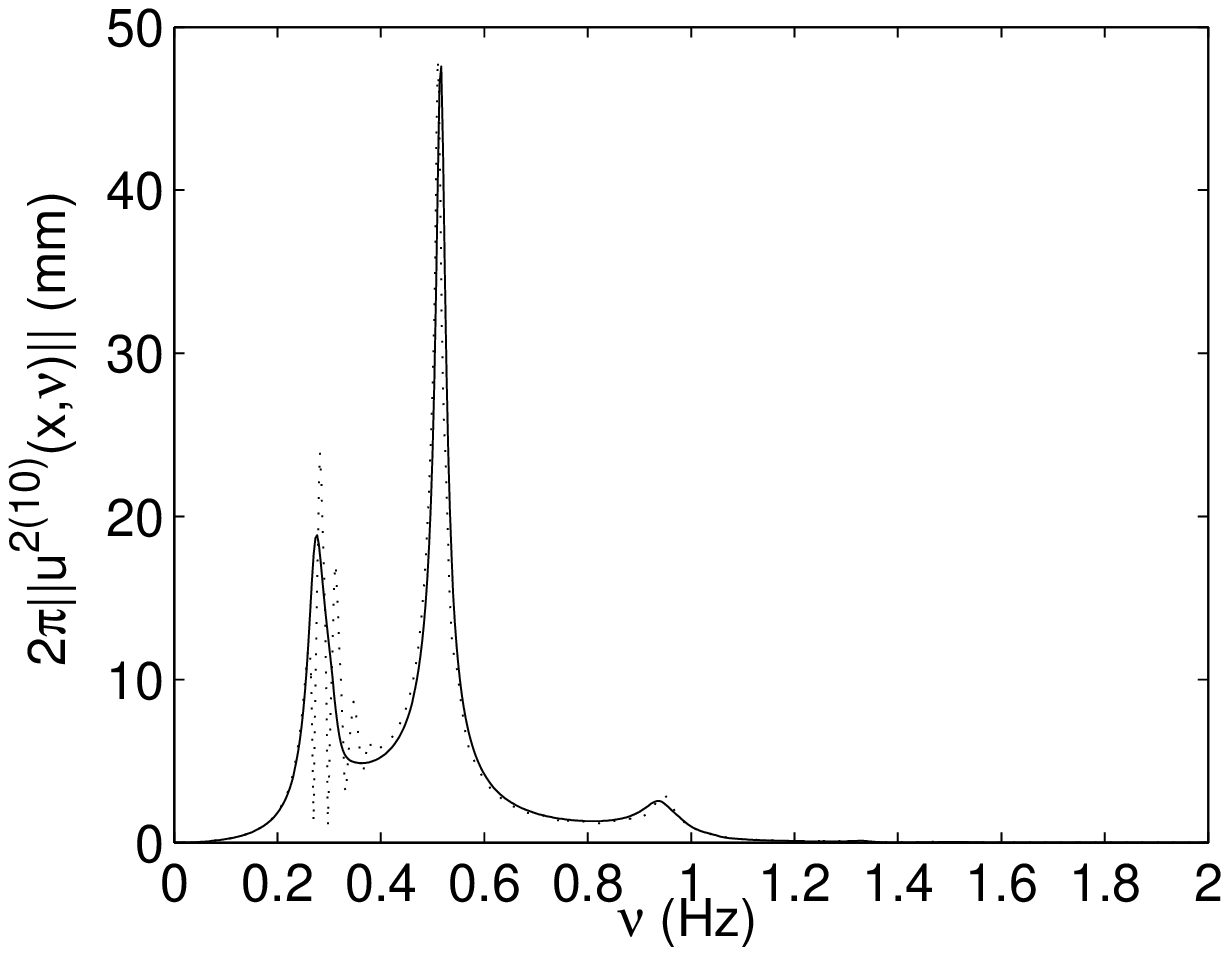}
\includegraphics[width=6.0cm] {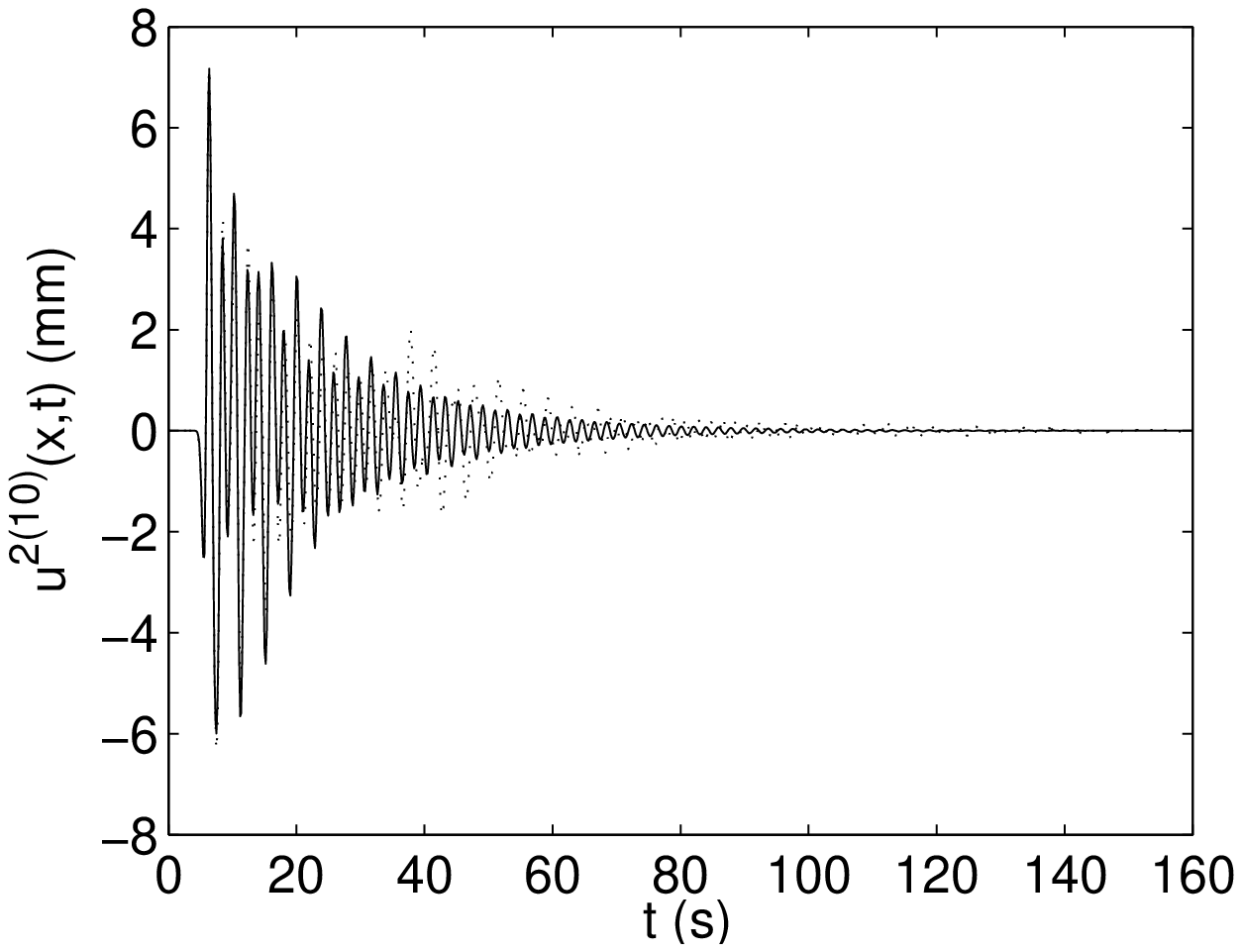}
\includegraphics[width=6.0cm] {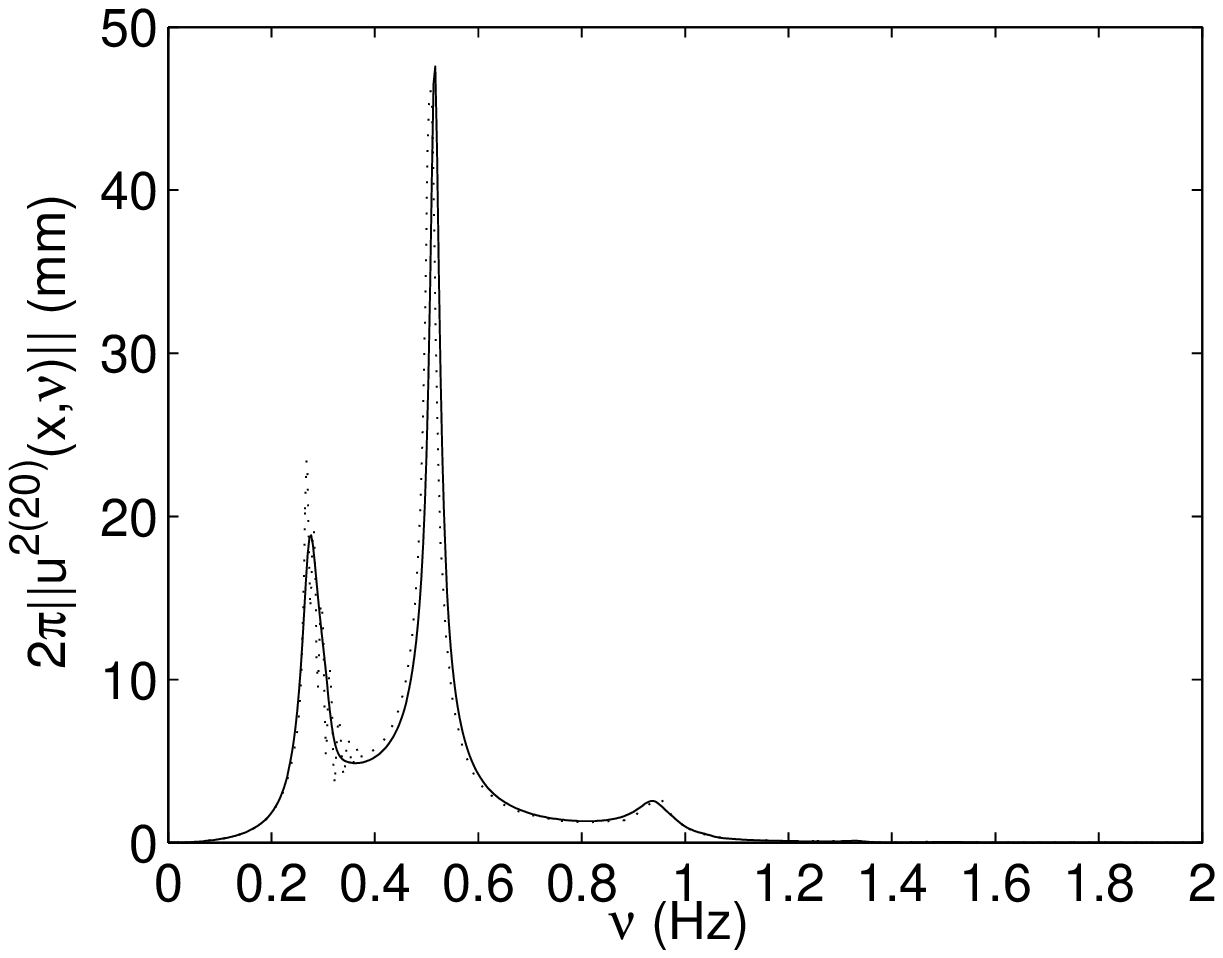}
\includegraphics[width=6.0cm] {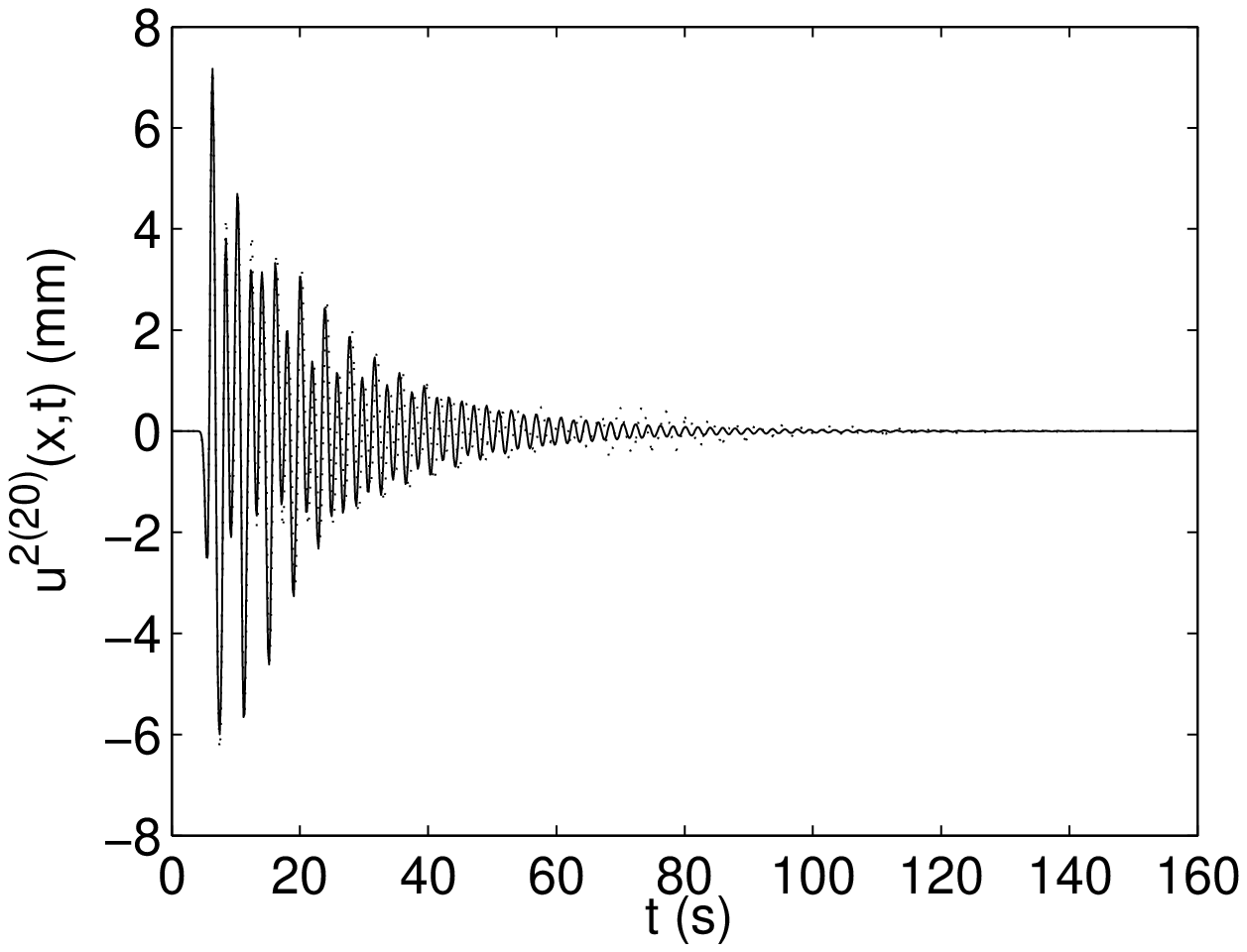}
\caption{Comparison of $2\pi$ times the  spectra (left panels) and
time histories (right panels) of the total displacement at  the
center of the top segment of a $50m\times 30m$ block of a
configuration with an infinite number of blocks (solid curves)
with the displacement at the same location of a central block
(dashed curves) of configurations having: i) ten such blocks (top
panels), ii) twenty such blocks (middle panels), and iii) fourty
such blocks. In all cases, the center-to-center spacing is
$d=65m$.} \label{CompMexicod65toptentwentfourt}
\end{center}
\end{figure}
In fig. \ref{CompMexicod65toptentwentfourt} we compare the spectra
and time histories of seismic response at the center of the {\it
top segment} of a centrally-located block in configurations with
10, 20, 40 and an infinite number of $50m\times 30m$ blocks
separated by $d=65$m. On the whole,  these displacement responses
are all the same, in both the frequency and time domains, marked
by relatively-long duration duration ($\approx 2$ min), and large
maximum and cumulative amplitudes. However, for the finite values
of $N_{b}$, there appears some splitting of the $N_{b}=\infty$ low
frequency resonance peak which gives rise to  beatings in addition
to those due to the presence of the two main resonance peaks.
\begin{figure}[ptb]
\begin{center}
\includegraphics[width=6.0cm] {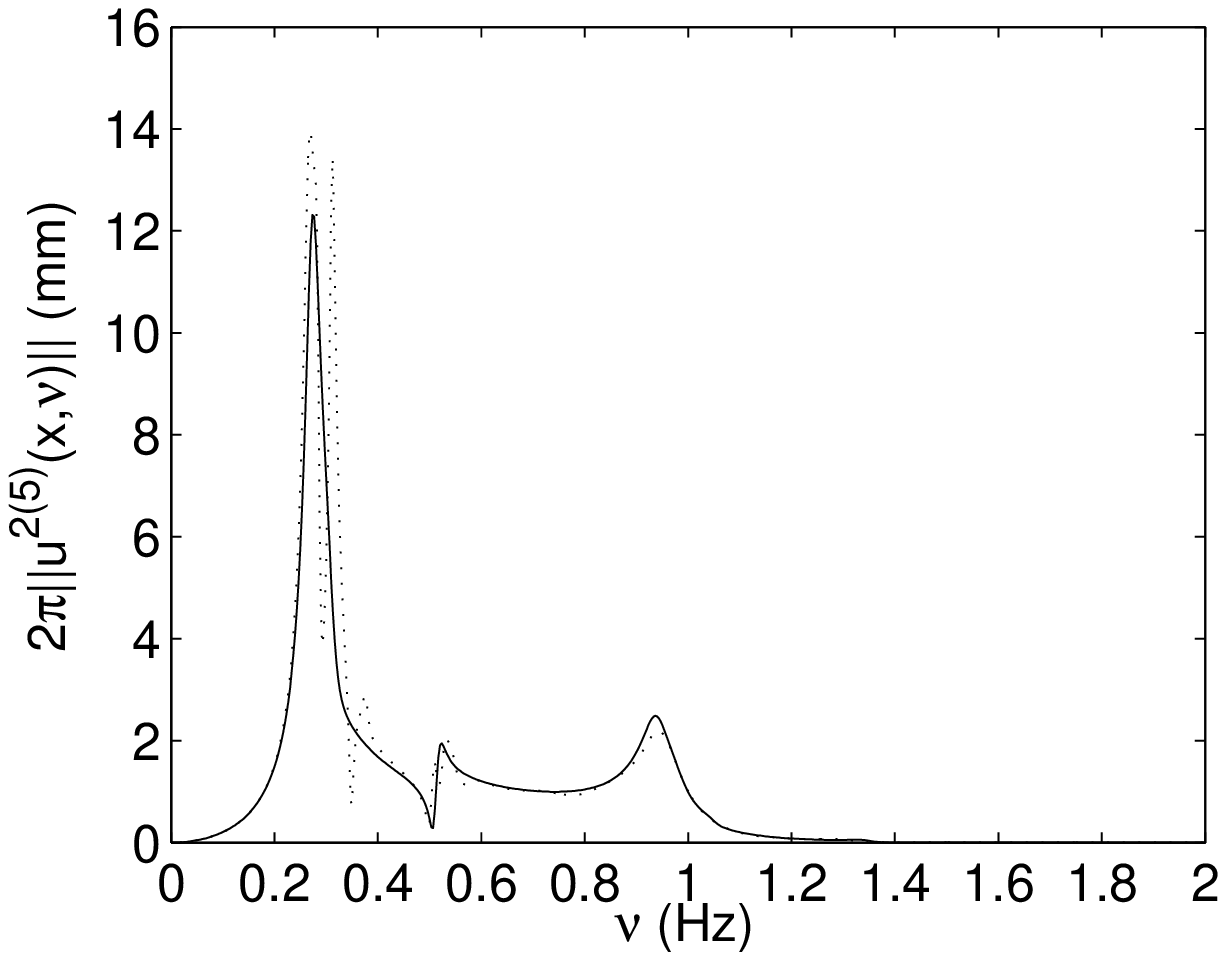}
\includegraphics[width=6.0cm] {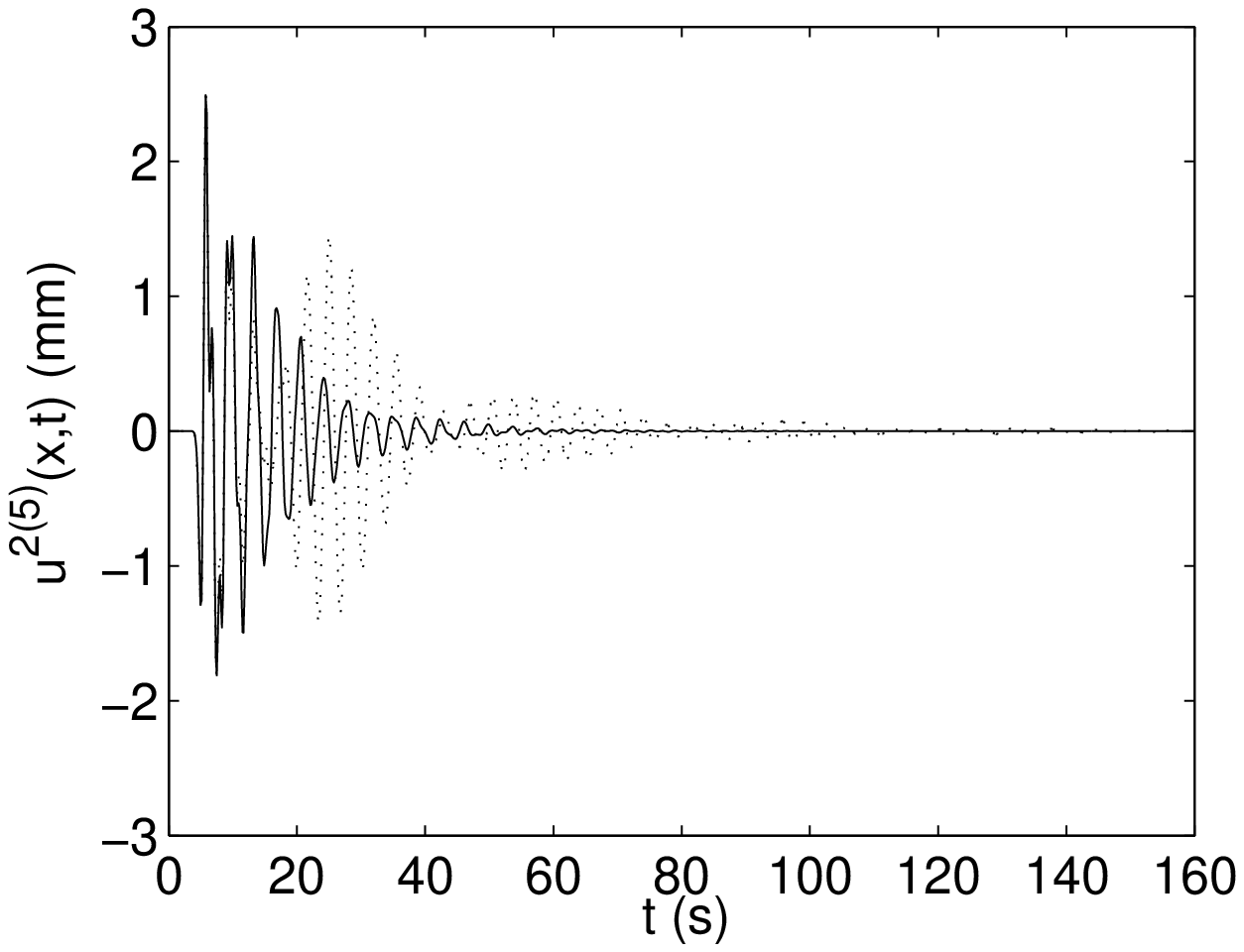}
\includegraphics[width=6.0cm] {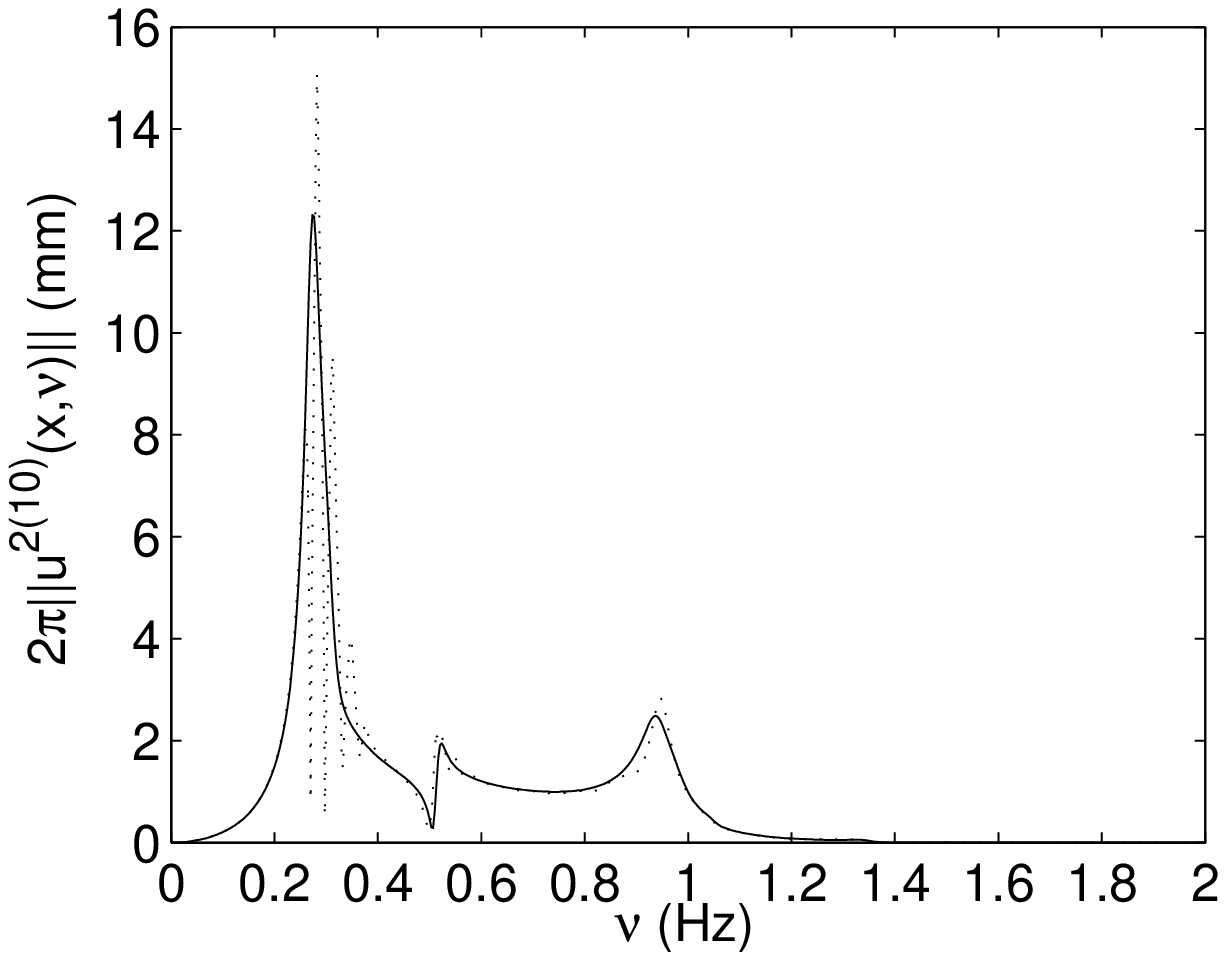}
\includegraphics[width=6.0cm] {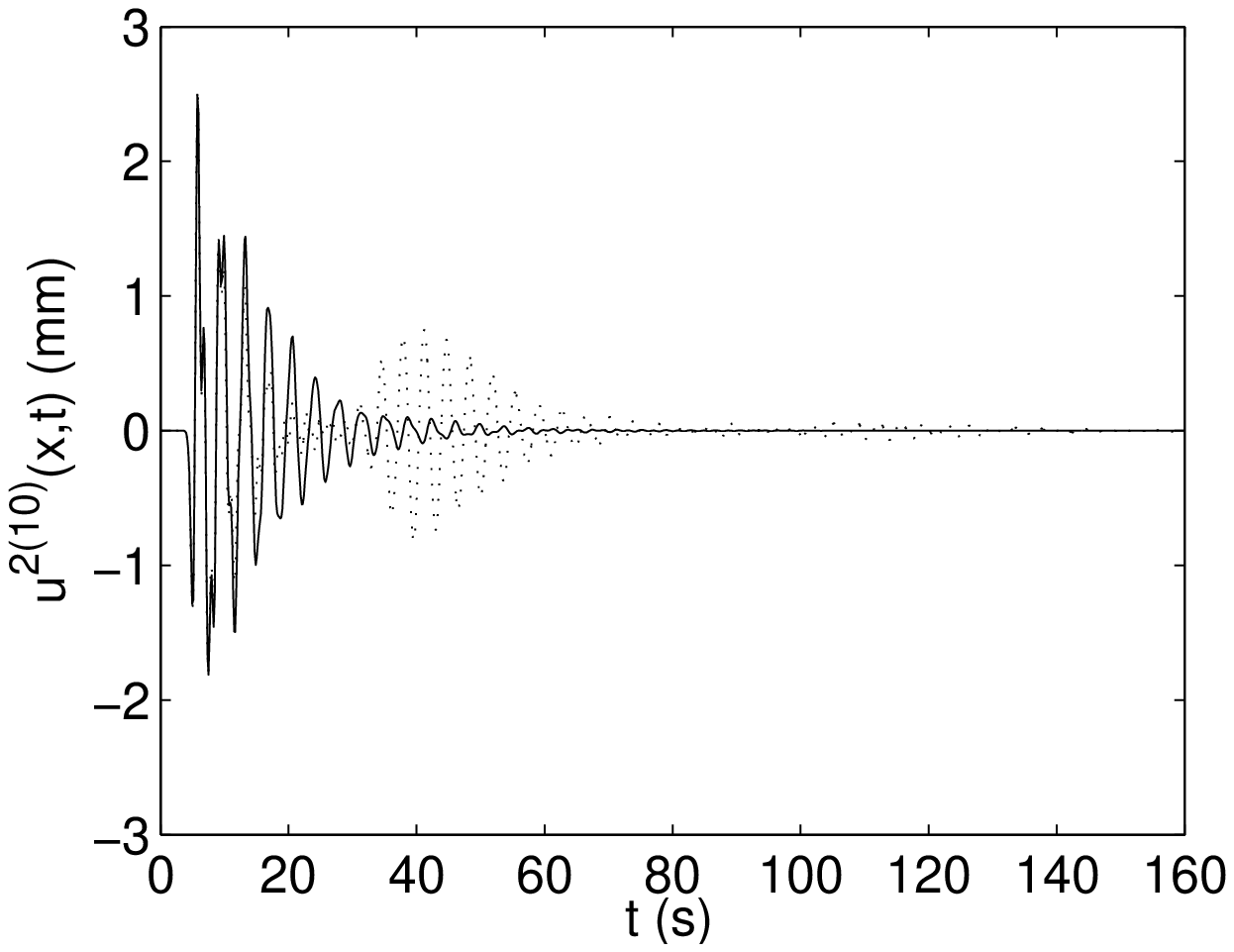}
\includegraphics[width=6.0cm] {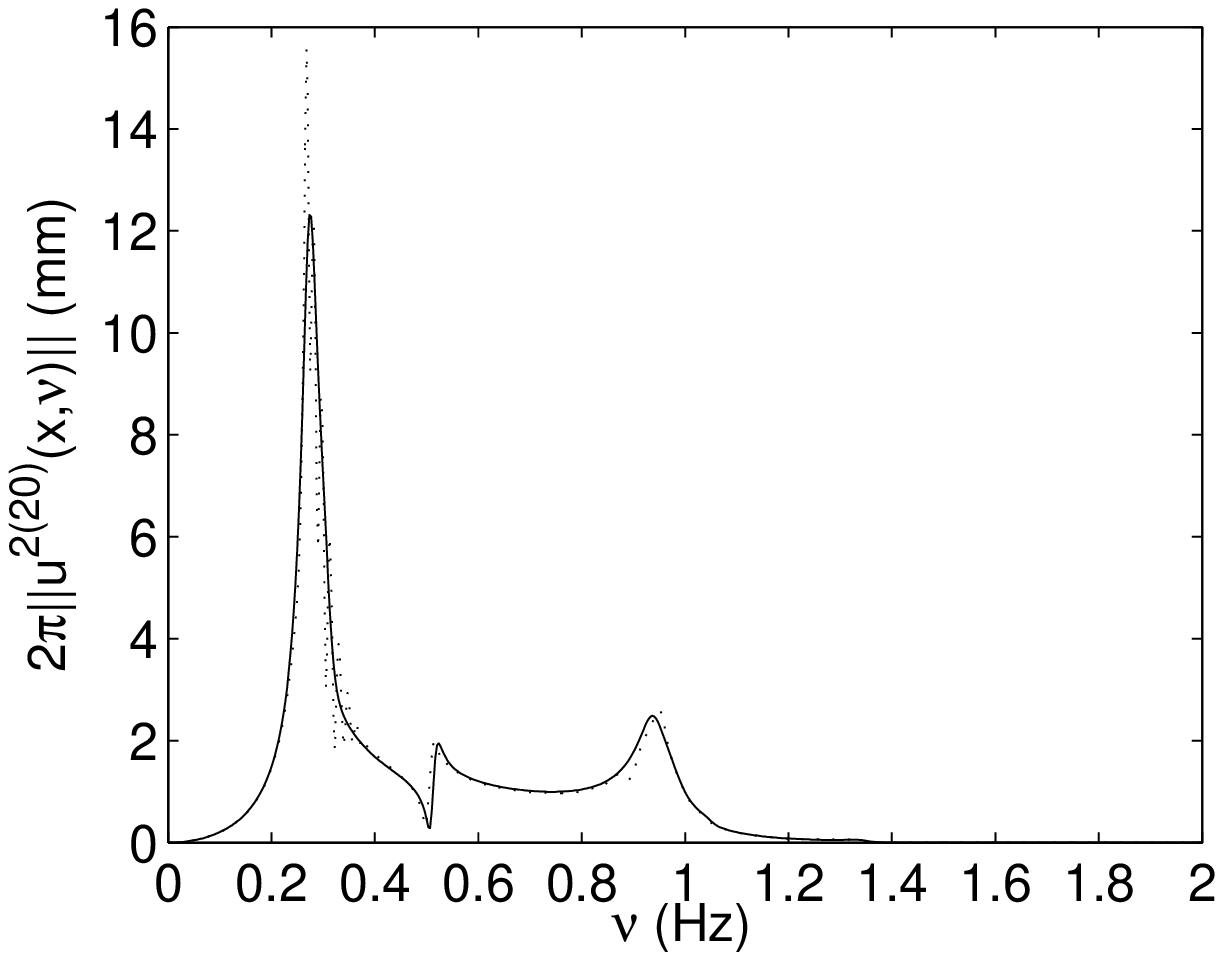}
\includegraphics[width=6.0cm] {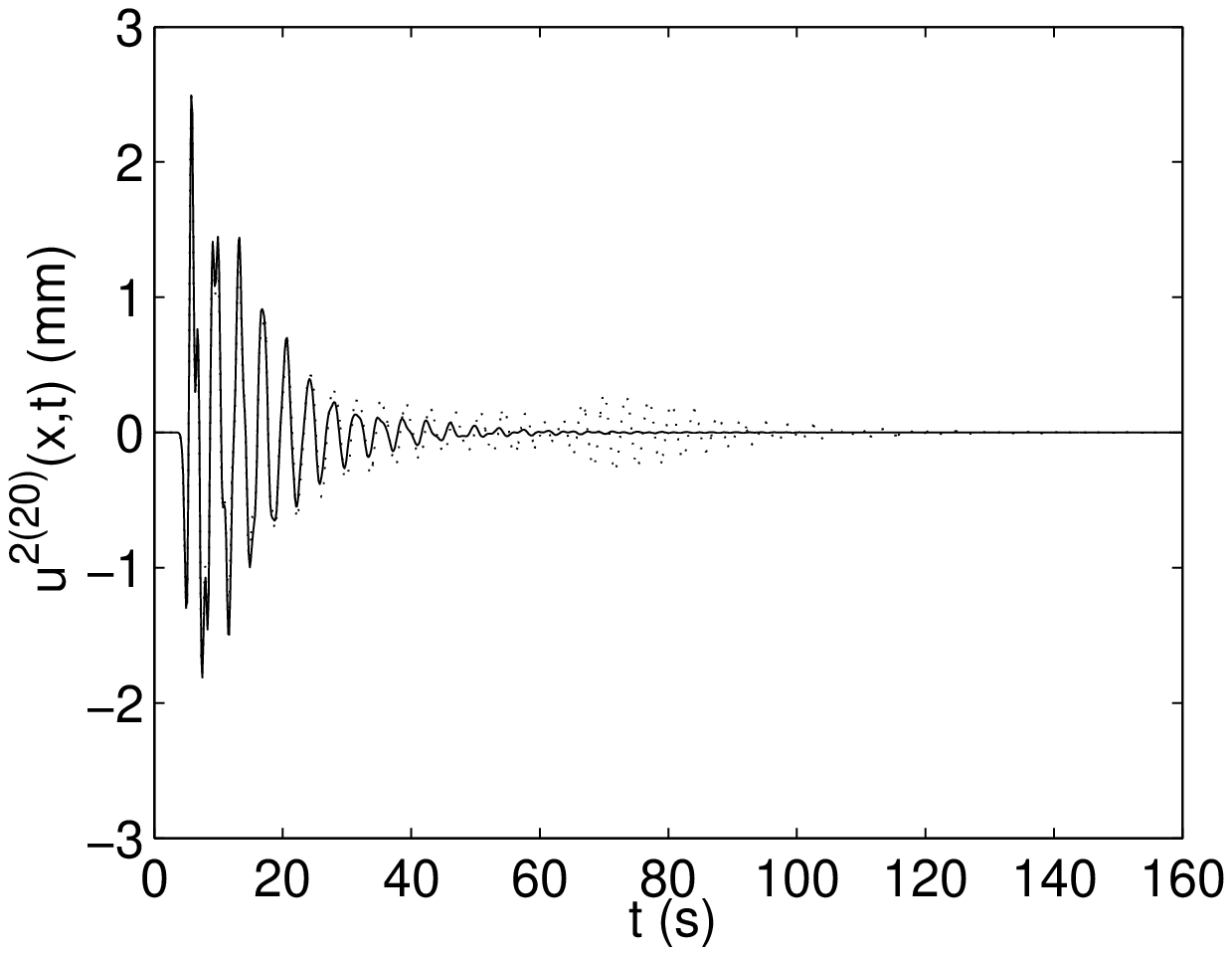}
\caption{Comparison of $2\pi$ times the  spectra (left panels) and
time histories (right panels) of the total displacement at  the
center of the {\it bottom segment} of a $50m\times 30m$ block of a
configuration with an infinite number of blocks (solid curves)
with the displacement at the same location of a central block
(dashed curves) of configurations having: i) ten such blocks (top
panels), ii) twenty such blocks (middle panels), and iii) fourty
such blocks. In all cases, the center-to-center spacing is
$d=65m$.} \label{CompMexicod65bottentwentfourt}
\end{center}
\end{figure}

In fig. \ref{CompMexicod65bottentwentfourt} we compare the spectra
and time histories of seismic response at the center of the {\it
bottom segment} of a central block in configurations with 10, 20,
40 and an infinite number of $50m\times 30m$ blocks separated by
$d=65$m. The splittings referred-to in the previous lines are now
more apparent and give rise to more pronounced beatings,
especially for the  10 and 20 block configurations. These beatings
are absent for the $N_{b}=\infty$ city. The durations are quite
long for the 10 and 20 block configurations, and, on the whole,
the signals are evocative of what has been often observed during
earthquakes in certain districts of Mexico City.
\begin{figure}[ptb]
\begin{center}
\includegraphics[width=6.0cm] {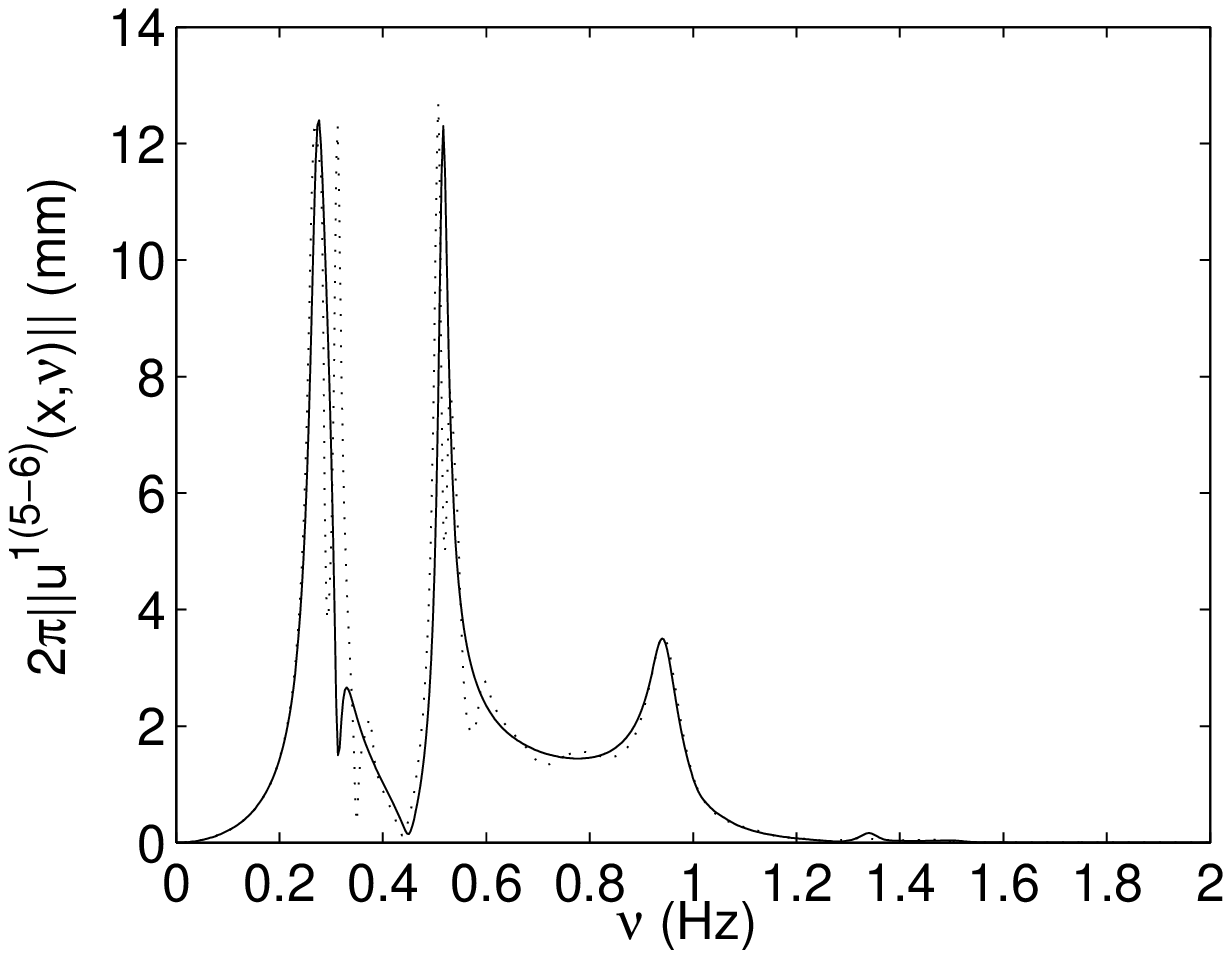}
\includegraphics[width=6.0cm] {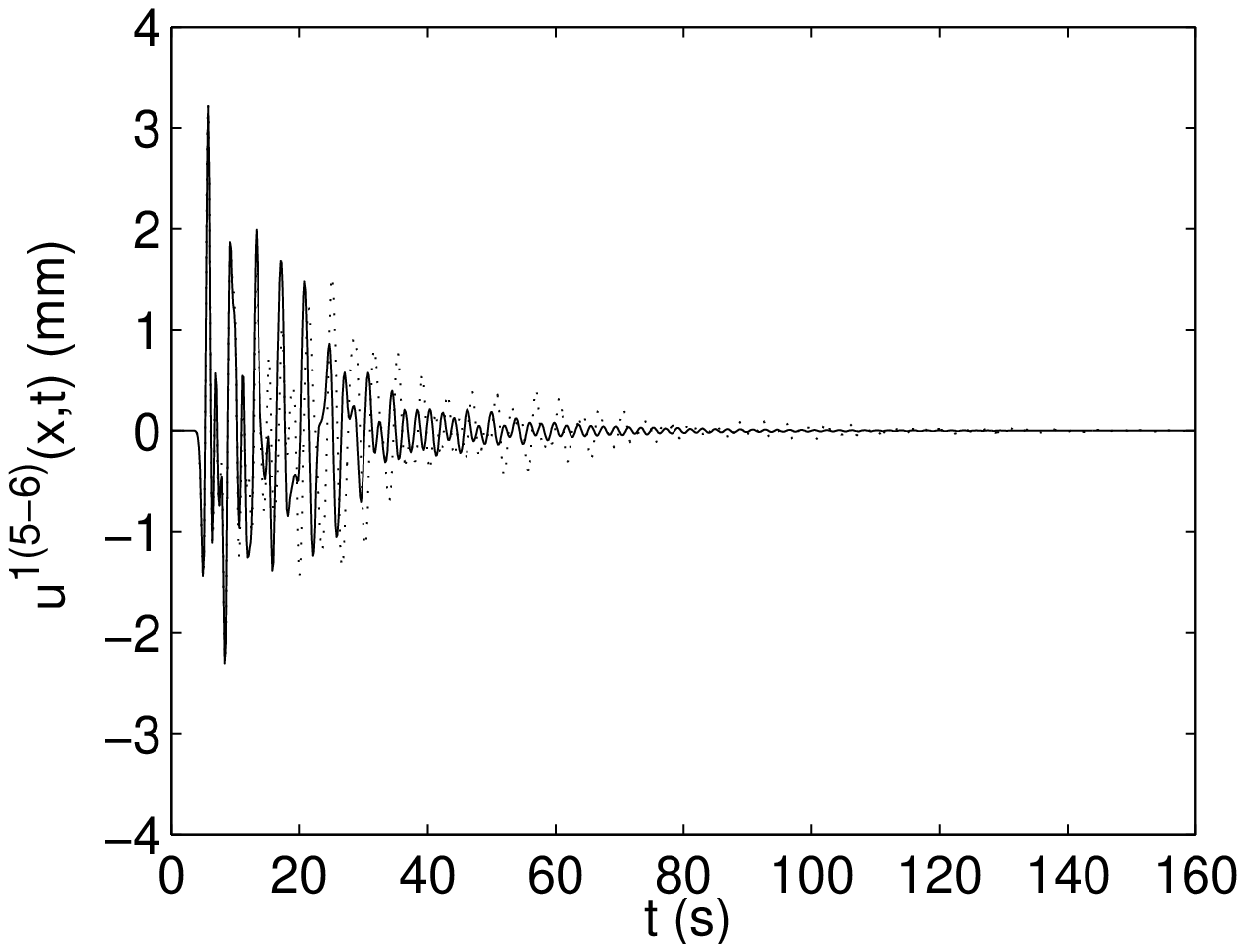}
\includegraphics[width=6.0cm] {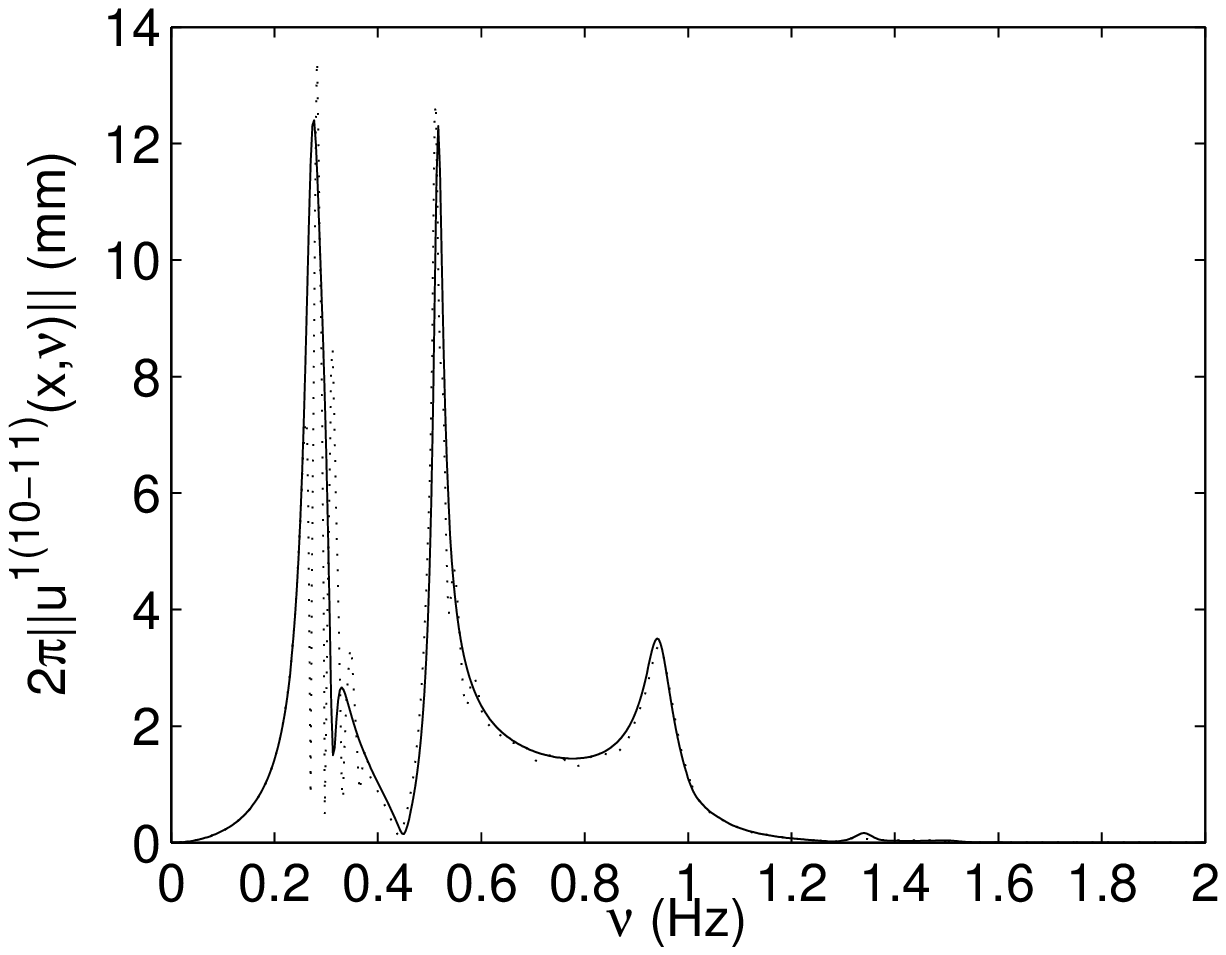}
\includegraphics[width=6.0cm] {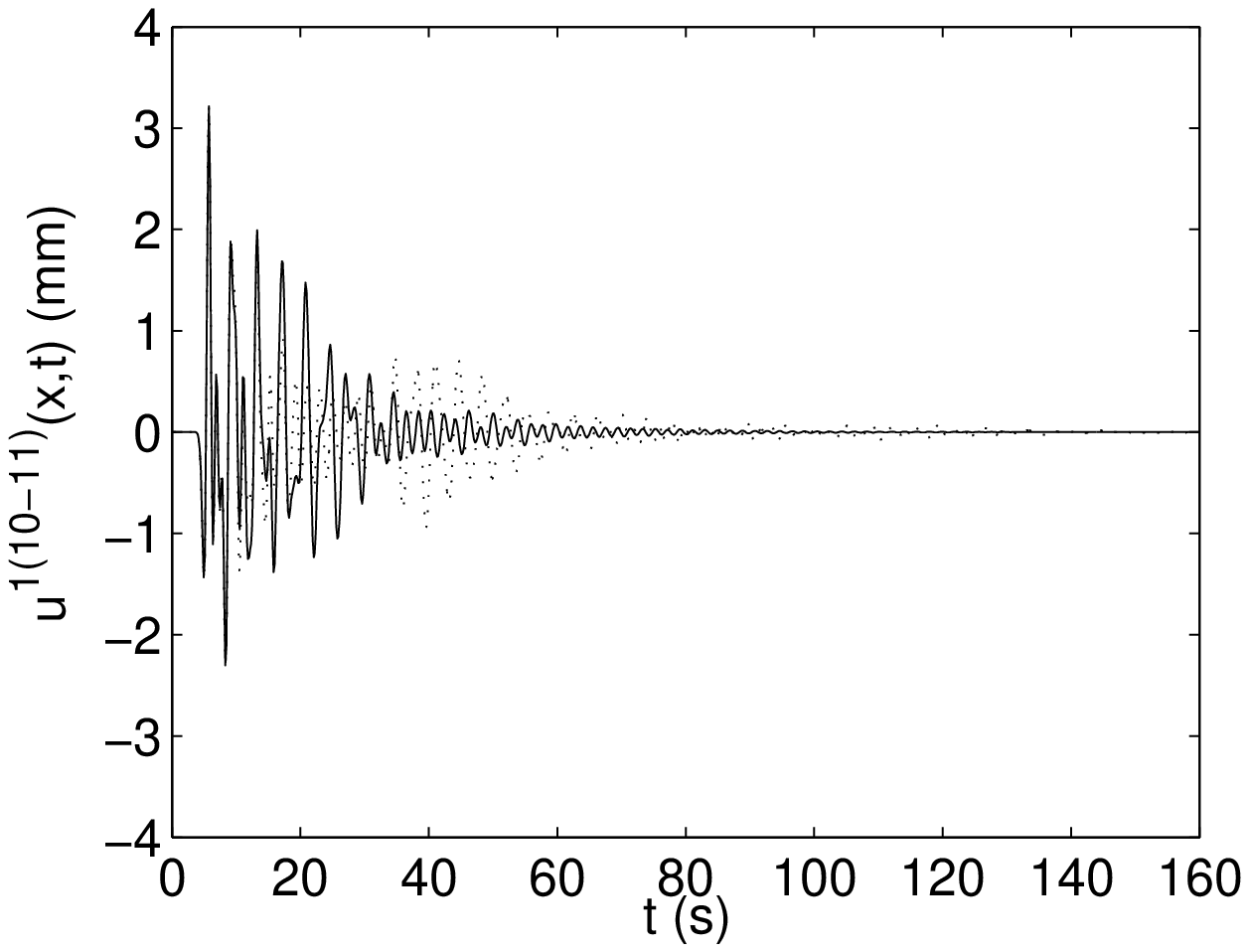}
\includegraphics[width=6.0cm] {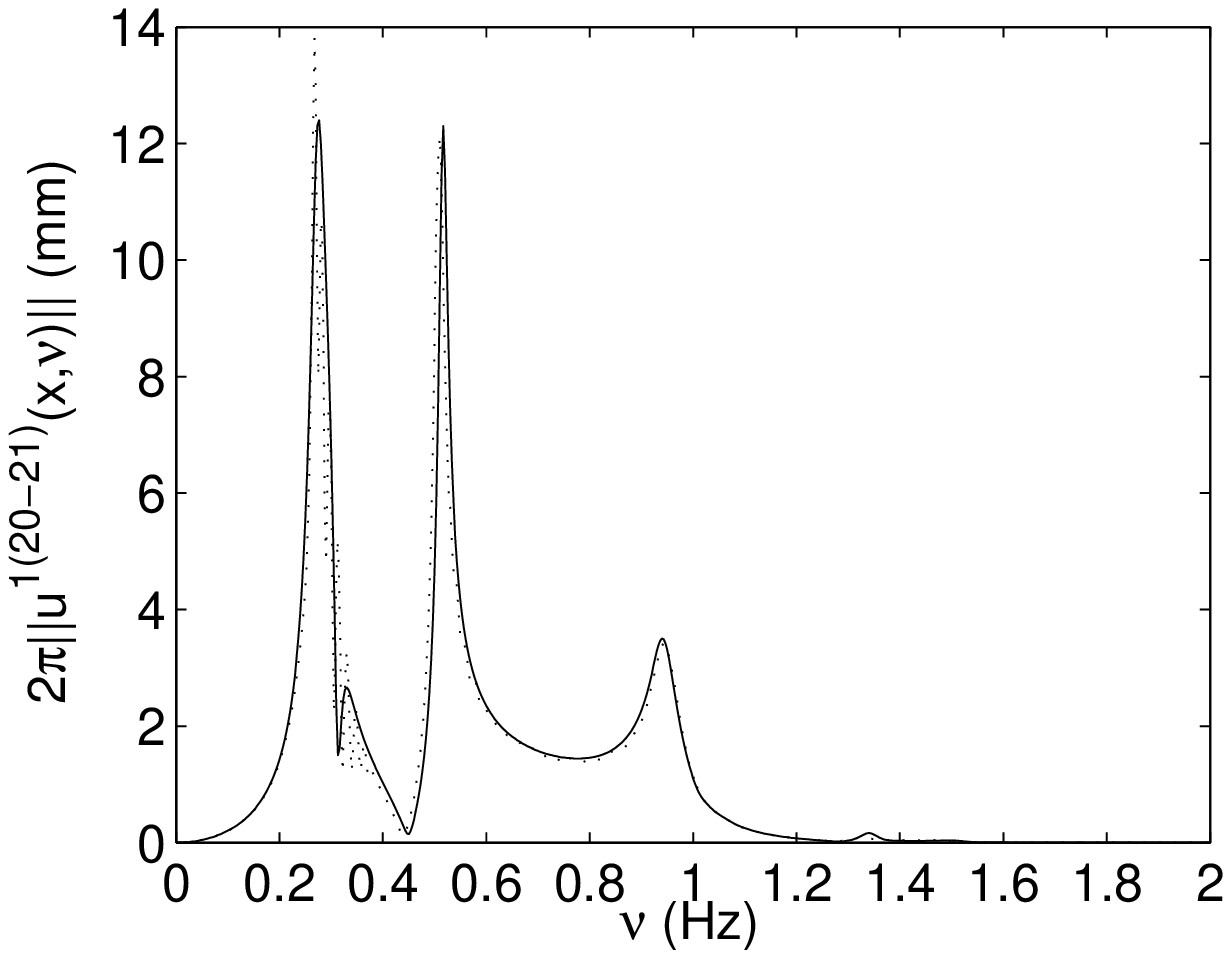}
\includegraphics[width=6.0cm] {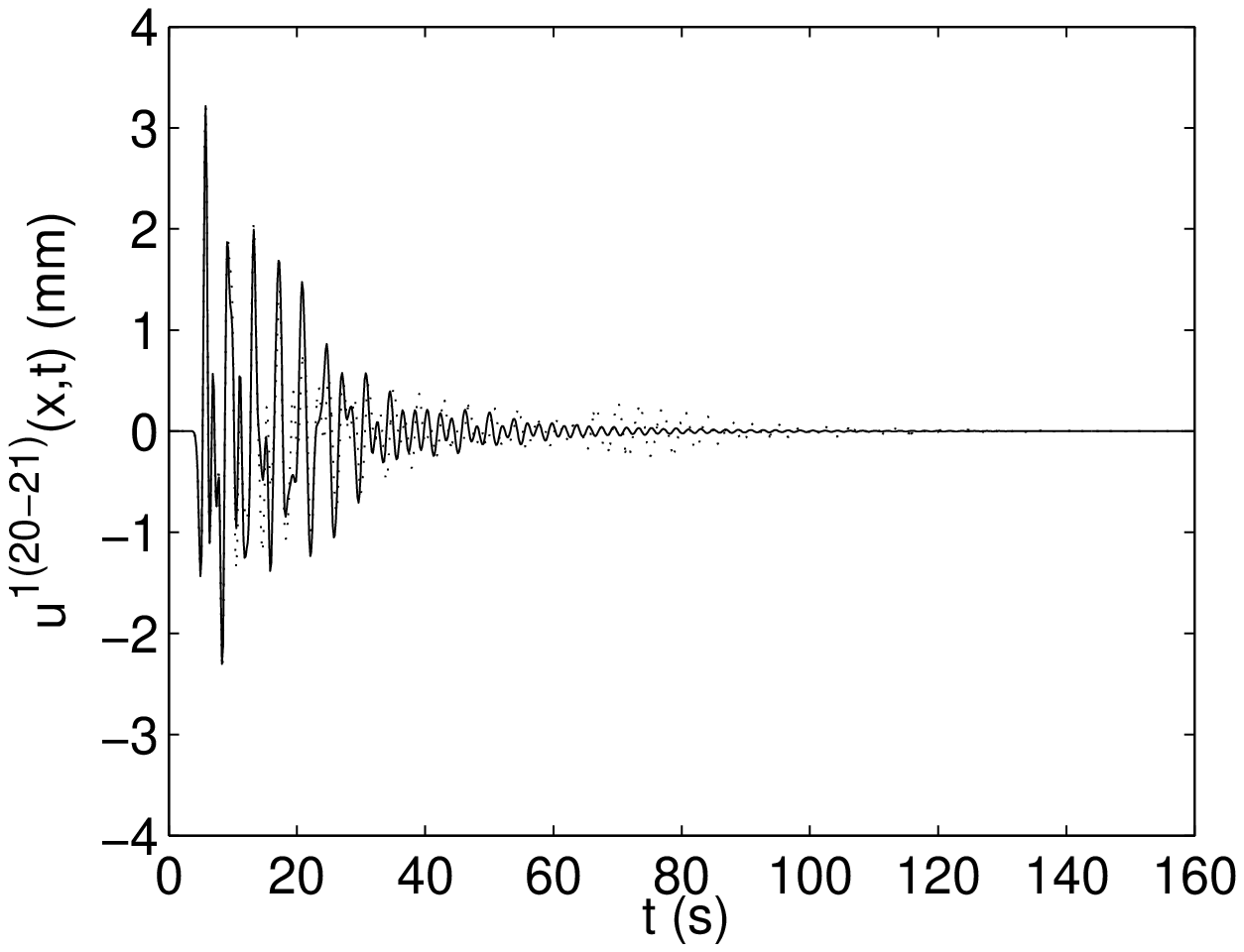}
\caption{Comparison of $2\pi$ times the  spectra (left panels) and
time histories (right panels) of the total displacement at  the
midpoint {\it on the ground between adjacent $50m\times 30m$
blocks} of a configuration with an infinite number of blocks
(solid curves) with the displacement at the same location  (dashed
curves) of a central region of configurations having: i) ten such
blocks (top panels), ii) twenty such blocks (middle panels), and
iii) fourty such blocks. In all cases, the center-to-center
spacing is $d=65m$.} \label{CompMexicod65miltentwentfourt}
\end{center}
\end{figure}

In fig. \ref{CompMexicod65miltentwentfourt} we compare the spectra
and time histories of seismic response at the midpoint {\it on the
ground between two adjacent centrally-located blocks} in
configurations with 10, 20, 40 and an infinite number of
$50m\times 30m$ blocks separated by $d=65$m. Again, we observe two
main resonance peaks for all $N_{b}$, giving rise to
characteristic beatings, to which are added other beatings due to
splittings of the first resonance peak, especially noticeable for
finite $N_{b}$.  This response is again quite evocative of ground
response observed  in the midst of certain districts of Mexico
City during many earthquakes that have affected this city
\cite{chba94}.

The presence of the additional low-frequency peaks in this set of
figures is linked either to the (finite) number of blocks
considered and/or to the total width $W$ of the finite
configuration (for $N=10,~20,~40$, $W$ is equal to
650m,~1300,~2600m respectively). These peaks cannot be
accounted-for in the dispersion relations written above, since
they  result from an analysis of configurations with an infinite
number of blocks (and for which $W=\infty$).
\subsubsection{Illustration of the spatial variability of response
in a configuration of ten blocks for center-to-center separations
$d=65$m}
Fig. \ref{Mexicod65snap} depicts snapshots of the displacement
field in the entire configuration containing ten blocks whose
center-to-center distance is $d=65$m.
\begin{figure}[ptb]
\begin{center}
\includegraphics[width=10.0cm] {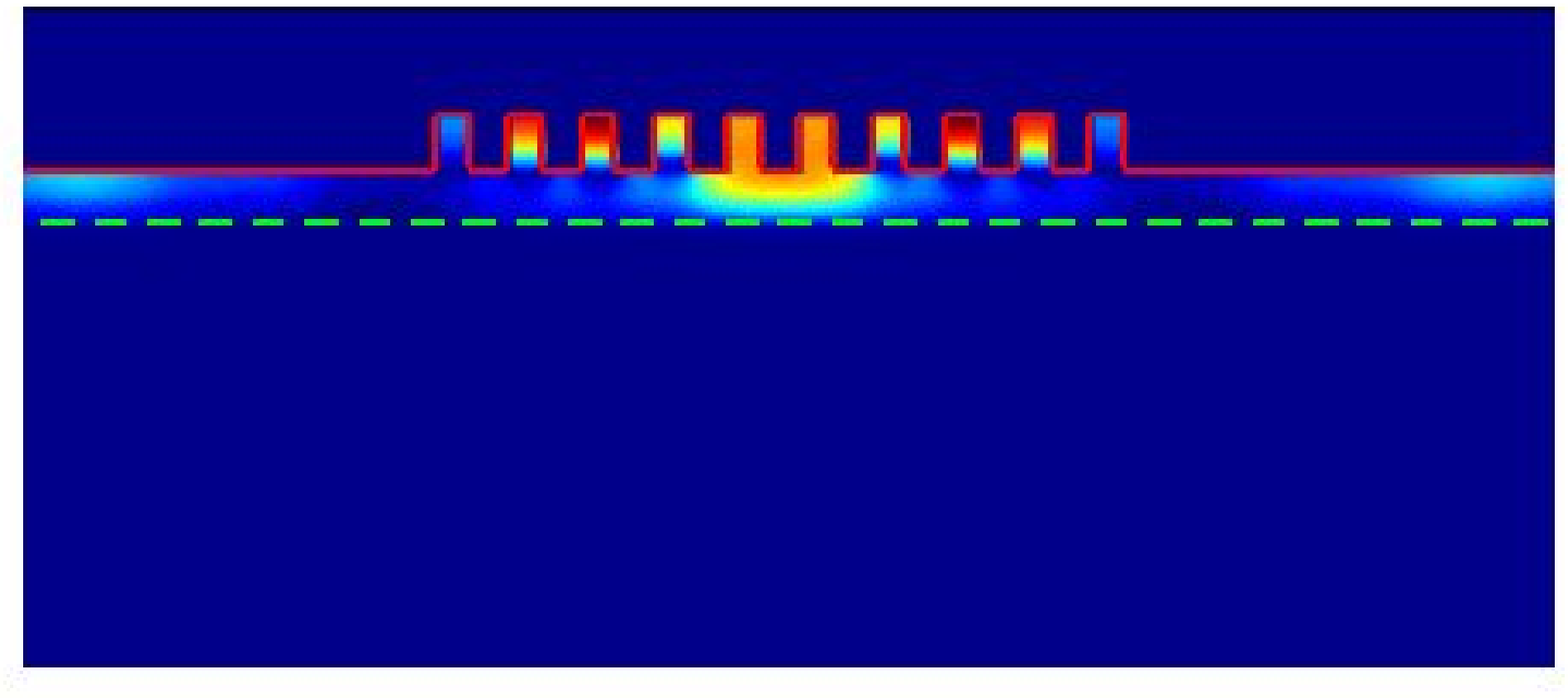}
\includegraphics[width=10.0cm] {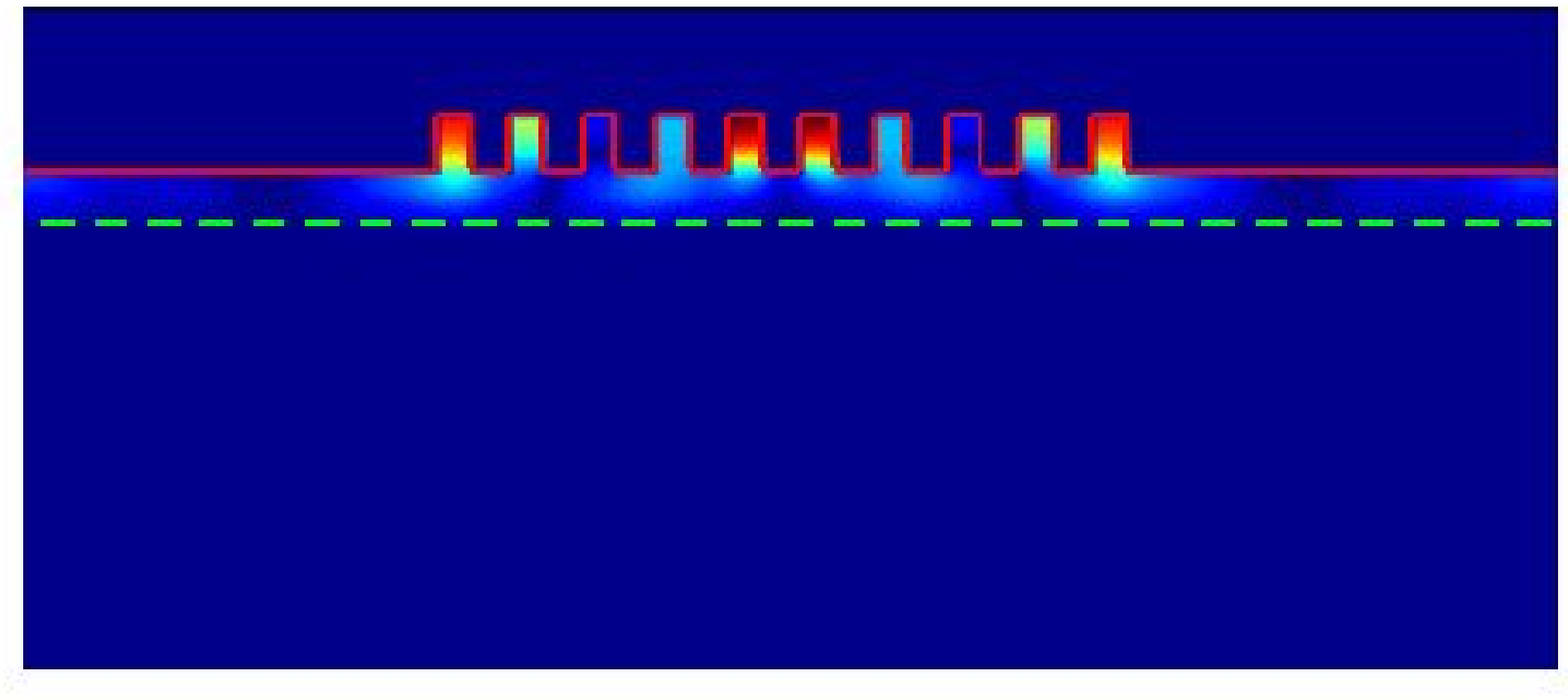}
\includegraphics[width=10.0cm] {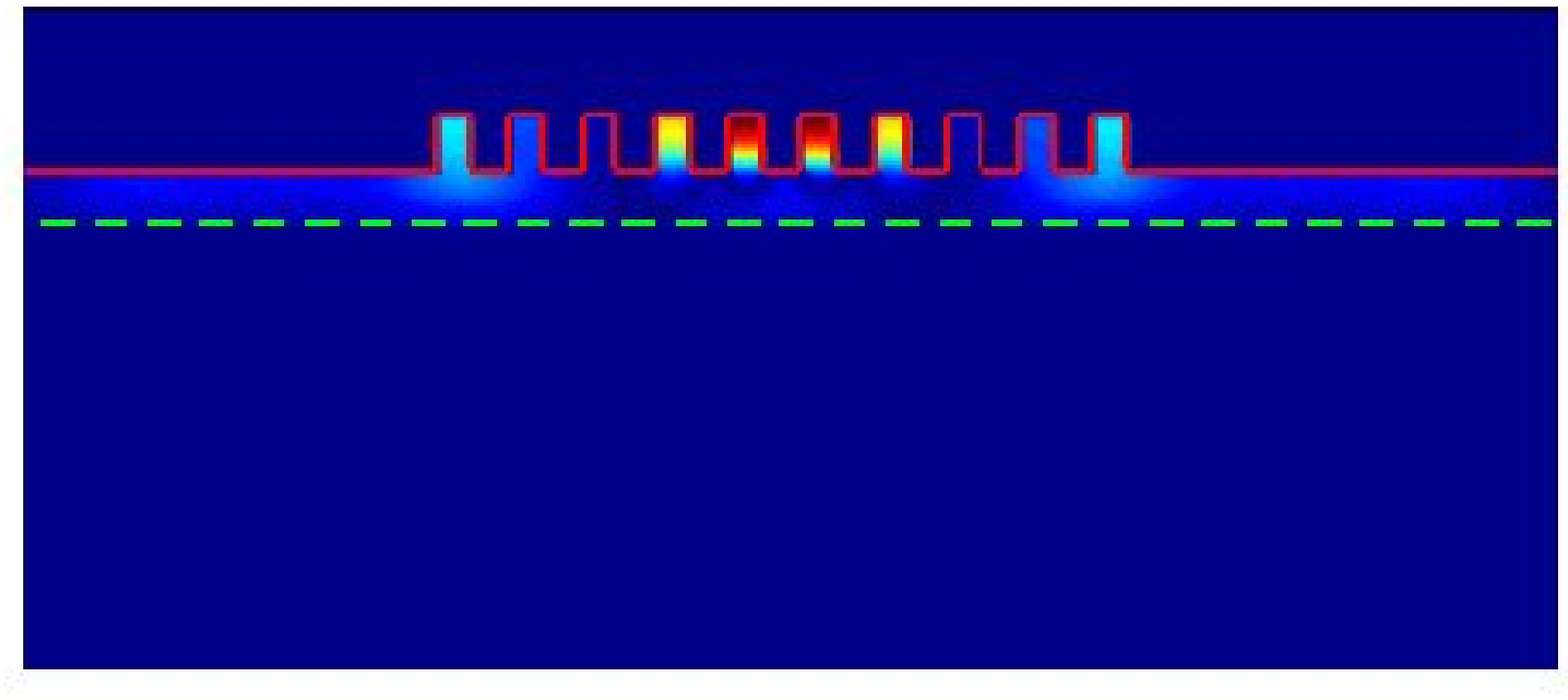}
\caption{Snapshots at  instants $t=25s$ (top panel), $t=32.5s$
(middle panel), $t=50s$ (bottom panel), of the total displacement
field for $10$ identical  $50m\times30m$  blocks whose
center-to-center spacing is $d=65$m.} \label{Mexicod65snap}
\end{center}
\end{figure}
It can be noticed that: i) the motions are still quite strong in
portions of the configuration some 50s after the arrival of the
initial pulse and ii) that the motion is quite variable spatially-
and temporally-speaking, iii) the spatial variability extends even
to within a given block.

The spatial variability of response is one of the characteristics
of the motion that has often been recorded within Mexico City
\cite{flno87,chba94}.
\clearpage
\subsubsection{Comparison of responses of the configuration without blocks
to the one with $N_{b}=\infty$
blocks for $d=300$m}
\begin{figure}[ptb]
\begin{center}
\includegraphics[width=6.0cm] {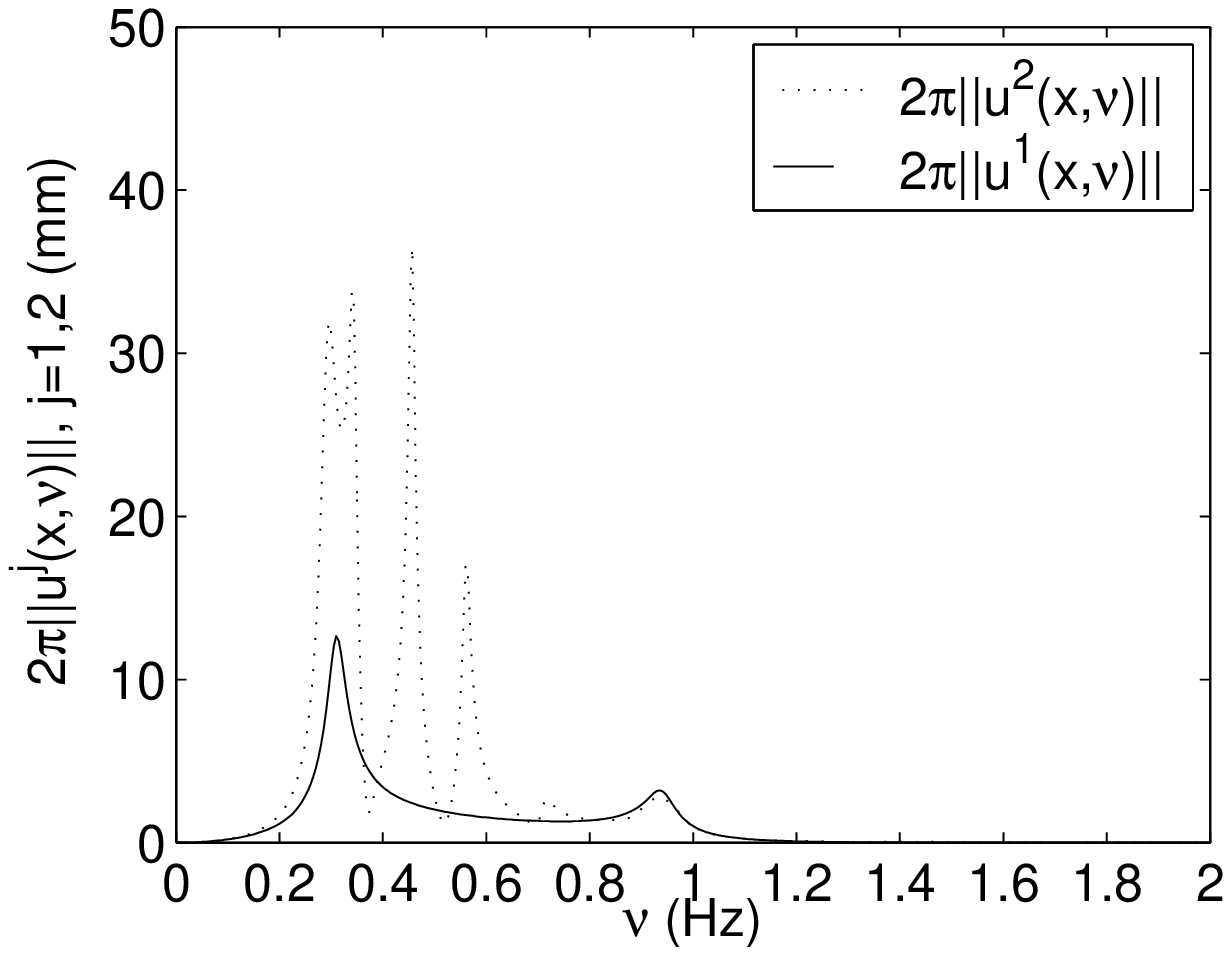}
\includegraphics[width=6.0cm] {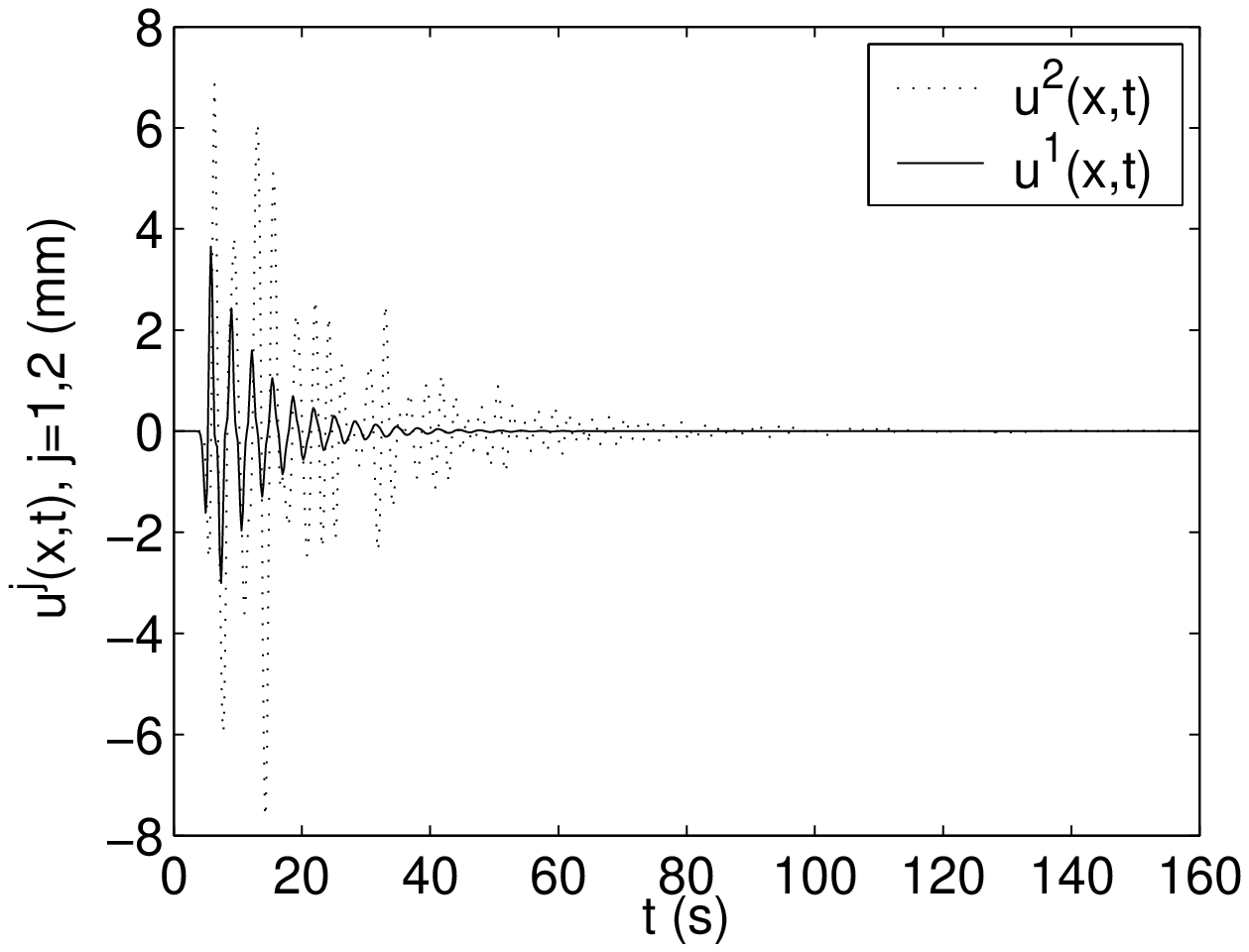}
\includegraphics[width=6.0cm] {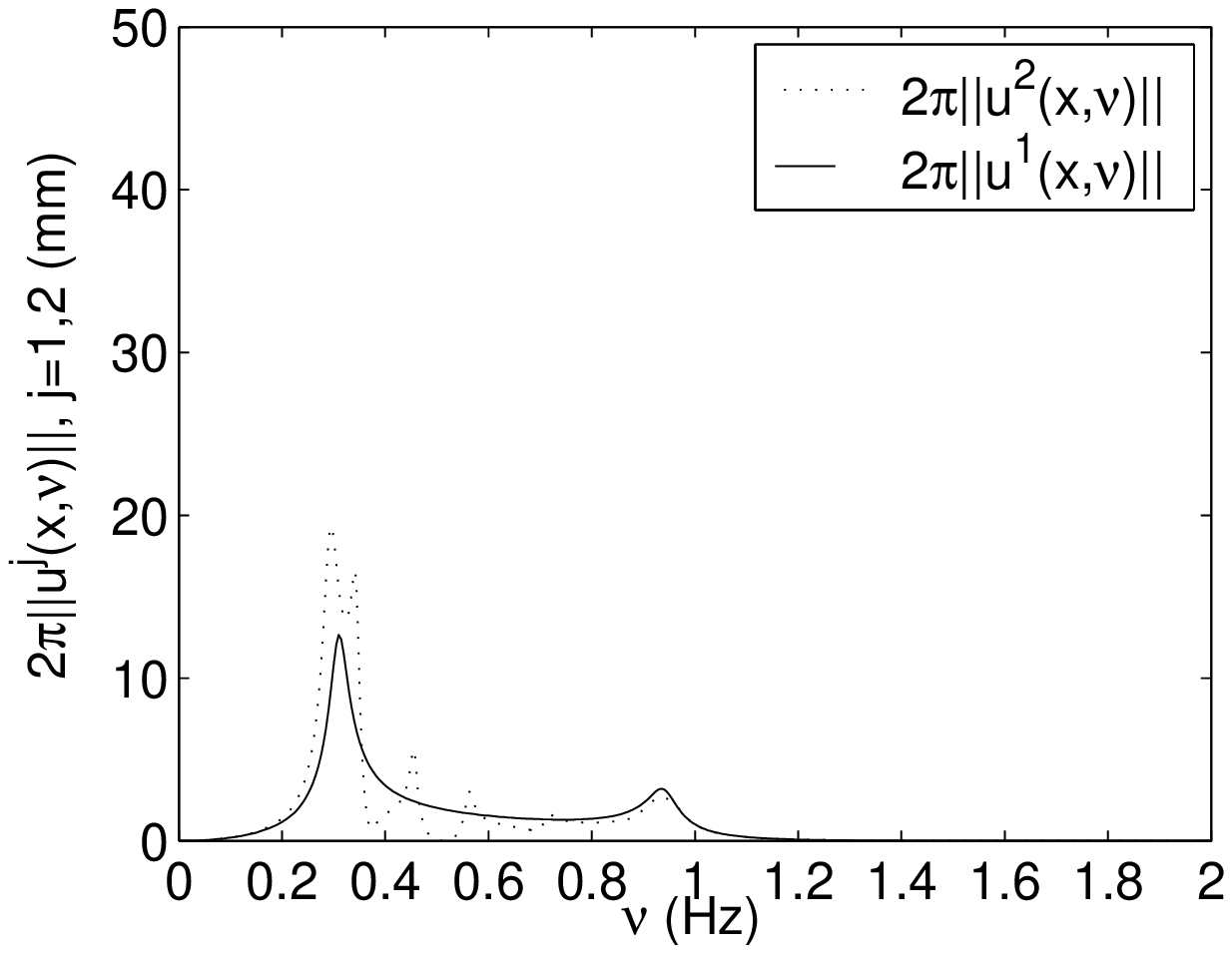}
\includegraphics[width=6.0cm] {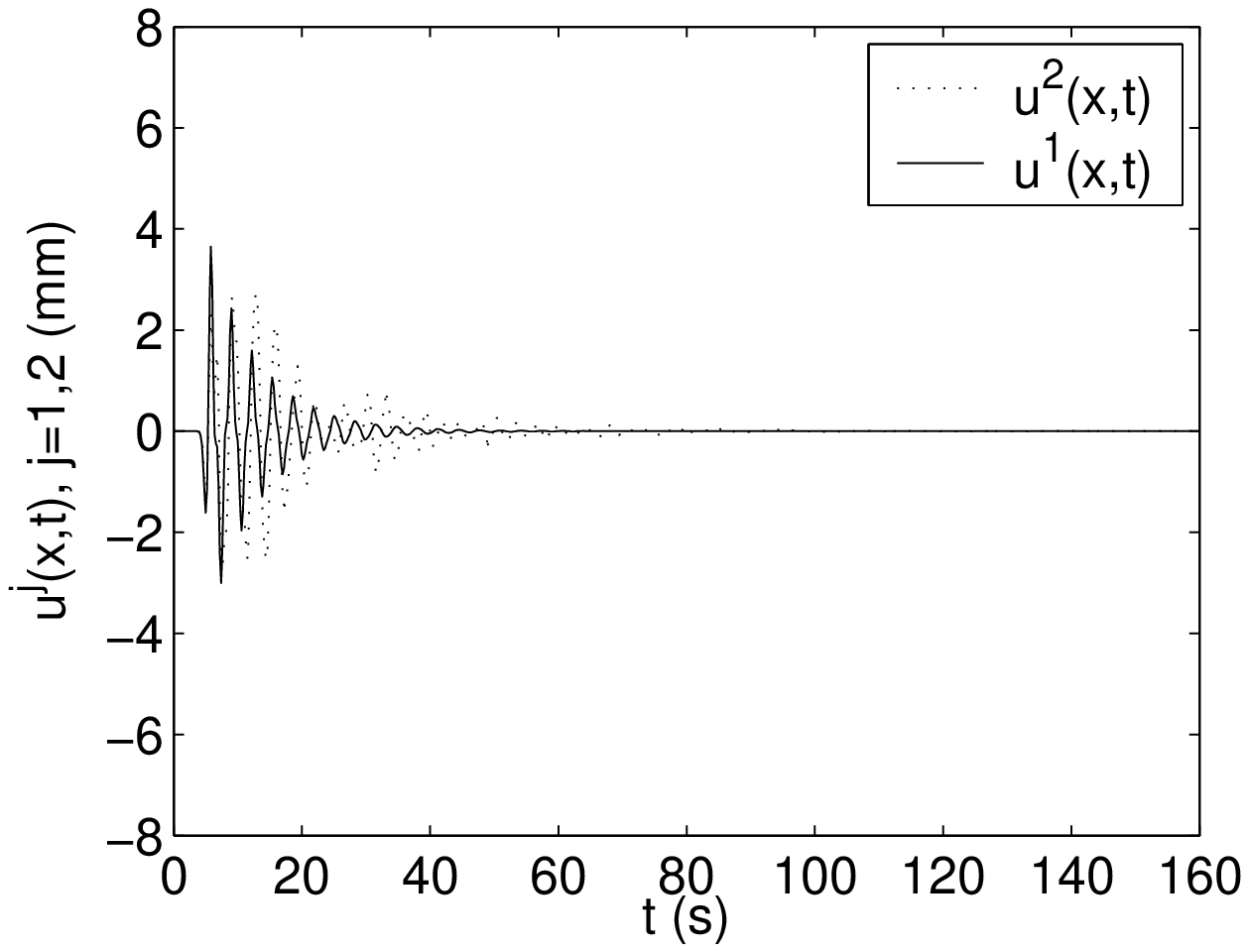}
\includegraphics[width=6.0cm] {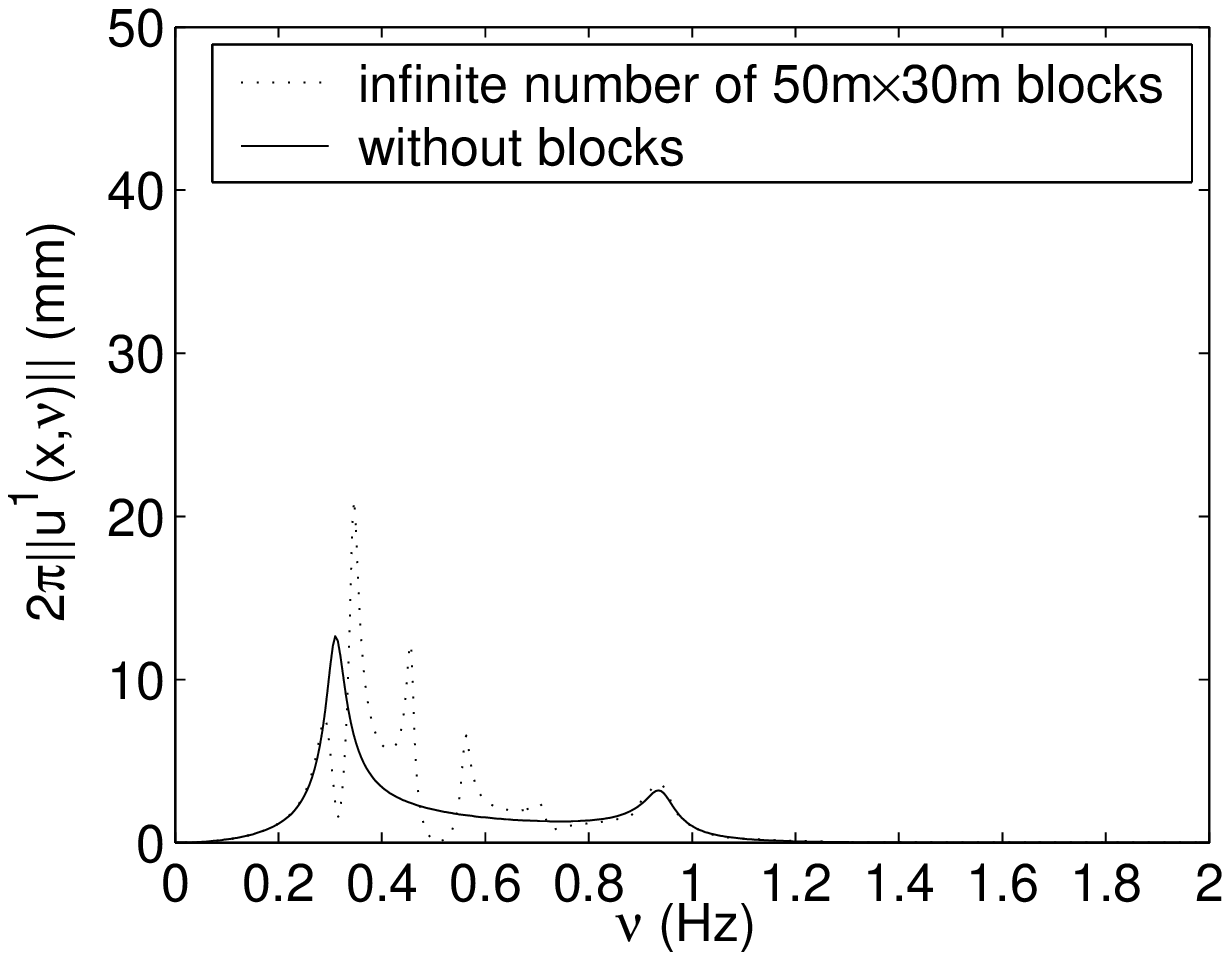}
\includegraphics[width=6.0cm] {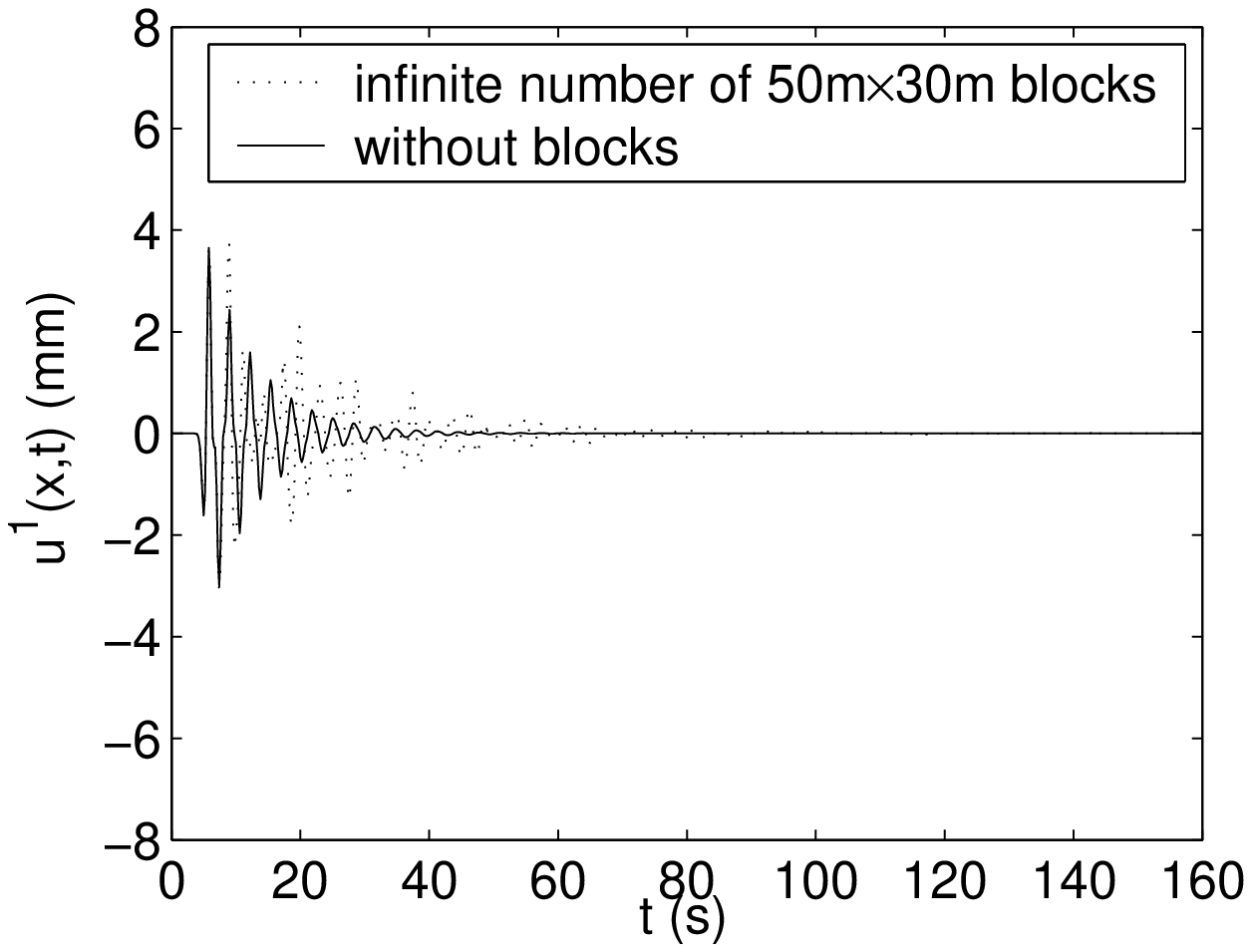}
\caption{Comparison of $2\pi$ times the  spectrum (left panels)
and time history (right panels) of the total displacement on the
ground in the absence of blocks (solid curves) with  the
displacements (dashed curves) at  the center of the top segment
(top panels), center of the bottom segment (middle panels) and
midpoint on the ground between adjacent  $50m\times 30m$ blocks,
of a $N_{b}=\infty$ configuration in which $d=300$m. }
\label{CompMexicod300}
\end{center}
\end{figure}
In fig. \ref{CompMexicod300}, we compare the spectra and time
histories in the presence  and in the absence of the blocks. When
the blocks are present, their number is infinite and their
center-to-center distance is 300m.

The excitation of the multi displacement-free base block mode
cannot be clearly distinguished because of the excitation of
quasi-Cutler modes whose existence is related to the periodic
nature of the block distribution. Note should be taken of the fact
that now (i.e., for $d=300$m) these modes are visible (with a
large quality factor) within the bandwidth of the solicitation,
whereas they were invisible for the $d=65$m periodic structure. As
previously, the soil-structure interaction appears as a resonance
peak associated with the excitation of the fundamental quasi-Love
mode.

These features show up at all three locations of the configuration
with blocks. They manifest themselves in the time domain by: i) a
larger duration (multiplied by $\approx 4$ on the top segment of
the blocks with respect to its value in the absence of the
blocks), ii) a larger peak amplitude (at the top of the blocks)
and larger cumulative motion, and iii) pronounced beatings at all
locations due to the periodic nature of the configuration.

These  features are once again evocative of those which have been
observed during earthquakes in certain districts Mexico City.
However, the excitation of the quasi-Cutler mode, which is
strongly-linked to the quasi-periodic or periodic nature of a
district of the city (these districts actually exist, as seen in
fig. \ref{figsite3}), can, at best, explain only part of the
features of response in this city, since the latter is not usually
solicited by a (normally-incident) plane wave.

It has been shown in the companion paper, and in
\cite{grobyetwirgin2005,grobyetwirgin2005II} that the correct
solicitation of a configuration for the study of the Michoacan
earthquake (and many other earthquakes affecting Mexico City) is a
cylindrical wave radiated by a shallow, laterally-distant source
that gives rise to Love waves after traveling within the crust
from the hypocenter to the city. In the same studies, it was shown
that this type of solicitation is another cause for the large
duration, large amplitude, and beatings of Mexico City response.
Unfortunately, source wave solicitation cannot be treated by the
quasi-modal analysis of periodic structures.
\clearpage
\subsubsection{Comparison of the MM and FEM results for a finite
set of blocks whose separation is $d=$300m}
 We now compare, in figs. \ref{Mexicod300b5}, \ref{Mexicod300b1}
 and \ref{Mexicod300b1side}, the results computed
by  the mode-matching technique (accounting for the fundamental
quasi-mode of the blocks), for an infinite number of
equally-spaced identical blocks, with those  we obtain by our
finite-element method, for a finite number $N_{b}=10$ of
equally-spaced identical blocks. In this case,
$\mathbb{B}=\{1,2,...,10\}$, with the block numbers $1$ and $10$
being located at the left and  right edges respectively of the
 finite configuration.
\begin{figure}[ptb]
\begin{center}

\includegraphics[width=6.0cm] {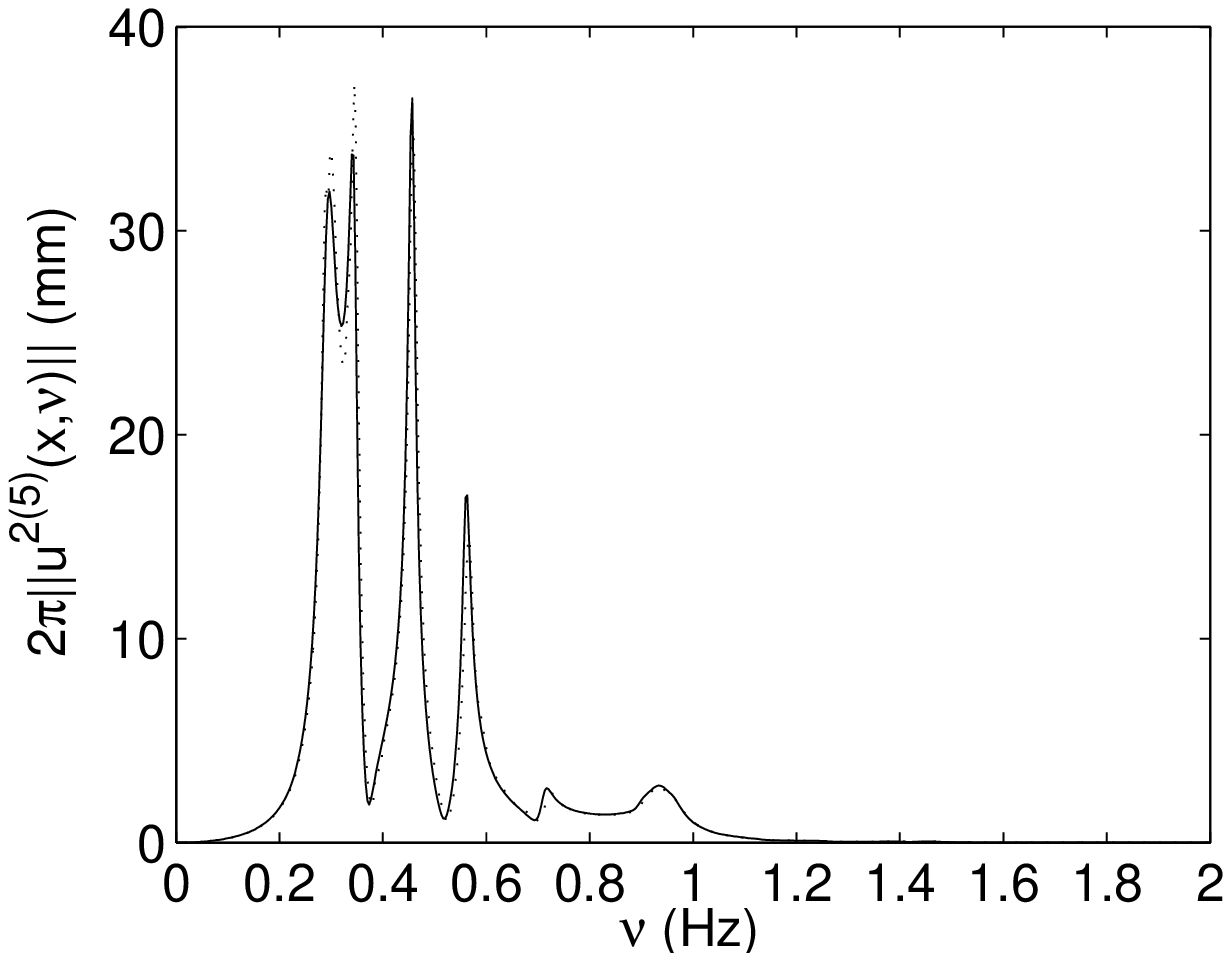}
\includegraphics[width=6.0cm] {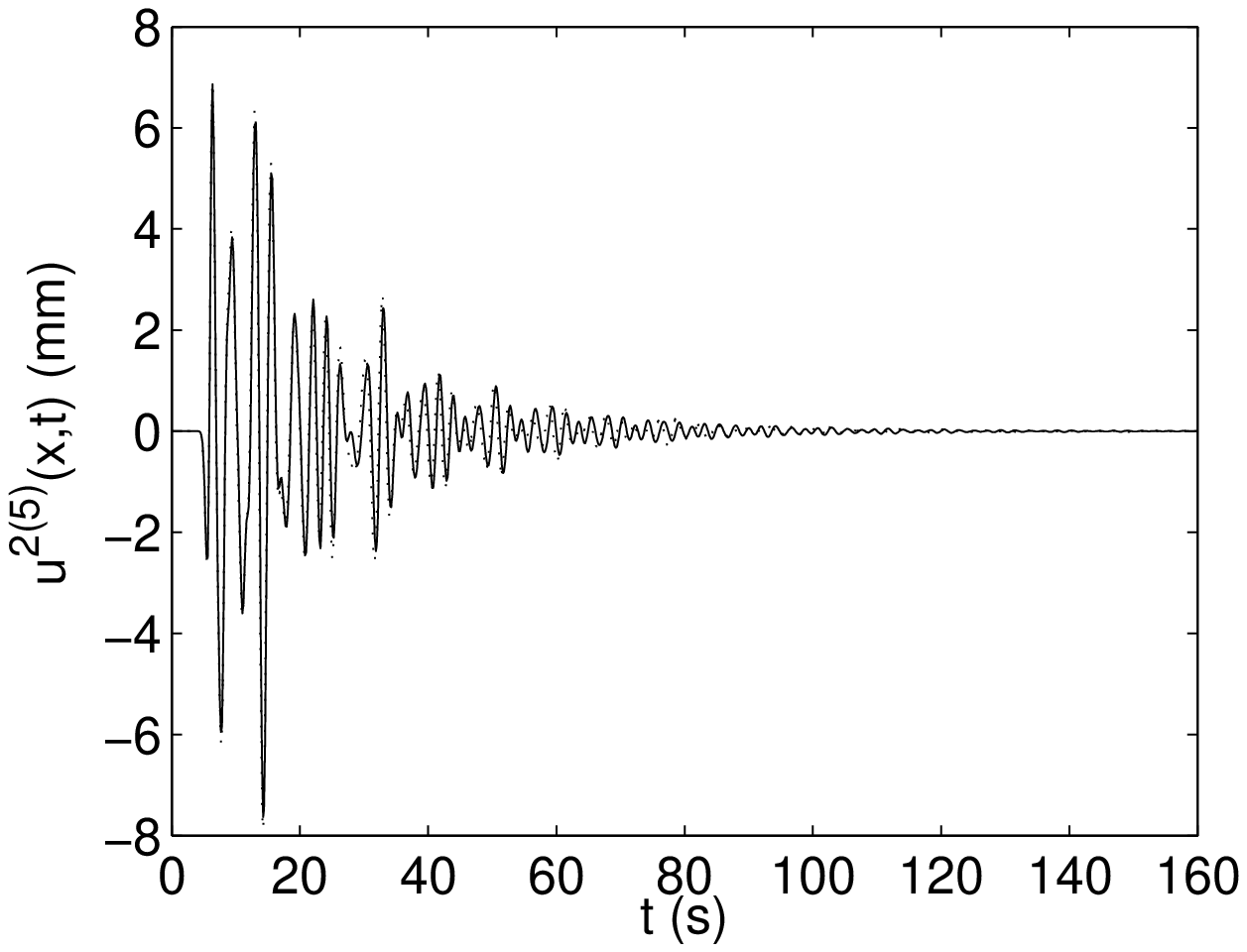}
\includegraphics[width=6.0cm] {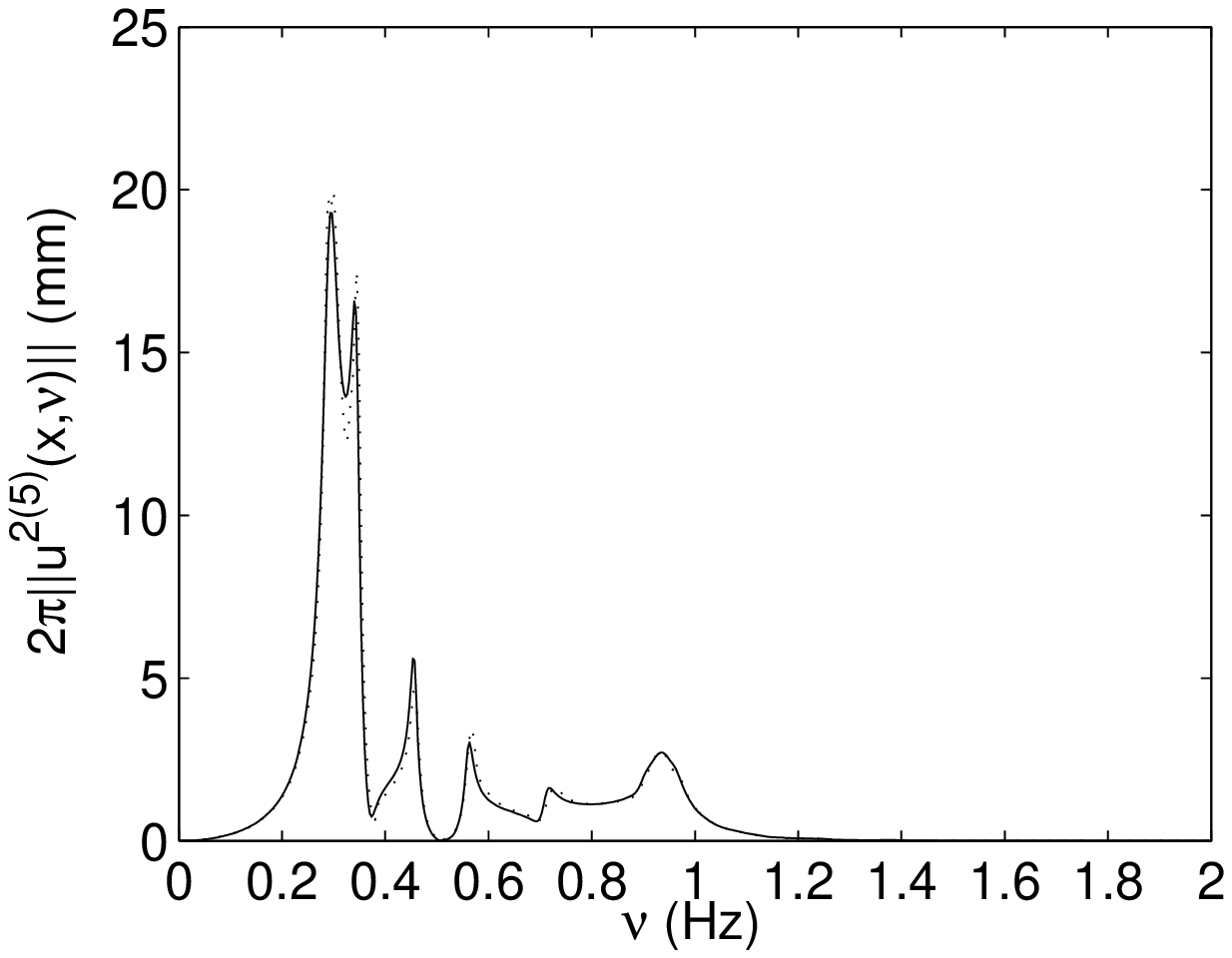}
\includegraphics[width=6.0cm] {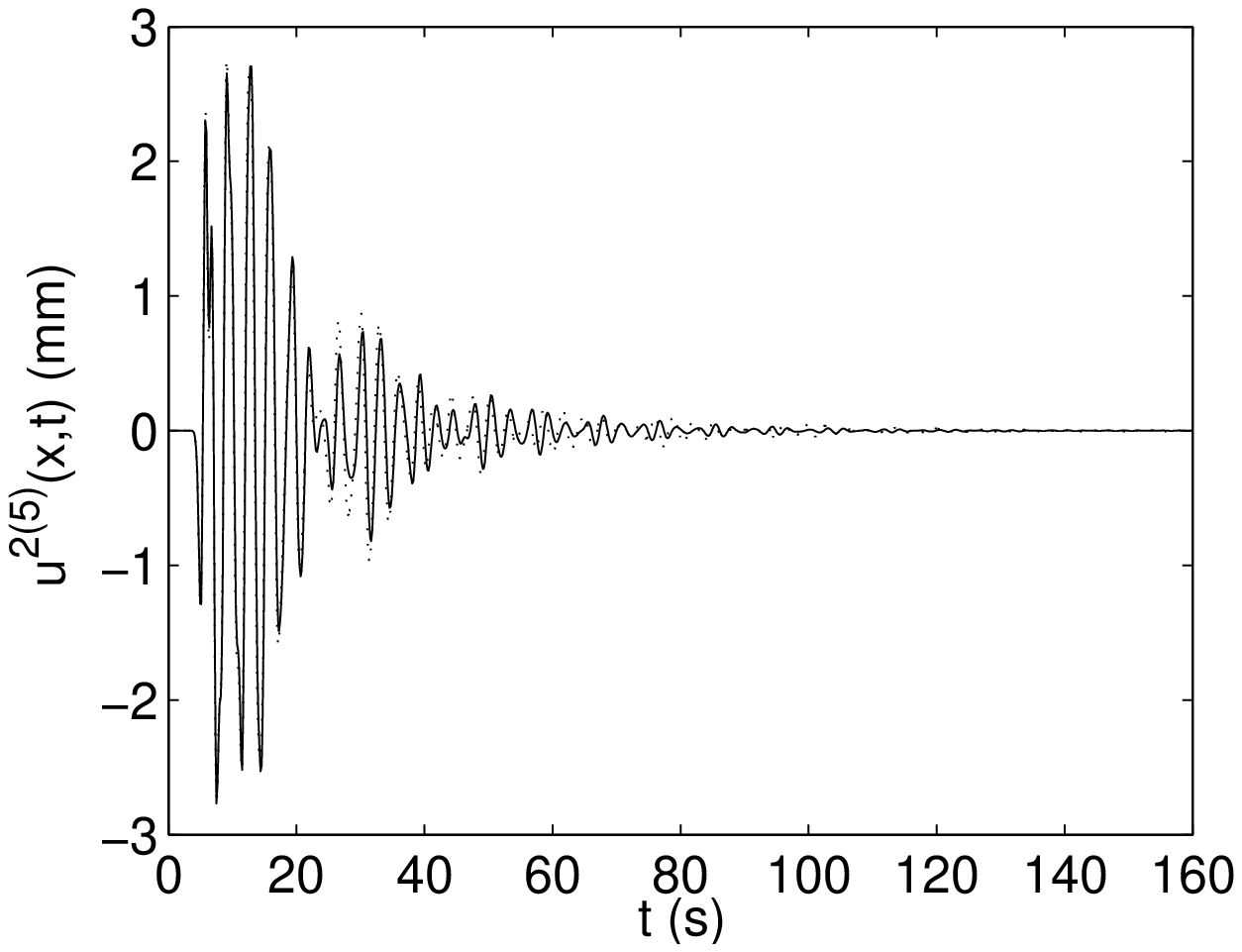}
\includegraphics[width=6.0cm] {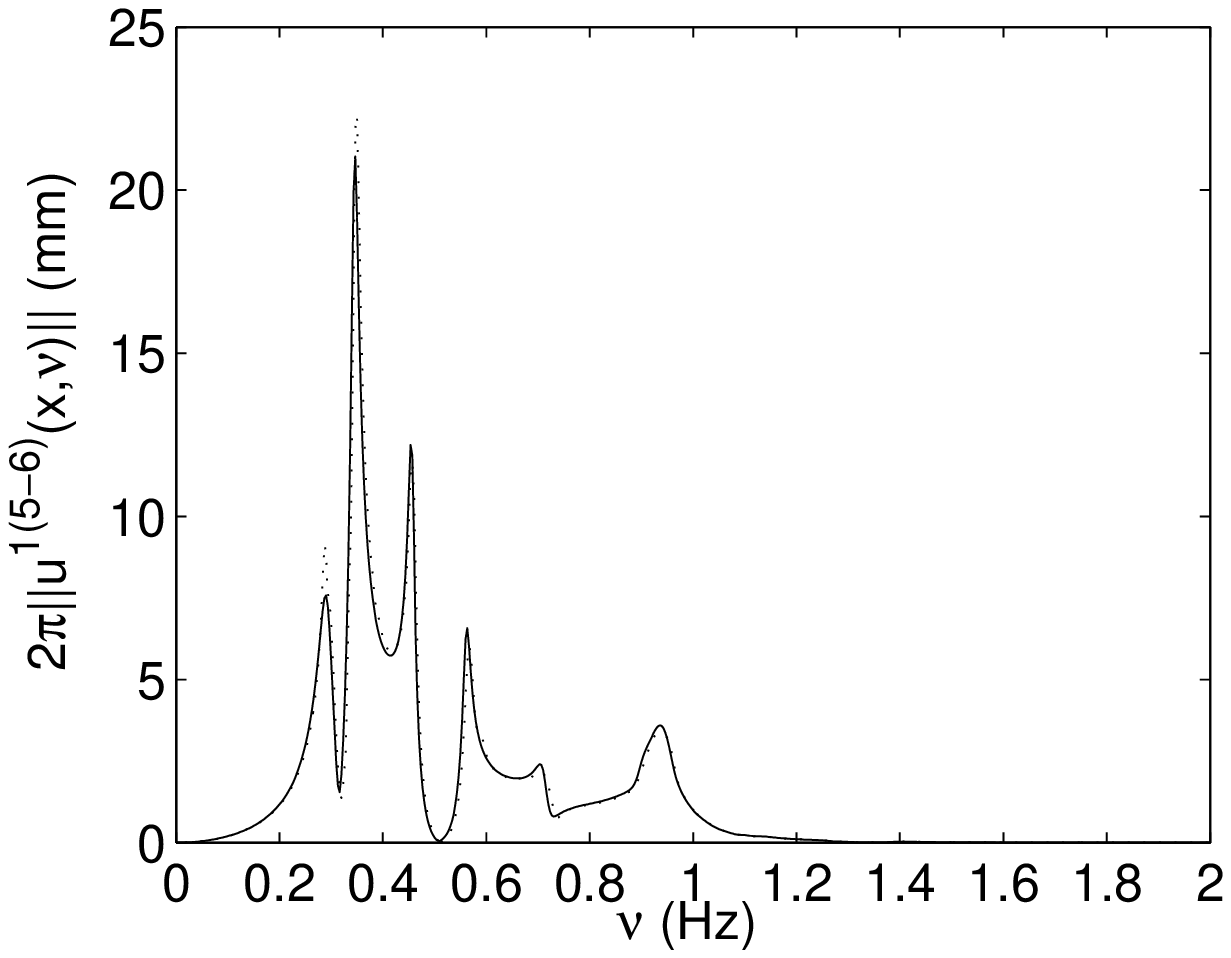}
\includegraphics[width=6.0cm] {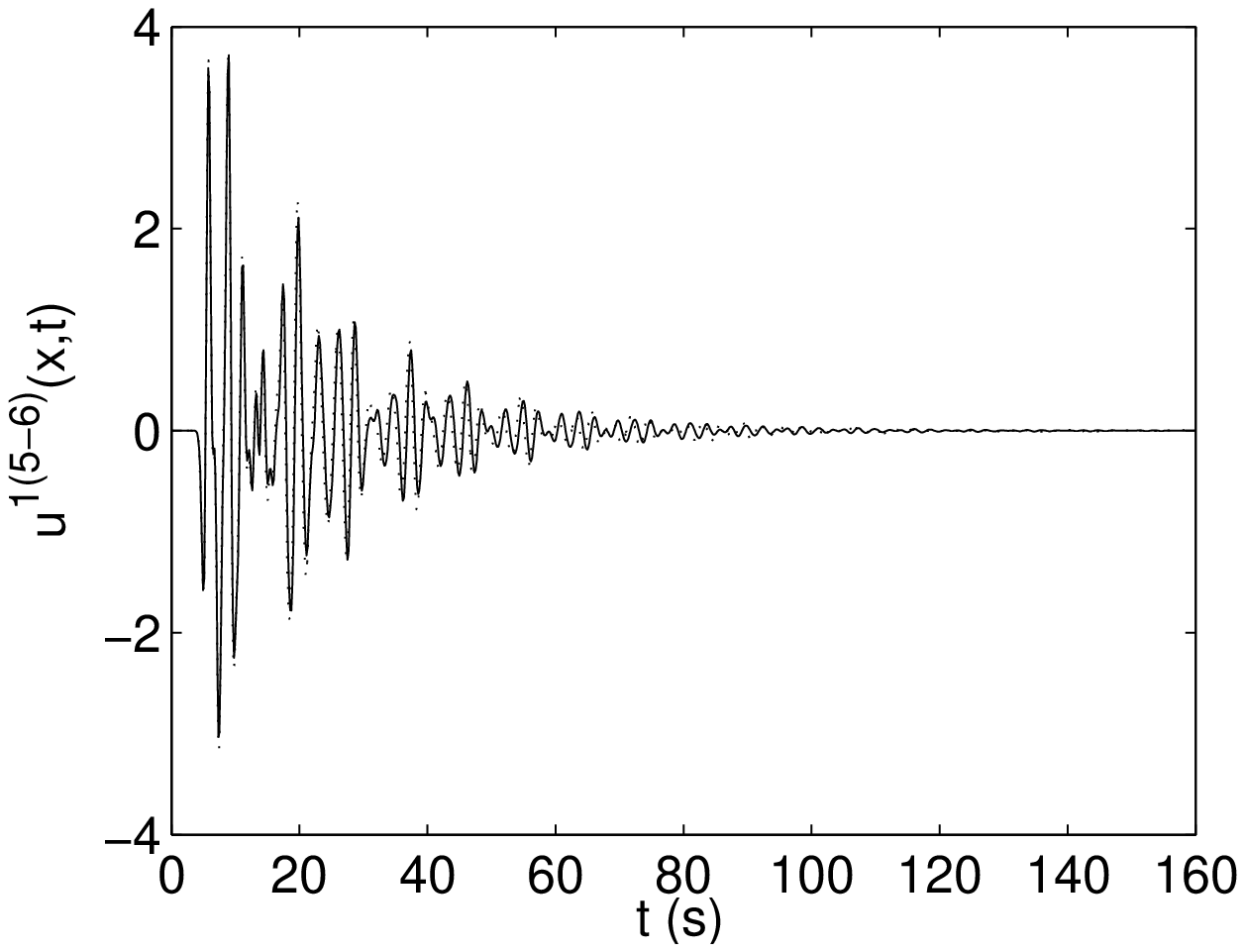}
\caption{Comparison of $2\pi$ times the spectrum (left panels) and
time history (right panels) of the total displacement as
calculated by the MM method for an infinite number of $50m\times
30m$ blocks separated by $d=300m$  (solid curves) with those
obtained by the FEM method in the {\it central portion} of a
configuration of $10$ identical $50m\times 30m$ blocks separated
by $d=300m$ (dotted curves): i) at the center of the top segment
of block number 5 (top panels), ii) at the center of the base
segment of block number 5, and iii) on the ground between  block
numbers 5 and 6.} \label{Mexicod300b5}
\end{center}
\end{figure}
\begin{figure}[ptb]
\begin{center}
\includegraphics[width=6.0cm] {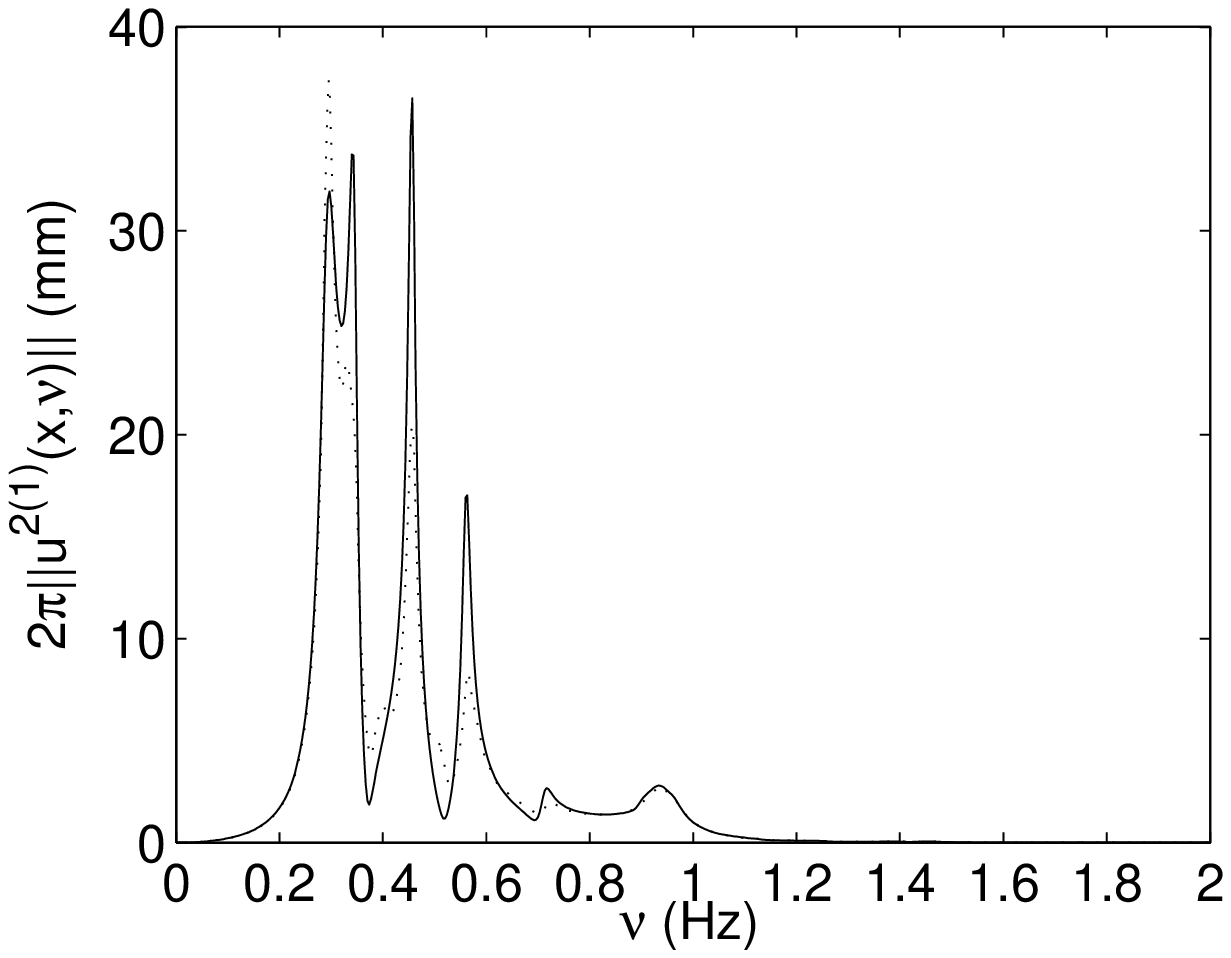}
\includegraphics[width=6.0cm] {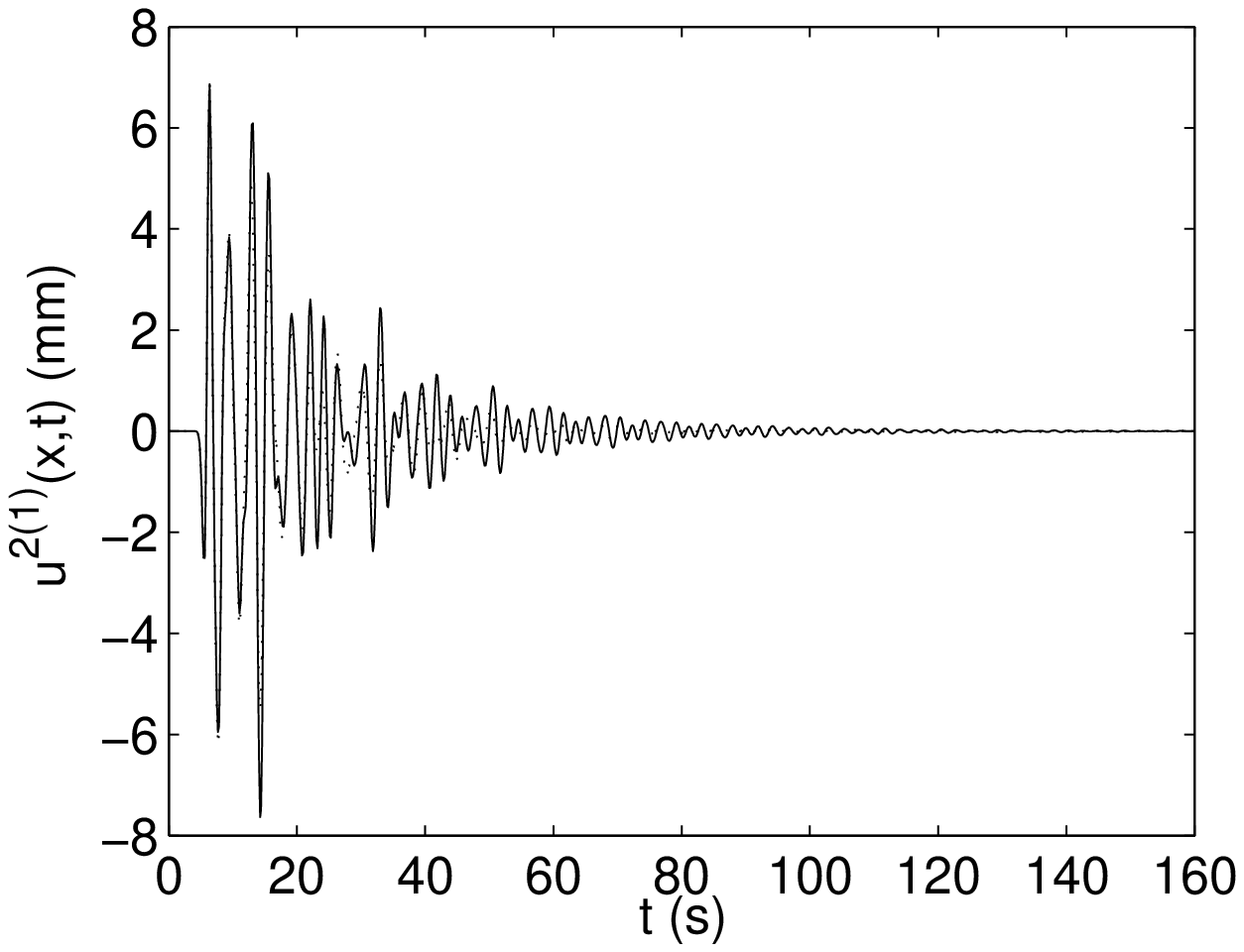}
\includegraphics[width=6.0cm] {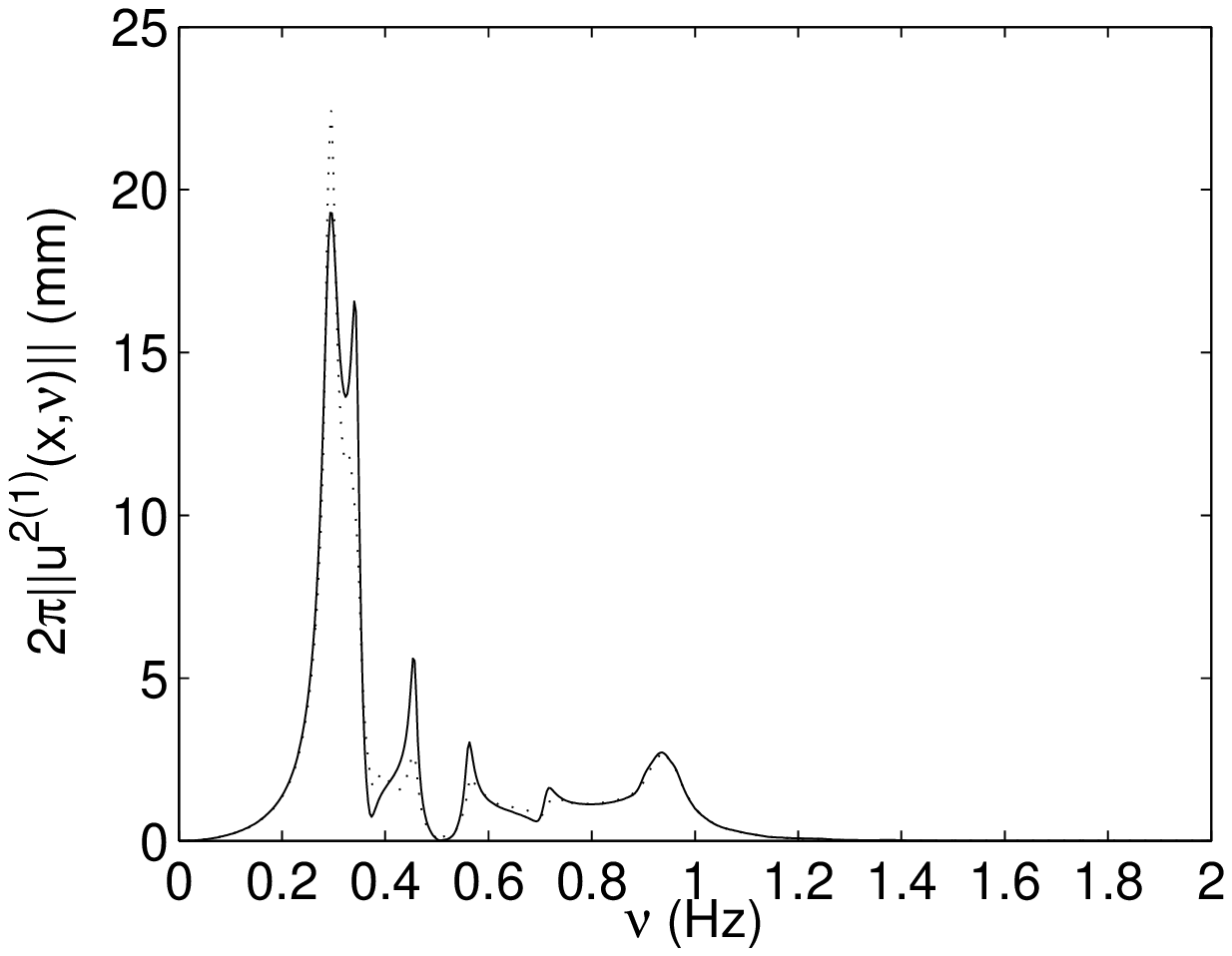}
\includegraphics[width=6.0cm] {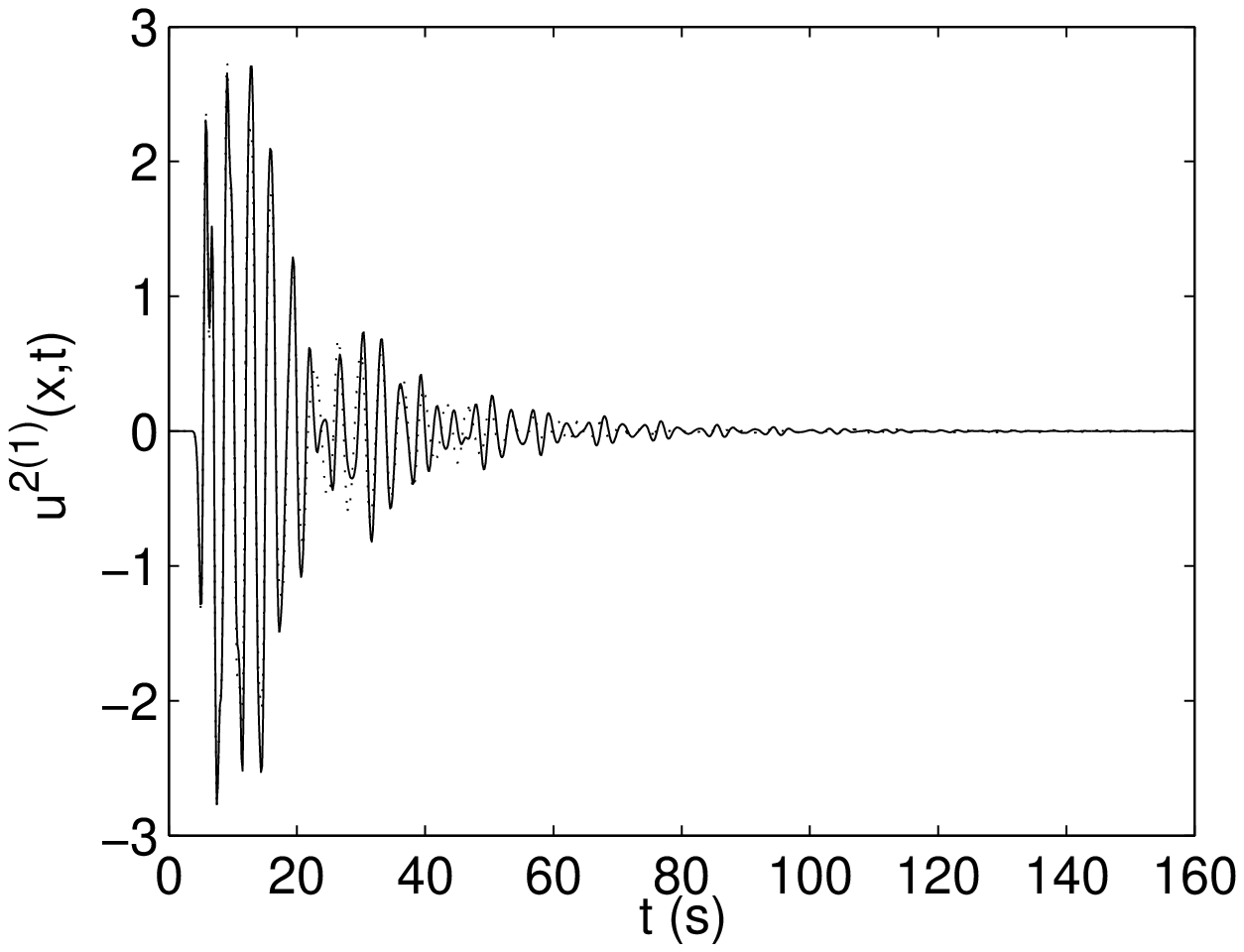}
\includegraphics[width=6.0cm] {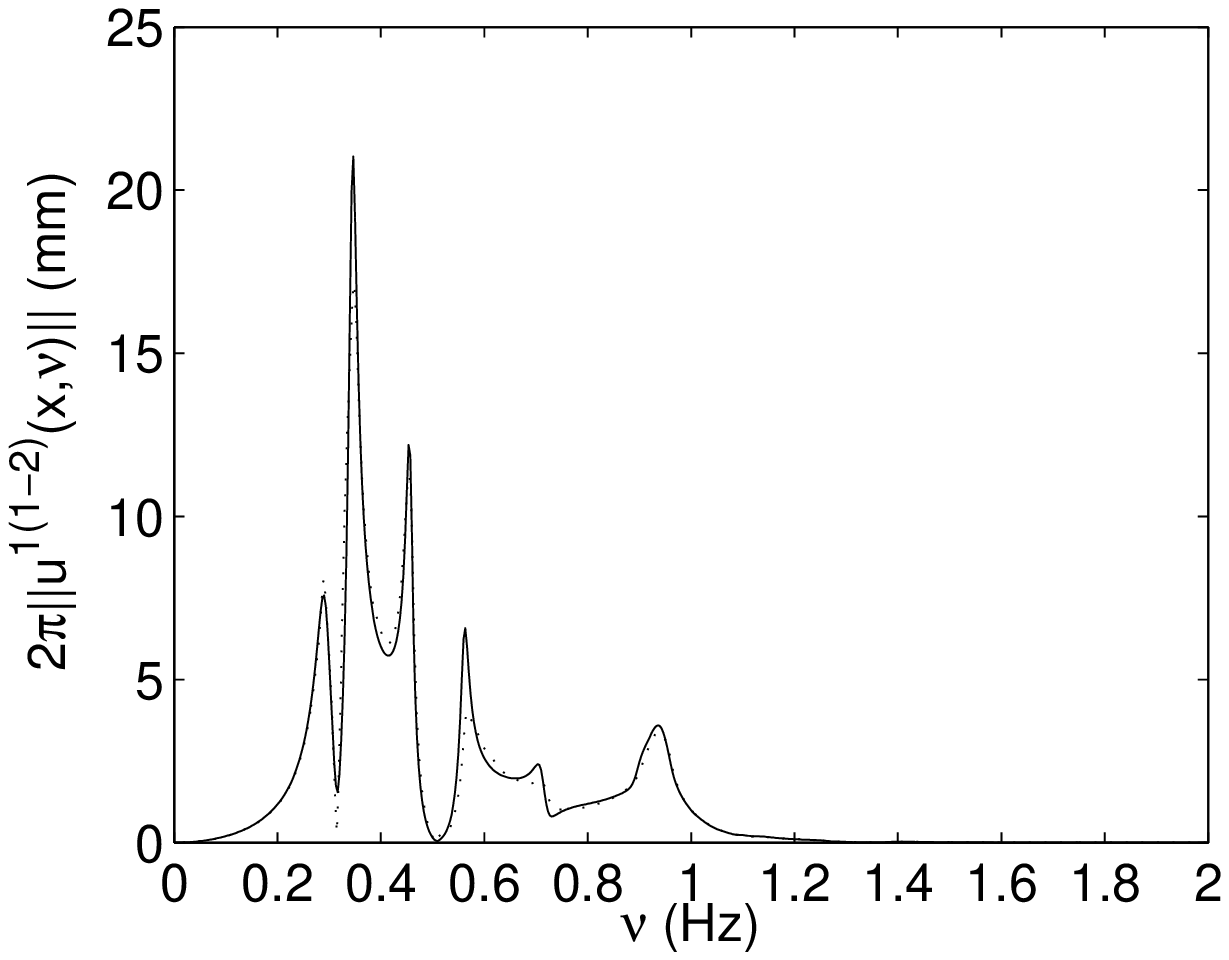}
\includegraphics[width=6.0cm] {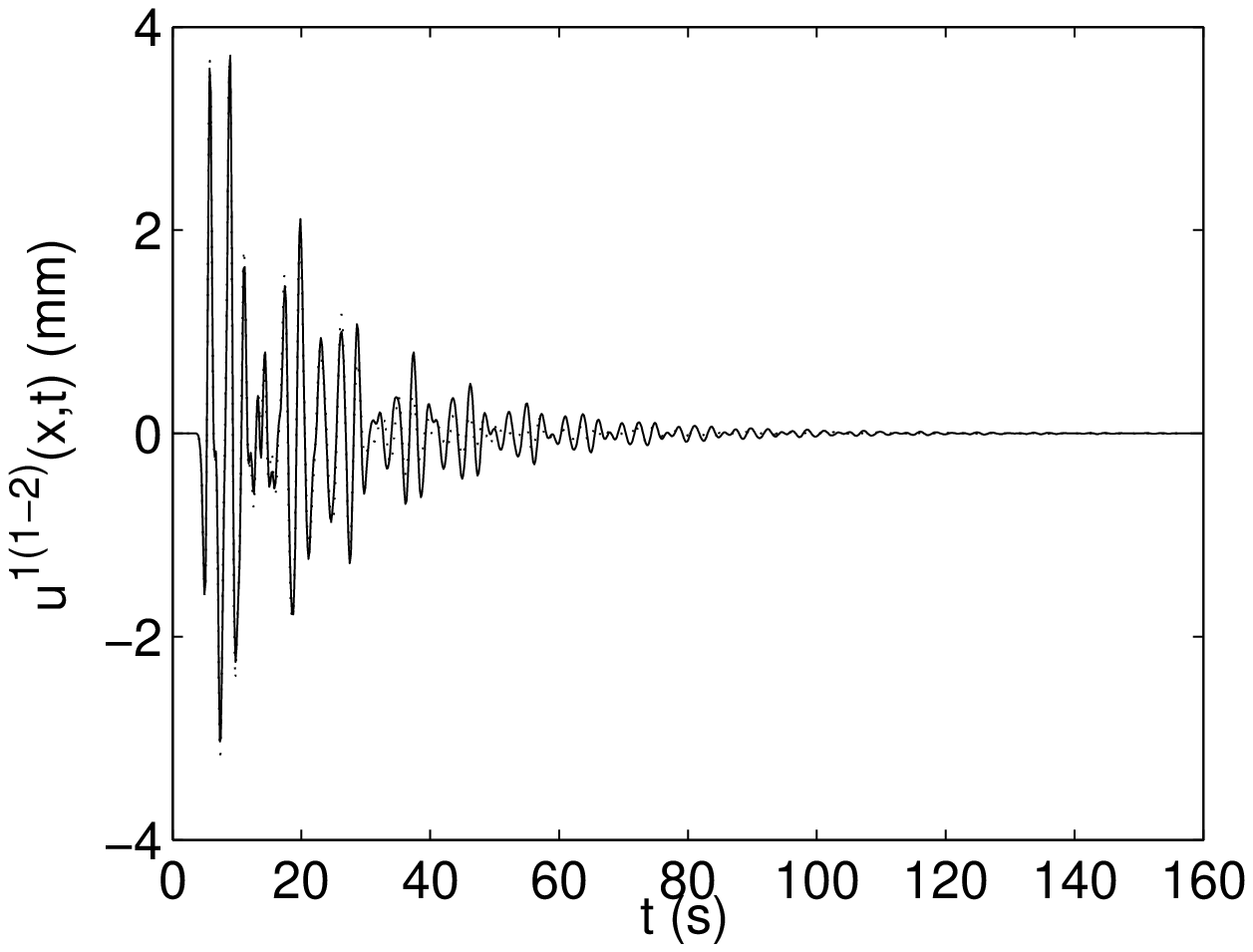}
\caption{Comparison of $2\pi$ times the spectrum (left panels) and
time history (right panels) of the total displacement as
calculated by the MM method for an infinite number of $50m\times
30m$ blocks separated by $d=300m$  (solid curves) with those
obtained by the FEM method in the {\it left-hand edge portion} of
a configuration of $10$ identical $50m\times 30m$ blocks separated
by $d=300m$ (dotted curves): i) at the center of the top segment
of block number 1 (top panels), ii) at the center of the base
segment of block number 1, and iii) on the ground between  block
numbers 1 and 2.} \label{Mexicod300b1}
\end{center}
\end{figure}
\begin{figure}[ptb]
\begin{center}
\includegraphics[width=6.0cm] {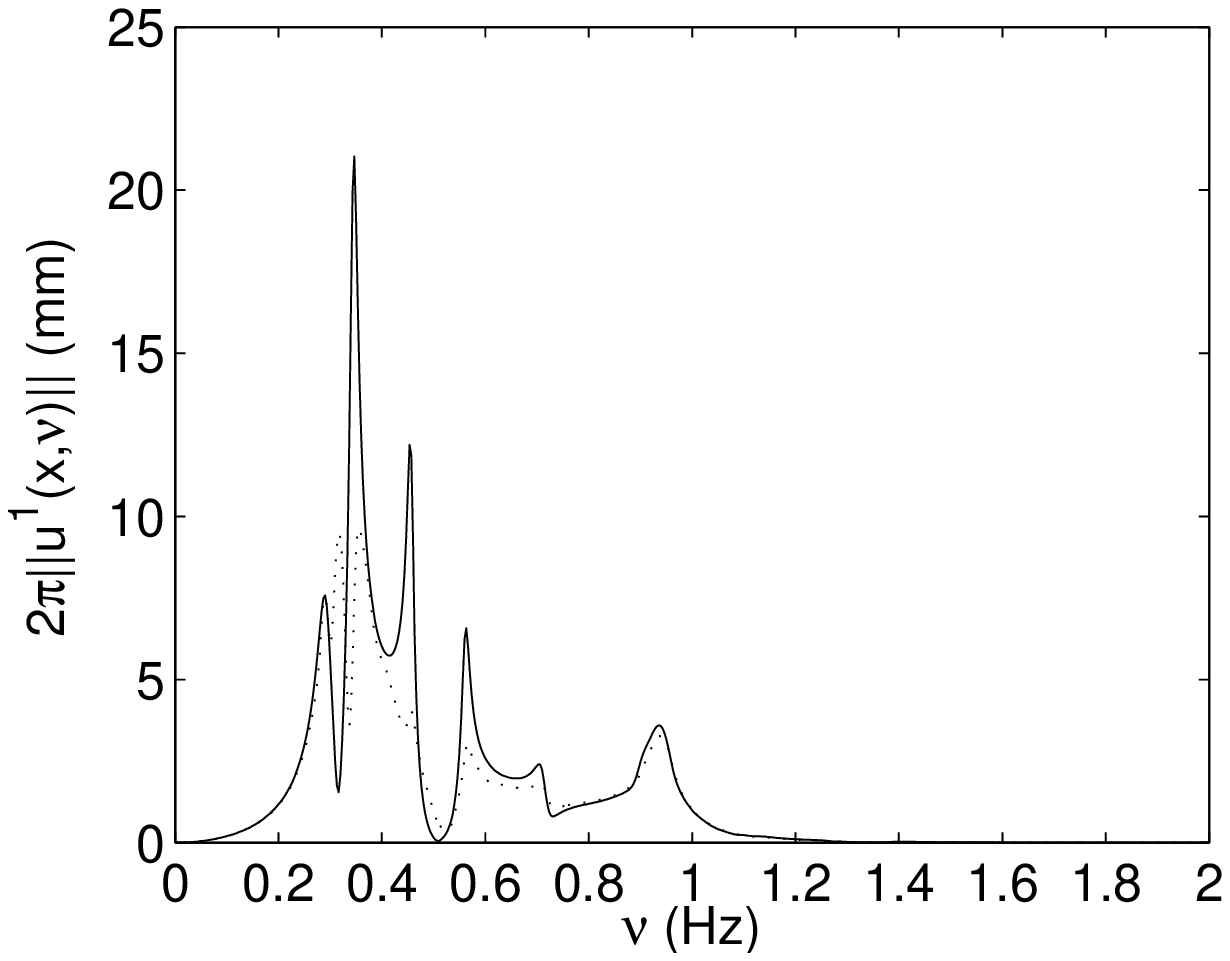}
\includegraphics[width=6.0cm] {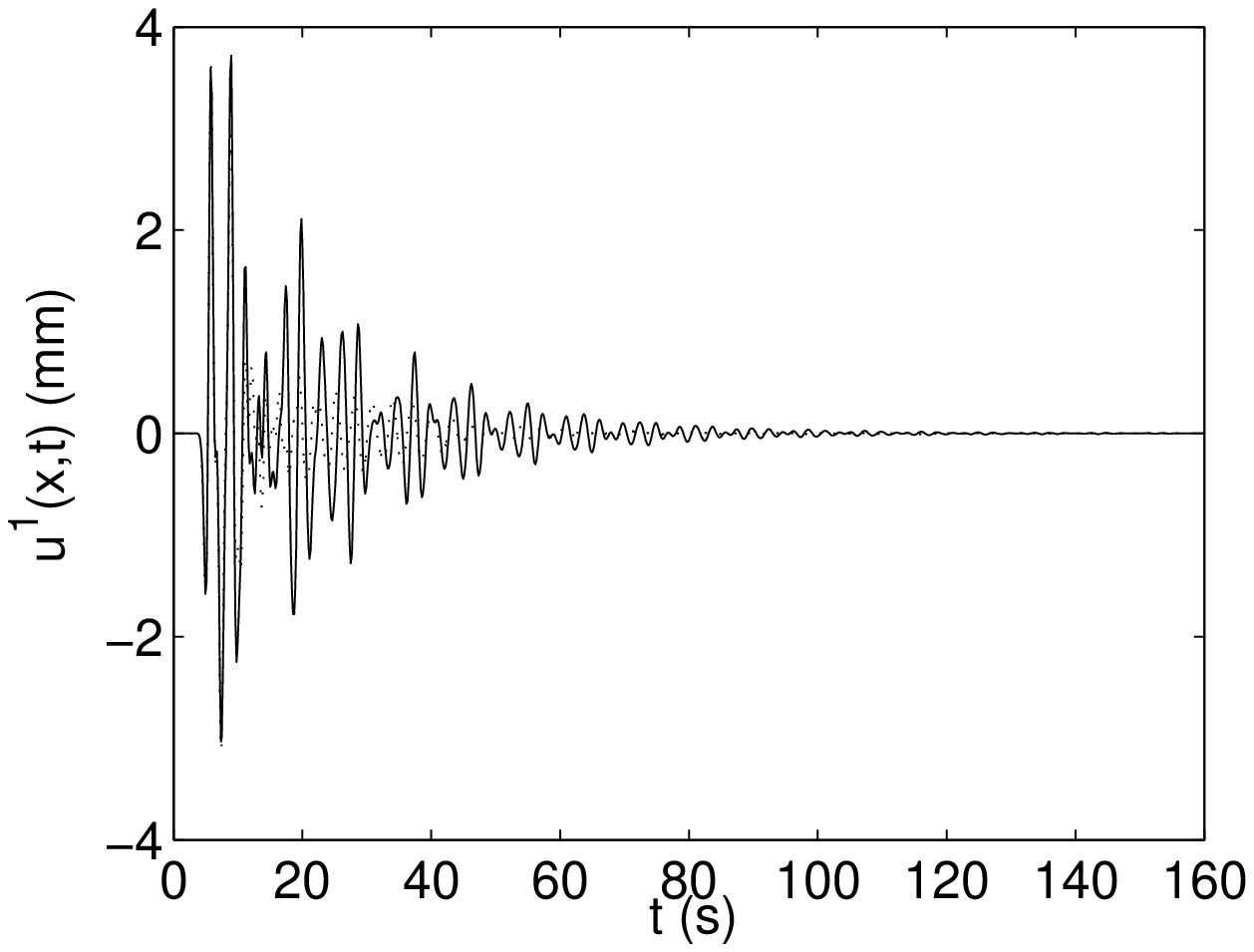}
\caption{Comparison of $2\pi$ times the spectrum (left panels) and
time history (right panels) of the total displacement as
calculated by the MM method for an infinite number of $50m\times
30m$ blocks separated by $d=300m$  (solid curves) with those
obtained by the FEM method in the left-hand edge portion of a
configuration of $10$ identical $50m\times 30m$ blocks separated
by $d=300m$ (dotted curves) at a point {\it on the ground  $150m$
to the left of the block $1$.}} \label{Mexicod300b1side}
\end{center}
\end{figure}

The width of the finite set  of blocks $W$ is  $3000m$ and
therefore closer (than in the previous case) to the infinite width
implicit in the MM theory. This may be the reason why the FE
results agree quite well with the MM results within the city. The
two computational modes match less well  at a point on the ground
outside the city (fig. \ref{Mexicod300b1side}), as one would
expect.

One observes that the response is spatially variable, of long
duration (attaining at some points $\approx 2$ min), with high
maximum and cumulative amplitudes, and characterized by beating
features, which is quite evocative of the  features of many
sismograms of Mexico City earthquakes
\cite{chba94,flno87,sior93,fuke98}. This suggests that rather
largely-spaced  blocks or buildings are more apt than
closely-spaced blocks or buildings to induce the large-scale
features (particularly the very long durations and beatings)
observed in this city. \clearpage
\subsubsection{Illustration of the spatial variablility of
response in a configuration of ten blocks separated by $d=300$m}
Fig. \ref{Mexicod300snap} contains three snapshots, at instants
$t$=25s, 32.5s and 50s, of the displacement field in the entire
configuration of 10 blocks with center-to-center separation of
300m.
\begin{figure}[ptb]
\begin{center}
\includegraphics[width=12.5cm] {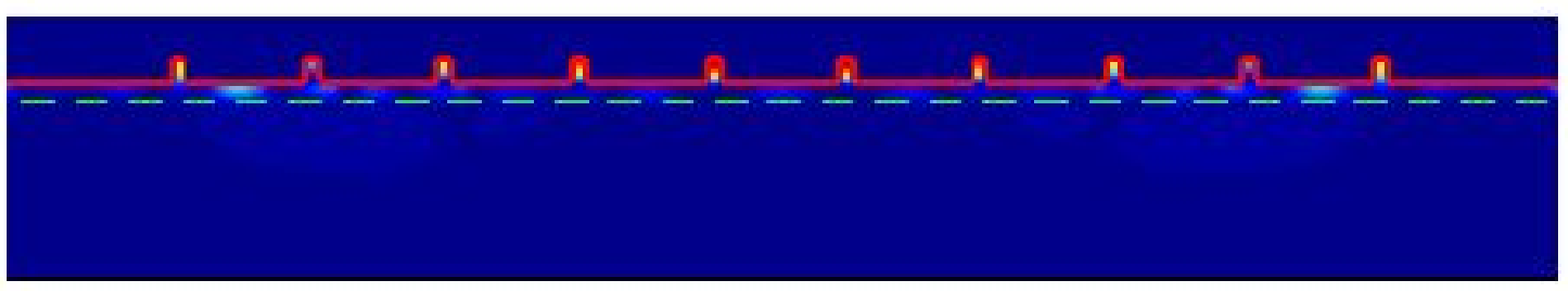}
\includegraphics[width=12.5cm] {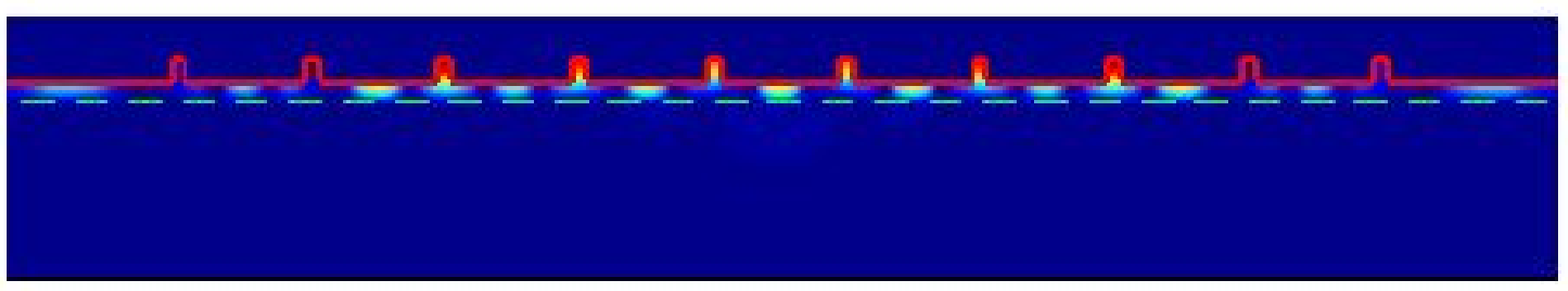}
\includegraphics[width=12.5cm] {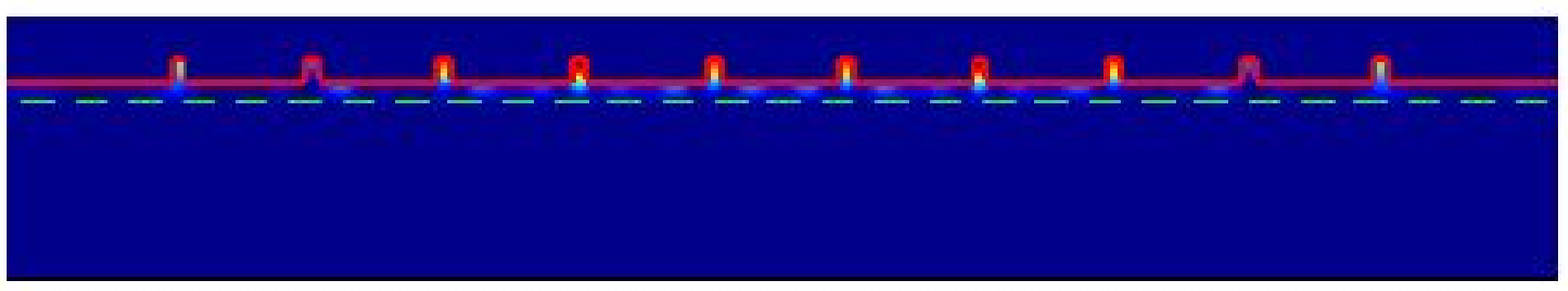}
\caption{Snapshots at various instants, $t=25s$, $t=32.5s$,
$t=50s$, of the total displacement field for $10$ identical
equally-spaced $50m\times30m$ relatively distant blocks
configuration, solicited by a normal incident plane wave. The
center-to-center spacing is $d=300m$.} \label{Mexicod300snap}
\end{center}
\end{figure}
Once again, one notes both the large- and small-scale
spatio-temporal variability of response in the city site.
\section{Conclusions}
The response, to a plane wave initially propagating
 in the substratum, of a finite set of
identical, equi-spaced blocks, each block modeling one, or a group
of buildings, in welded contact with a soft layer overlying a hard
half space, was investigated in a theoretical manner via the mode
matching technique.

The capacity of this technique to account for the complex
phenomena provoked by the presence of blocks on the ground was
demonstrated by comparison of the numerical results to which it
leads to those obtained by a finite element method.

It was shown that the presence of the blocks induces a
modification of the phenomena that are produced by the
configuration without blocks, or of a configuration of closed
blocks disconnected from the geophysical half-space. In
particular, the blocks modify the dispersion relation of what, in
the absence of the blocks, constitutes the Love modes. Moreover,
the periodic nature of the urban site is responsible for the
existence of another vibrational mode which is a variant of the
Cutler mode encountered in electromagnetic waveguide structures.

The dispersion relation for the periodic configurations with an
infinite number of blocks is very complex, but an approximation of
this relation lends itself to a fairly-explicit theoretical
analysis, inspired notably by the method first adopted in the
electromagnetic wave community. The general features of this
dispersion relation, revealed by the theoretical analysis (and
manifested by the existence of several types of vibrational
modes), were shown to actually exist by means of the numerical
study.

The excitation of the vibrational modes was then studied in the
particular case of city districts with 10 and an infinite number
of identical, periodically-disposed blocks. A common feature of
the influence of the blocks is the excitation of the quasi-Love
mode, which occurs even for solicitation by a plane wave (recall
that, for this type of solicitation, it is not possible to excite
ordinary Love modes in a flat ground (i.e., no blocks)/soft
horizontal layer/hard substratum configuration
\cite{grobyetwirgin2005}). The trace of quasi-Love mode excitation
in the frequency domain was shown to be a shift to lower frequency
and an increase of the amplitude of the first (lowest-frequency)
peak of the response.

The change of the phenomena provoked by plane wave solicitation,
from a configuration without blocks (for which there exist only
bulk waves in the geophysical structure), to one with blocks (for
which there exist quasi-Love modes characterized by a field in the
substratum that is predominantly a surface wave in the substratum)
is a manifestation of  the so-called \textit{soil-structure
interaction}.

Multi displacement-free base block modes were shown to be excited
in all the configurations and to correspond to a coupled mode.

The modifications of the response due to the presence of blocks
separated by 65m were found to be rather modest in the Nice-like
site.

The modifications of the frequency-domain response (and less so of
the time-domain response) were found to be fairly substantial for
structures involving blocks separated by 65m in the Mexico
City-like site. In particular,   10 or 20 blocks separated by 65m,
were shown to give rise to anomalous features (amplifications of
peak and cumulative motion, large durations and beatings) that are
even closer to those observed in Mexico City than configurations
with a larger number (40, $\infty$) of blocks.

All the anomalous features found for the Mexico City-like site
with 10 or an infinite number of blocks were found to be closer to
the actually-observed anomalous features observed during
earthquakes in Mexico City when the separation between blocks is
300m rather than 65m. This was found to be due to the fact that
the structure with the larger period enables the quasi-Cutler mode
to make itself felt within the range of frequencies of the source.

The theoretical findings and numerical results of the present
study cannot account for all the anomalous phenomena observed in
many of the earthquakes in Mexico City (notably the
exceptionally-long durations) for the obvious reasons that the
model adopted herein, i.e.,  SH plane  wave solicitation, 2D,
periodic geometries, simple underground (horizontal homogeneous
soft layer of infinite lateral extent overlying a homogeneous
lossless hard substratum), simple homogenized building blocks,
linear soil behavior, etc., is incomplete, and in some respects,
rather far removed from reality. Nevertheless, the results of our
study indicate that it is possible that the excitation of
vibrational modes, whose structure is closely-related to those
described herein, was responsible for at least  part of the
large-scale, anomalous mechanical effects that have caused so much
damage in past earthquakes in urban areas.

The most important finding of this work, which substantiates those
obtained in \cite{wiba96,tswi03,grts05}, is that the presence of
groups of buildings (i.e., city blocks) can modify substantially
the seismic motion in a city. Moreover, provided the blocks are
arranged quasi-periodically with a sufficiently large  period $d$,
the seismic motion is of longer duration, and of higher
 cumulative (sometimes peak) amplitude on the ground (and, of course,
 in the buildings) than when the built structures
 are not present.

Therefore, it is advisable to integrate (as is starting to be done
 \cite{febi06})  the presence, composition and layout,
of buildings, together with, and to the same extent as, the
features of the underground structure and composition, into the
large-scale computer codes that are being employed \cite{wagr98}
to predict the level and durations of shaking in highly-populated,
economically-important, earthquake-prone areas.

\end{document}